\documentstyle[rose,twoside]{article}

\newcommand{\clif}{{\cal C}\!\ell}
\newcommand{\e}{\mbox{e}} 

\makeatletter
\@addtoreset{equation}{section}
\makeatother

\begin{document}

\setcounter{page}{355} 

\thispagestyle{firstheadings} 

\def\firstrightmark{\sit Random Oper. and
Stoch. Equ.,  \srm Vol. 4, No. 4, pp. 355--400  (1996) \\
 \copyright \   VSP 1996}

\mbox{}

\vspace{1in} 

\noindent{\bbf  A unified theory for construction of  
arbitrary speeds ($0
\leq v < \infty$) solutions of the 
relativistic wave equations}

\bigskip 

\noindent{WALDYR A.\ RODRIGUES, Jr.$^{1}$ and  JOS\'E E.\
MAIORINO$^{2}$}

\medskip 

\noindent 
$^1${\msl Instituto de Matem\'atica, Estat\'\i stica e 
Computa\c c\~ao Cient\'{\i}fica \\
IMECC-UNICAMP; CP 6065, 13081-970, Campinas, SP, Brazil} 

\noindent
$^2${\msl Instituto de F\'{\i}sica ``Gleb Wataghin'' \\ 
IFGW-UNICAMP; CP 6165, 13081-970, Campinas, SP, Brazil } 

\medskip 

\noindent {\srm Received for {\ssl ROSE} 14 March 1996} 

\bigskip 

\noindent {\sbf Abstract---}{\srm Representing the relativistic
physical fields as sections of the Clifford Bundle (or of the
Spin-Clifford Bundle) of Minkowski spacetime we show that all the
relativistic wave equations satisfied by these fields possess solutions
traveling with arbitrary speeds $0 \leq v < \infty$. By giving rigorous
mathematical definitions of reference frames and of the Principle of
Relativity (PR) we prove that physical realizations of the $v>1$
solutions of, {\em e.g.\/}, the Maxwell equations imply in a breakdown
of the PR, but in no contradiction at all with known physical facts.  }

\bigskip 

\section{INTRODUCTION}
  
In this paper we present  methods for constructing
solutions with arbitrary speeds ($0\leq v < \infty$)\footnote{We use
units such that $c=1$, where $c$ is the so-called speed of light in
vacuum.} of the main relativistic wave equations of physics, namely the
(scalar) homogeneous wave equation (HWE) and Maxwell, Weyl,
Klein-Gordon and Dirac equations. Several examples are worked in detail. 

In Section \ref{sec1} we show how to represent all fields mentioned
above as sections of the Cliford bundle $\clif (M)$ or of the
Spin-Clifford bundle $\clif_{Spin_+ (1,3)}(M)$ of Minkowski
spacetime \cite{1,2,3}. We then show that if the HWE and the
Klein-Gordon equation (KGE) have solutions with arbitrary speeds $0
\leq v < \infty$, it then follows that also Maxwell, Weyl and Dirac
equations have such solutions.

The solutions we are going to exhibit for the relativistic wave
equation are families of {\em undistorted progressive waves\/}
(UPWs). By UPW, following Courant and Hilbert \cite{4} we mean that
the waves are distortion-free, {\em i.e.\/} that they are
translationally invariant and thus do not spread, or that they
reconstruct their original form after a certain period of time.

In Section \ref{sec2} we present $0 \leq v < \infty$ UPW solutions of
the HWE and of the KGE. The meaning of the velocity of propagation of a
UPW and  the concepts of phase and group velocities are discussed in
details. Misconceptions regarding the velocity of propagation of energy
of a given wave are clarified.  In Section \ref{sec3} we present $v \neq
1$ UPW solutions of Maxwell equations, discussing the remarkable
characteristics of these solutions in contrast to the $v=1$ solutions.
One of the principal novelties is that the $v \neq 1$ solutions have in
general non-null field invariants and are not transverse waves.  We also
discuss in this section the important problem of the velocity of
transport of energy for the $v>1$ UPWs solutions of Maxwell equations.

It must be said that there are experimental data \cite{5,6}  obtained
with
techniques developed by Lu and Greenleaf \cite{7,8}  showing the
existence of pressure waves moving with speeds\footnote{Here $c_s$ is the speed of sound in water.} $v < c_s$ and $v >
c_s$ which
confirm the theoretical predictions, showing in particular that the energy associated with the pressure waves can travel with speed $v_\varepsilon > c_s$. In \cite{6} it is discussed the
possibility of designing physical devices for launching in physical
space the $v \neq 1$ solutions of Maxwell equations.

If this is indeed possible we must investigate the status of the
Principle of Relativity (PR). This is done in Section \ref{sec4},
where we give a thoughtful and rigorous mathematical definition of
this principle and its relation with Lorentz invariance. We prove that
the $v>1$ solutions of Maxwell equations imply necessarily in a
breakdown of the PR, which is necessary in order to avoid logical
contradictions. In Section \ref{sec5} we present our conclusions.

\markboth{W.\ A.\ Rodrigues Jr.\ and J.\ E. Maiorino}{On UPW solutions 
of relativistic wave equations} 

\section{A UNIFIED THEORY FOR CONSTRUCTION OF
UPW
SOLUTIONS OF MAXWELL, DIRAC AND WEYL EQUATIONS} \label{sec1} 

To fix the notations we recall here the main results
concerning the theory of Clifford algebras (and bundles) and their
relationship with the Grassmann algebras (and bundles). In particular a
self-consistent presentation of the so called spacetime and Pauli
algebras
is given. Also the concept of Dirac-Hestenes spinors and their
relationship with the usual Dirac spinors used by physicists is
clarified. We introduce moreover the concept of the Clifford bundle
of spacetime and the Clifford calculus. As we shall
see, this formalism provides a unified theory for the construction of
UPW subluminal, luminal and superluminal solutions of Maxwell, Dirac
and Weyl equations once we have arbitrary speed ($0 \leq v < \infty$)
solutions of the homogeneous wave equation (HWE) and of the
Klein-Gordon equation {KGE}. More details on these topics  can
be found in \cite{1,2,3}.

\subsection{Exterior, Grassmann and Clifford algebras} \label{bs1}

Let $V$ be a $n$-dimensional real\footnote{Here
$I\!\!R$
denotes the real field.} vector space, $V^*$ its dual space, $T^r
V$ the space of $r$-contravariant tensors over $V (r \geq 0, T^0
V \equiv I\!\!R, \, T^1 V = V)$ and let $TV$ be the tensor algebra of
$V$.

We recall that the {\it exterior algebra} of $V$ is the quotient
algebra
\begin{equation} \label{be1}
\mbox{$\bigwedge$} V = T V/J
\end{equation}
where $J$ is the bilateral ideal in $TV$ generated by elements of
the form $u \otimes v + \linebreak v \otimes u, u, v \in V$. The
elements of
$\mbox{$\bigwedge$} V$ will be called multivectors, or multiforms if
they are
elements of $\mbox{$\bigwedge$} V^*$.

Let $\rho : TV \rightarrow \mbox{$\bigwedge$} V$ be the  canonical
projection of
$TV$ onto $\mbox{$\bigwedge$} V$. Multiplication in $\mbox{$\bigwedge$}
V$ will be denoted
as usual by $\wedge : \mbox{$\bigwedge$} V \rightarrow \mbox{$\bigwedge$}
V$ and called exterior (or
wedge or Grassmann) product. We have
\begin{equation} \label{be2}
A \wedge B = \rho(A \otimes B).
\end{equation}
We recall that $\mbox{$\bigwedge$} V$ is a $2^n$-dimensional associative
algebra
with unity.\footnote{\mbox{$\bigwedge V$} is what
old
physics textbooks call the algebra of  antisymmetric tensors.}
In addition it is a $Z$-graded algebra, {\em i.e.\/},
\begin{equation} \label{be3}
\mbox{$\bigwedge$} V = \bigoplus^n_{r=0} \mbox{$\mbox{$\bigwedge$}^r V;
\ \ \ \mbox{$\bigwedge$}^r V\wedge \mbox{$\bigwedge$}^s
V \subset \mbox{$\bigwedge$}^{r+s}$} V \; ;
\end{equation}
$r,s \geq 0$, where $\mbox{$\bigwedge$}^r V = \rho (T^r V)$ is the
${n \choose r}$- dimensional subspace of $r$-vectors,
$\mbox{$\bigwedge$}^0 V = I\!\!R$, $\mbox{$\bigwedge$}^1
V=V$; $ \mbox{$\bigwedge$}^r V = \{\phi\}$ if $r > m$. If $A \in
\mbox{$\bigwedge$}^r V$
for some $r(r=0,\ldots, n)$ then $A$ is said to be homogeneous,
otherwise
it is said to be inhomogeneous. For $A_p \in \mbox{$\bigwedge$}^p V$ and
$B_q \in
\mbox{$\bigwedge$}^q V$ we have
\begin{equation} \label{be4}
A_p \wedge B_q = (-1)^{pq} B_q \wedge A_p .
\end{equation}

Let $(E_1, \ldots, E_n)$ be a basis for $V$. Then a basis for
$\mbox{$\bigwedge$}
V$ is
{\small 
\begin{equation} \label{be5} 
\{1, E_1, \ldots E_n, E_1 \wedge E_2, \ldots, E_1 \wedge E_n, \ldots,
E_{n-1} \wedge
E_n, E_1 \wedge E_2 \wedge E_3, \ldots, E_1 \wedge E_2 \wedge \ldots
\wedge E_n\} 
\end{equation} 
} 
Then, if $A \in \mbox{$\bigwedge$} V$, we can write
\begin{equation}
A = S + A^i E_i + \frac{1}{2!} A^{ij} E_i \wedge E_j + \frac{1}{3!}
A^{ijk} E_i \wedge E_j \wedge E_k + \ldots + P E_1 \wedge E_2 \wedge
\ldots\wedge E_n,
\end{equation}
where $S, A^{ij}, A^{ijk}, \ldots, P \in I\!\!R$ and $A^{ij} =
-A^{ji}$,
etc... The element
\begin{equation}
E_{n+1} = E_1 \wedge E_2 \ldots \wedge E_n
\end{equation}
is called the pseudoscalar of the algebra $\mbox{$\bigwedge$} V$. (The
analogous
element for $\mbox{$\bigwedge$} V^*$ is also called the volume
element).
We define the projector $\langle\rangle_k : \mbox{$\bigwedge$} V
\rightarrow \mbox{$\bigwedge$}^k V$ by
$\mbox{$\bigwedge$} V \ni A \mapsto A_k$, for $A =
{\displaystyle\sum^n_{k=0}} A_k, A_k \in
\mbox{$\bigwedge$}^k V$.

Now let $g \in T^2 V^*$ be a metric for $V$ of signature $(p,q)$,
{\em i.e.\/}, $g : V \times V \rightarrow I\!\!R$ and let $g^{-1} \in T^2 V,
g^{-1} e: V^* \times V^* \rightarrow I\!\!R$ be the metric of the dual
space. If $u,v \in V$ and $\alpha, \beta \in V^*$ such that
$\alpha(u)=1,
\beta(v)=1$ we have
\begin{equation}
g(u,v)=g^{-1} (\alpha, \beta) \; .
\end{equation}
We can use $g$ to induce a scalar product on $\mbox{$\bigwedge$} V$, $G
: \mbox{$\bigwedge$} V
\times \mbox{$\bigwedge$} V \rightarrow I\!\!R$. We define
\begin{equation}
G(A,B) = \det (g(u_i,v_j))
\end{equation}
for homogeneous multivectors $A = u_1 \wedge \ldots \wedge u_r \in
\mbox{$\bigwedge$}^r V$;
 $ B = v_1 \wedge \ldots \wedge v_r \in \mbox{$\bigwedge$}^r V$, $u_i,
 v_j \in \mbox{$\bigwedge$}^1 V$,
 $i,j=1,\ldots,r$.
This scalar product is extended to all $V$ due to linearity and
associativity. $G(A,B) = 0$ if $A \in \mbox{$\bigwedge$}^r V$, $B \in
\mbox{$\bigwedge$}^s V$, $r
\neq s$. When both $A, B \in \mbox{$\bigwedge$}^0 V$, $ G(A,B)$ means
the product
$AB$. The algebra $\mbox{$\bigwedge$} V$ endowed with this scalar
product is
called Grassmann algebra and will be denoted
$\mbox{$\bigwedge$}(V,g)$.

On $\mbox{$\bigwedge$} V$ and $\mbox{$\bigwedge$} (V, g)$ there are two
important involutive
morphisms:

(i) Main automorphism $\widehat A : \mbox{$\bigwedge$} V \rightarrow
\mbox{$\bigwedge$} V$, 
\begin{equation}
\begin{array}{c}
(A\wedge B)\widehat{~} = \widehat A \wedge \widehat B, \ A,B \in
\mbox{$\bigwedge$} V \; ; \\
\widehat A = A \ \ {\rm if} \ \ A \in \mbox{$\mbox{$\bigwedge$}^0$} V,
\ \widehat A = -A
\ \
{\rm if}  \ \  A \in \mbox{$\mbox{$\bigwedge$}^1$} V \; .
\end{array}
\end{equation}

(ii) Reversion $\sim \, : \mbox{$\bigwedge$} V \rightarrow
\mbox{$\bigwedge$} V$,
\begin{equation}
\begin{array}{c}
(A\wedge B)^{\sim} = \widetilde B \wedge \widetilde A, \ A,B \in
\mbox{$\bigwedge$} V \; ; \\
\widetilde A = A \ \ {\rm if} \ \ A \in \mbox{$\bigwedge$}^0 V  \oplus
\mbox{$\bigwedge$}^1 V \; .
\end{array}
\end{equation}
We define also:

(iii) Conjugation: $-$: $\mbox{$\bigwedge$} V \rightarrow
\mbox{$\bigwedge$} V$, 
\begin{equation} 
\overline A=(\widehat
A)^\sim = (\widetilde A)^\wedge, \ \forall A \in \bigwedge V. 
\end{equation} 

We introduce now the important concepts of left contraction $\rfloor
: \mbox{$\bigwedge$} V \times \mbox{$\bigwedge$} V \rightarrow
\mbox{$\bigwedge$} V$ and right contraction $\lfloor :
\mbox{$\bigwedge$} V \times \mbox{$\bigwedge$} V \rightarrow
\mbox{$\bigwedge$} V$ through the definitions
\setcounter{equation}{12}
\begin{equation}
G(A \rfloor B, C) = G(B, \widetilde A \wedge C); \ G(A\lfloor B, C) =
G(A, C
\wedge \widetilde B), \ \forall C \in \mbox{$\bigwedge$} V \; .
\end{equation}
$\rfloor$ and $\lfloor$ satisfy the rules
\begin{equation} \label{be14} 
\begin{array}{l}
1. \ x \rfloor y = x.y = g(x,y); \ x \lfloor y = x.y = g(x,y) \; ; \\
2. \ x \rfloor (A \wedge B) = (x \rfloor A) \wedge B + \widehat A
\wedge (x \rfloor B) \; ;  \\
3. \ (A \wedge B) \lfloor x = A \wedge(B \lfloor x) + (A\lfloor x)
\wedge \widehat B \; ; \\
4. \ (A \wedge B) \rfloor C = A \rfloor (B \rfloor C);\ \ \ A\lfloor (B
\wedge C) = (A\lfloor B)
\lfloor C \; ;
\end{array}
\end{equation}
where $x,y \in \mbox{$\bigwedge$}^1 V$, $A,B,C \in \mbox{$\bigwedge$}
V$.

The notation $A.B$ will be used for contractions when it is clear
from the context which factor is the contractor and which factor is
being contracted. When just one of the factors is homogeneous, it is
understood to be the contractor. When both factors are homogeneous we
agree that the one with the lowest degree is the contractor, so that
for $A \in \mbox{$\mbox{$\bigwedge$}^r$} V, B \in \mbox{$\bigwedge$}^s V$
we have $A.B = A\rfloor B$ if
$r \leq s$, $A.B = A \lfloor B$, if $r\geq s$. From the definitions and
eq.(\ref{be14})
we easily verify that
\begin{equation}
s \rfloor A =0, \ A \lfloor s = 0 \ \ \forall s \in I\!\!R, A \in
\mbox{$\bigwedge$} V \; .
\end{equation}

We are now ready to present the definition of the real {\it Clifford}
algebra ${\cal C} \ell (V, g)$ associated with the pair $(V,g)$. In
order to do that we define the {\it Clifford product} (denoted by
juxtaposition of symbols) between $x \in V$ and $A \in
\mbox{$\bigwedge$} V$ by

\[
xA = x \rfloor A + x \wedge A = x . A + x \wedge A
\]
and extend this product by linearity and associativity to all of
$\mbox{$\bigwedge$} V$.

Equipped with the Clifford product $\mbox{$\bigwedge$} V$ becomes
isomorphic to
the Clifford algebra ${\cal C}\ell (V,g)$.\footnote{
We
can show that ${\cal C}\ell(V,g) = TV/J$ where $J$ is the bilateral
ideal on $TV$ generated by elements of the form $a \otimes b + b
\otimes a
- 2g(a,b), a,b \in V \subset TV $\ \protect\cite{1}.} Observe that
$\mbox{$\bigwedge$} V$ equipped with the exterior product and ${\cal
C}\ell(V,g)$
equipped with the Clifford product are, of course, not isomorphic as
algebras. However, $\mbox{$\bigwedge$} V$ and ${\cal C}\ell (V,g)$ are
isomorphic as
linear spaces over $I\!\!R$.

Consider the basis of $\mbox{$\bigwedge$} V$ given by (\ref{be5}) 
and suppose that 
\begin{equation} \label{be16} 
g(E_i,E_j) = \left\{\begin{array}{ll}
+1, & i=j = 1,\ldots,p ; \\
-1, & i=j=p+1,\ldots,p+q ; \\
0 , & {\rm otherwise}.
\end{array}\right.
\end{equation}

Then it is clear that $E_i E_j = E_i . E_j + E_i \wedge E_j = E_i
\wedge
E_j$ for all $i \neq j$. Since $\mbox{$\bigwedge$} V$ and ${\cal C}\ell
(V,g)$ are
isomorphic as linear spaces we can write for $X \in {\cal C}\ell (V,g)$
\begin{equation}
X = S + X_i E^i + \frac{1}{2} X_{ij} E^i E^j + \frac{1}{3!} X_{ijk} E^i
E^j E^k + \cdots P E^1 E^2 E^3 \ldots E^n,
\end{equation}
where $E^i . E_j = \delta^i_j$ and $\{E^i\}$, $i=1, \ldots, n$ is
called
the reciprocal basis of $V$. Also $S, X_i, X_{ij}, \ldots, P \in
I\!\!R$
and $X_{ij} = -X_{ji}$, etc.

For $g$ of signature $(p,q)$ as in eq.(\ref{be16}) ${\cal C}\ell (V,g)$ is
denoted  ${\cal C}\ell_{p,q}$. Using the projector operator
$\langle\rangle_k$
defined above we can show that the contraction $A_r . B_s$, for $A_r
\in \mbox{$\bigwedge$}^rV \subset {\cal C}\ell_{p,q} $, $ B_s \in
\mbox{$\bigwedge$}^sV \subset {\cal C}\ell_{p,q}$
is
given by
\begin{equation} \label{be18} 
A_r . B_s = \left\{\begin{array}{l}
\langle A_r B_s\rangle_{|r-s|} \ \ {\rm if} \ \ r,s > 0 \\
0 \ \ {\rm if} \ \ r=0 \ \ {\rm or} \ \ s=0
\end{array}\right. \; .
\end{equation}
Eq.(\ref{be18}) defines then an inner product in ${\cal C}\ell_{p,q}$.

We now define the Hodge star operator $\star : {\cal C}\ell (V, g)
\rightarrow
{\cal C}\ell (V,g)$ by
\begin{equation}
\star A = \widetilde A\, E_{n+1}, \quad A \in {\cal C}\ell (V,g).
\end{equation}

A simple calculation shows that $\star|_{\mbox{$\bigwedge$}^p}$ maps
$\mbox{$\bigwedge$}^pV \rightarrow
\mbox{$\bigwedge$}^{n-p} V$ for $p=0,1,\ldots,n$. We observe that ${\cal
C}\ell_{p,q}$ is
a $Z_2$-graded algebra. This means the following. Let
${\cal C}\ell^+_{p,q} ({\cal C}\ell^-_{p,q})$ denote the set of even
(odd)
multivectors of ${\cal C}\ell_{p,q}$, {\em i.e.\/}, elements of
$\mbox{$\bigwedge$}^{2r} V \subset
{\cal C}\ell_{p,q} (\bigwedge^{2r+1} \subset {\cal C}\ell_{p,q})$. We have
${\cal C}\ell^+_{p,q} {\cal C}\ell^+_{p,q} \subset {\cal
C}\ell^+_{p,q}$,
 ${\cal C}\ell^-_{p,q} {\cal C}\ell^-_{p,q} \subset {\cal
 C}\ell^+_{p,q}$,
${\cal C}\ell^+_{p,q} {\cal C}\ell^-_{p,q} \subset {\cal
C}\ell^-_{p,q}$,
${\cal C}\ell^-_{p,q} {\cal C}\ell^+_{p,q} \subset {\cal
C}\ell^-_{p,q}$.  The
${\cal C}\ell^+_{p,q}$ is a sub-algebra of ${\cal C}\ell_{p,q}$, called
the even
sub-algebra of ${\cal C}\ell_{p,q}$. All Clifford algebras ${\cal
C}\ell_{p,q}$
are isomorphic to matrix algebras over the fields $I\!\!R$,
$\;\mbox{{\sf I}}\!\!\!C$ or
$I\!\!H$, respectively the real, complex and quaternion fields. We can
find in \cite{9} tables giving the representations of ${\cal
C}\ell_{p,q}$
as matrix algebras. For what follows we need to know the following
representations:
\begin{equation}
\begin{array}{lcl}
\mbox{Complex numbers}  & -&{\cal C}\ell_{0,1} \simeq \;\mbox{{\sf
I}}\!\!\!C \\
\mbox{Quaternions}  & -&{\cal C}\ell_{0,2} \simeq I\!\!H \\
\mbox{Pauli algebra}  & -&{\cal C}\ell_{3,0} \simeq M_2(\;\mbox{{\sf
I}}\!\!\!C) \\
\mbox{Spacetime algebra}  & -&{\cal C}\ell_{1,3} \simeq M_2(I\!\!H) \\
\mbox{Majorana algebra}  & -&{\cal C}\ell_{3,1} \simeq M_4(I\!\!R) \\
\mbox{Dirac algebra}  & -&{\cal C}\ell_{4,1} \simeq M_4 (\;\mbox{{\sf
I}}\!\!\!C)
\end{array}
\end{equation}

Since it is a theorem that ${\cal C}\ell^+_{p,q} \simeq {\cal
C}\ell_{q,p-1}$
for $p \geq 1$ and ${\cal C}\ell^+_{p,q} \simeq {\cal C}\ell_{p,q-1}$
for $q \geq
1$ we
have the following useful {\it identifications} to be used later:
\begin{equation} \label{be21} 
{\cal C}\ell^+_{1,3} \simeq {\cal C}\ell^+_{3,1} \simeq {\cal
C}\ell_{3,0}; \;
{\cal C}\ell^+_{4,1} \simeq {\cal C}\ell_{1,3}; \; {\cal C}\ell^+_{3,0}
\simeq I\!\!H ;
\;
I\!\!H^+ \simeq \;\mbox{{\sf I}}\!\!\!C .
\end{equation}
A very important result is that the Dirac algebra is the tensor
complexification of both ${\cal C}\ell_{1,3}$ and ${\cal C}\ell_{3,1}$,
{\em i.e.\/},
\begin{equation} \label{be22} 
{\cal C}\ell_{4,1} \simeq \;\mbox{{\sf I}}\!\!\!C \otimes {\cal
C}\ell_{1,3} ; \ \ {\cal C}\ell_{4,1} \simeq
\;\mbox{{\sf I}}\!\!\!C \otimes {\cal C}\ell_{3,1}
\end{equation}
Since it is a well known result that $I\!\!H$ is represented by a
subset
of invertible two by two complex matrices belonging to
$M_2(\;\mbox{{\sf I}}\!\!\!C)$,
eq.(\ref{be21}) and eq.(\ref{be22}) show that ${\cal C}\ell_{1,3}$ has also a
complex
$4
\times 4$ matrix representation which can be made identical to the
$M_4(\;\mbox{{\sf I}}\!\!\!C)$ representation of ${\cal C}\ell_{4,1}$.

Before ending this section we recall that the Clifford product
between two general elements $A, B \in {\cal C}\ell_{p,q}$ can be
written
\begin{eqnarray}
AB &=& \sum_{r,s} \langle A\rangle_r \langle B\rangle_s = \sum_{r,s} A_r
B_s \nonumber \\  
&=& \sum_{r,s} (\langle A_r B_s\rangle_{|r-s|} + \langle A_r
B_s\rangle_{|r-s|+2} +
\cdots + \langle A_r B_s\rangle_{r+s}) .
\end{eqnarray}
We define also the norm of a multivector $A \in {\cal C}\ell_{p,q}$ by
\begin{equation}
|A|^2 = \langle \widetilde A A\rangle_0.
\end{equation}
If $A \in {{\cal C}\!\ell}_{3,0}$ is homogeneous and $|A|^2 \neq 0$,
the inverse of $A$ is\footnote{The calculation of $A^{-1}$ (when it
exists) for a general $A \in {{\cal C}\!\ell}_{p,q} $ is not so
simple.}
\begin{equation}
A^{-1} = \widetilde A/|A|^2, \ \ A^{-1} A = AA^{-1} = 1 .
\end{equation}

\subsection{The spacetime and  Pauli algebras} \label{bs2} 

We call $I\!\!R^{1,3} = (I\!\!R^4,g)$ where $g : I\!\!R^4\times
I\!\!R^4 \rightarrow I\!\!R$ is a
Lorentzian metric of signature $(1,3)$. $I\!\!R^{1,3}$ is called the
Minkowski vector space. Let $\{E_\mu\}$, $\mu = 0,1,2,3$ be a basis of
$I\!\!R^4$; we have
\begin{equation}
g(E_\mu, E_\nu) = \eta_{\mu\nu} = \left\{\begin{array}{l}
+1 \ \ \mu=\nu=0; \\
-1 \ \ \mu=\nu=1,2,3; \\
\ 0 \ \ {\rm otherwise.}
\end{array}\right.
\end{equation}

The fundamental relation generating the {\bf spacetime algebra}
${\cal C}\ell_{1,3}$ is then
\begin{equation} \label{be27}
E_\mu E_\nu + E_\nu E_\mu = 2\eta_{\mu\nu} .
\end{equation}
Eq.(\ref{be27}) is identical to the relation satisfied by the famous
Dirac
(gamma) matrices and indeed we know from Section \ref{bs1}  that the $E_\mu$
have
a complex matrix representation in $M_4(\;\mbox{{\sf I}}\!\!\!C)$.
Naturally, $\dim
{\cal C}\ell_{1,3} = 16$. The pseudoscalar of ${\cal C}\ell_{1,3}$ will
be
denoted by $E_5=E_0 E_1 E_2 E_3$ and $E^2_5 = -1$. $E_5$ anticommutes
with odd multivectors and commutes with even multivectors.
We call $\{E^\mu\}$, $\mu=0,1,2,3$, such that $E^\mu .
E_\nu=\delta^\mu_\nu$ the reciprocal basis to $\{E_\mu\}$.

We call the pair $I\!\!R^{1,3*} = (I\!\!R^4, g^{-1})$ the dual space of
$(I\!\!R^4,g)$ and call $\{\Gamma_\mu\}$, $\mu = 0,1,2,3$ the dual
basis to
$\{E_\mu\}$. Analogously, $\{\Gamma^\mu\}$, $\mu=0,1,2,3$ such that
$g^{-1}(\Gamma^\mu,
\Gamma_\nu) = \delta^\mu_\nu$ is the reciprocal basis to
$\{\Gamma_\mu\}$. The
Clifford algebra associated to $I\!\!R^{1,3*}$ will be denoted
$*{\cal C}\ell_{1,3} \simeq {\cal C}\ell_{1,3} \simeq M_2(I\!\!H)$.
Of course we have the fundamental relation
\begin{equation}
\Gamma^\mu\Gamma^\nu + \Gamma^\nu\Gamma^\mu = 2\eta^{\mu\nu},
\end{equation}
where $\eta^{\mu\nu} = \eta_{\mu\nu}$.

The {\bf Pauli algebra} ${\cal C}\ell_{3,0}$ is the Clifford algebra of
$(I\!\!R^3, g_E)$, {\em i.e.\/} of Euclidean space equipped with the Euclidean
metric
$g_E$. If $\{\vec\Sigma_i\}$, $i=1,2,3$ is an orthonormal basis of
$I\!\!R^3$, {\em i.e.\/} $\vec\Sigma_i . \vec\Sigma_j = g_E
(\vec\Sigma_i,\vec\Sigma_j)=\delta_{ij}$ then the Clifford algebra
${\cal C}\ell_{3,0}$ is generated by the fundamental relation
\begin{equation} \label{be29} 
\vec\Sigma_i \vec\Sigma_j + \vec\Sigma_j \vec\Sigma_i = 2\delta_{ij} .
\end{equation}
$I = \vec\Sigma_1 \vec\Sigma_2 \vec\Sigma_3$ is the pseudoscalar of
${\cal C}\ell_{3,0}$.
 We verify that $I^2=-1$ and that $I$ commutes with all $X \in
{\cal C}\ell_{3,0}$, so that $I$ is like ${ \rm i}=\sqrt{-1}$. A basis for
${\cal C}\ell_{3,0}$ is $(1,\vec\Sigma_i, \vec\Sigma_i \vec\Sigma_j, \
\vec\Sigma_1 \vec\Sigma_2 \vec\Sigma_3)$. Taking into  account that
\begin{equation} \label{be30} 
\star(\vec\Sigma_1\vec\Sigma_2) = \vec\Sigma_2\vec\Sigma_1 I =
\vec\Sigma_3;
 \  \star (\vec\Sigma_1\vec\Sigma_3)=\vec\Sigma_3\vec\Sigma_1 I =
 -\vec\Sigma_2;
 \ \star
(\vec\Sigma_2\vec\Sigma_3) = \vec\Sigma_3 \vec\Sigma_2 I =
\vec\Sigma_1;
\end{equation}
we see that $\forall \ X \in {\cal C}\ell_{3,0}$ can be written as
\begin{equation} \label{be31} 
X = (s+Ip) + (\vec A+I\vec B) \ \ s,p\in I\!\!R; \; \vec A, \vec B \in
I\!\!R^3
;
\end{equation}
{\em i.e.\/}, $X$ is ``formally the sum" of a ``complex number" and a
``complex vector". From eq.(\ref{be21}) we see that ${\cal C}\ell_{3,0}
\simeq
{\cal C}\ell_{1,3}^+$. We can exhibit this isomorphism by identifying
$\Sigma_k = E_k E_0 \simeq \vec \Sigma_k$,  $k=1,2,3$ and where $E^2_0
= 1$ is timelike. We define in ${\cal C}\ell_{3,0}$ the operator of
spatial inversion. For $X\in{\cal C}\ell_{3,0}$ as in eq.(\ref{be31}),
\begin{equation}
\ast : X\mapsto X^* = (s+ Ip) - (\vec A + I \vec B).
\end{equation}

With the above identification $p=p^\mu E_\mu \in
\mbox{$\bigwedge$}^1 (R^{1,3}) \subset {\cal C}\ell_{1,3}$ can be
represented in
${\cal C}\ell^+_{1,3}
\simeq {\cal C}\ell_{3,0}$ by
\begin{equation}
p \mapsto p E_0 = p^0 + p^i \Sigma_i = p^0 + \vec p.
\end{equation}
For $f=\frac{1}2 f^{\mu\nu} E_\mu E_\nu \in \mbox{$\bigwedge$}^2
(I\!\!R^{1,3}) \subset
{\cal C}\ell_{1,3}$ where
\begin{equation}
f^{\mu\nu} = \eta^{\mu\alpha} \eta^{\nu\beta} f_{\alpha\beta} =
\left(\begin{array}{cccc}
  0 & -{\cal E}_1 & -{\cal E}_2 & - {\cal E}_3 \\
{\cal E}_1 &  0   & -{\cal B}_3 & {\cal B}_2 \\
{\cal E}_2 & {\cal B}_3  &  0   & -{\cal B}_1 \\
{\cal E}_3 & -{\cal B}_2 & {\cal B}_1  & 0
\end{array}\right)
\end{equation}
we can write
\[
f=-f^{0i} E_iE_0 + f^{ij} E_iE_j = - f^{0i} \Sigma_i + f^{ij} \Sigma_i
\Sigma_j
\]
and taking into account (\ref{be29}) and (\ref{be30}) we can write 
\begin{equation} \label{be35} 
f = \vec {\cal E} + I\vec {\cal B} \qquad I = \Sigma_1 \Sigma_2
\Sigma_3 .
\end{equation}
Eq.(\ref{be35}) shows that any bivector $\vec F \in \mbox{$\bigwedge$}^2
(I\!\!R^{1,3}) \subset {\cal C}\ell_{1,3}$ can be represented in ${\cal
C}\ell_{3,0}
\simeq {\cal C}\ell^+_{1,3}$ by a sum of a vector and a Pauli bivector
or
a ``complex vector".

We define next the Spin groups ${\rm Spin}_+(3,0) \simeq SU(2)$ and
${\rm Spin}_+(1,3)\simeq SL(2,\;\mbox{{\sf I}}\!\!\!C)$, which are
respectively the covering
groups of
$SO_+(3)$, the special rotation group and $SO_+(1,3) \simeq
{\cal L}^\uparrow_+$, the
restricted orthochronous Lorentz group. We have
\begin{equation} \label{be36a}
\mbox{Spin}_+(3,0) = \{ R \in {\cal C}\ell^+_{3,0} \; |
\quad 
|R|=1\},
\end{equation}
\begin{equation} \label{be36b}
\mbox{Spin}_+(1,3) = \{U \in {\cal C}\ell^+_{1,3} | \quad 
|U|=1\}.
\end{equation}

An arbitrary Lorentz rotation is given 
for $a \in {\cal C}\ell_{1,3}$. 
by $a \mapsto U a \widetilde U =
U a
U^{-1}$, $U \in {\rm Spin}_+ (1,3)$ 
We
can prove that any $U \in {\rm Spin}_+(1,3)$ can be written in the
form $U = \pm \e^f, f \in \mbox{$\bigwedge$}^2 (R^{1,3})$, and the choice
of the sign
can always be positive except when $U = -\e^f$ with $f^2=0$. When
$f^2>0$, $U$ is a boost and when $f^2 < 0$,  $U$ is a spatial
rotation.
We end this Section with  the definitions of minimal left {\it ideals}
of
${\cal C}\ell_{p,q}$ and  of {\it geometrically equivalent}
ideals.

We say that $e \in {\cal C}\ell_{p,q}$ is {\it idempotent} if $\e^2 =
e$;
it is called a {\em primitive} idempotent if it {\it cannot} be written as
a sum of two mutually annihilating idempotents, {\em i.e.\/}, $e \neq e' +
e''$, with $(e')^2 = e'$, $(e'')^2 = e''$, $e' e'' = e''e'=0$.

The sub-algebra $I_e \subset {\cal C}\ell_{p,q}$ is called a {\em left
ideal}
of
${\cal C}\ell_{p,q}$ if $\forall \psi \in I_e$ and $\forall X \in {\cal
C}\ell_{p,q}$
we have
$X\psi \in I_e$ (a similar definition exists for right ideals).

An ideal is said to be {\it minimal} if it contains only {\it
trivial} sub-{\rm i}deals. It can be shown that the minimal left ideals of
${\cal C}\ell_{p,q}$ are of the form ${\cal C}\ell_{p,q}e$, where $e$
is a
primitive idempotent.

Consider now ${\cal C}\ell_{1,3}$ and the orthonormal bases $ \Sigma =
\{E_\mu\}$ and $\dot \Sigma = \{\dot E_\mu\}$ where $\dot E_\mu = U
E_\mu
\widetilde
U$, $U \in {\rm Spin}_+ (1,3)$. We can easily verify that the following
elements are primitive idempotents of ${\cal C}\ell_{1,3}$:
\begin{equation}
\begin{array}{ccc}
e_\Sigma = \frac{1}{2} (1+E_0); &   e_\Sigma' = \frac{1}{2}
(1+E_3 E_0); &  e_\Sigma'' = \frac{1}{2} (1+E_1 E_2 E_3);\\
e_{\dot \Sigma} = \frac{1}{2} (1+\dot E_0) ; & e_{\dot\Sigma}' =
\frac{1}{2} (1+\dot E_3 \dot E_0) ; & \ e_{\dot\Sigma}'' =
\frac{1}{2} (1+\dot E_1 \dot E_2 \dot E_3).
\end{array}
\end{equation}
It is trivial to verify that $e_\Sigma$ and $e_{\dot\Sigma}$ are
related by
\begin{equation} \label{be38}
e_{\dot\Sigma} = U e_\Sigma U^{-1}, \qquad u \in {\rm Spin}_+ (1,3).
\end{equation}
There is no element $U \in {\rm Spin}_+(1,3)$ relating, {\em e.g.\/}
$e_{\Sigma}'$ with $e_\Sigma$. Consider now the ideals $I_\Sigma =
{\cal C}\ell_{1,3} e_\Sigma$ and $I_{\dot\Sigma} = {\cal C}\ell_{1,3}
e_{\dot\Sigma}$. We say that $I_\Sigma$ and $I_{\dot\Sigma}$ are {\it
geometrically equivalent} if $e_\Sigma$ and $e_{\dot\Sigma}$ are
related
by eq.(\ref{be38}). Since $I_\Sigma = {\cal C}\ell_{1,3} e_\Sigma$ and
$I_{\dot \Sigma} =
{\cal C}\ell_{1,3} e_{\dot\Sigma}$ and since ${\cal C}\ell_{1,3} U
\simeq
{\cal C}\ell_{1,3} \ \forall U \in {\rm Spin}_+(1,3)$, we can write
\begin{equation} \label{be39}
I_{\dot\Sigma} = I_\Sigma U^{-1}
\end{equation}
Eq.(\ref{be39}) defines a correspondence between elements of ideals
that are
geometrically equivalent. The quotient set $\{I_\Sigma\}/{\cal R}$
where
${\cal R}$
is
the equivalence relation given by eq.(\ref{be39}) is called the space
of the
Dirac algebraic spinors $\Sigma$. Of course in the basis
$\dot\Sigma$ the spinor is represented by $\psi_{\dot\Sigma}$ and
\begin{equation}
\psi_{\dot\Sigma} = \psi_\Sigma U^{-1}
\end{equation}

Section \ref{bs4}, where we introduce the {\it fundamental} concept of
Dirac-Hestenes spinors, will clarify the meaning of the above
definitions.

\subsection{Dirac algebra ${\cal C}\ell_{4,1}$, its relation
with
${\cal C}\ell_{1,3}$ and Dirac-Hestenes spinors} \label{bs3}

Consider the vector space $I\!\!R^{4,1} = (I\!\!R^5,g)$ and let
$\{E_a\}$,  $a
=
0,1,\ldots,4$ be an orthonormal basis:
\begin{equation}
g(E_a, E_b) = \left\{\begin{array}{cl}
1, & a=b = 1,2,3,4; \\
-1, & a=b = 0; \\
\ 0, & \mbox{otherwise}.
\end{array}\right.
\end{equation}
Let ${\cal C}\ell_{4,1}$ be the Clifford algebra of $I\!\!R^{4,1}$ and
$I =
E_0 E_1 E_2 E_3 E_4$ the corresponding pseudoscalar.
Note that $I^2=-1$ and that
$E_a I = IE_a$, and thus $I\!\!R \oplus II\!\!R$ is the center of
${\cal C}\ell_{4,1}$
and the pseudoscalar $I$ plays therefore the role of the imaginary
unit (as $I$ in the case of the Pauli algebra).

Let us define
\begin{equation} \label{be42}
\Gamma_\mu = E_\mu E_4, \ \ \ \mu=0,1,2,3.
\end{equation}

Then,
\begin{equation} \label{be43}
\Gamma_\mu \Gamma_\nu + \Gamma_\nu \Gamma_\mu = 2\eta_{\mu\nu}.
\end{equation}
One can easily see from (\ref{be42}), (\ref{be43}) and with $I$ playing the role
of the imaginary unity $({\rm i}=\sqrt{-1})$ that ${\cal C}\ell_{4,1}$ is
isomorphic to the complexified spacetime algebra, {\em i.e.\/},
\begin{equation} \label{be44}
{\cal C}\ell_{4,1} \simeq \;\mbox{{\sf I}}\!\!\!C \otimes {\cal
C}\ell_{1,3}.
\end{equation}
Indeed, each $X \in {\cal C}\ell_{4,1}$ can be written
\begin{eqnarray} \label{be45}
 X &=& a+X^a E_a + \frac{1}2 X^{ab} E_a E_b + \frac{1}{3!}
X^{abc} E_a
E_b  E_c + \frac{1}{4!} X^{abcd} E_a E_b E_c E_d + Ib  \nonumber \\
& = &(a+Ib) + X^\mu E_\mu - (Y^\mu E_\mu I) - IX^S E_0 E_1 E_2 E_3 +
\cdots
\nonumber \\
& = &  (a+Ib) + (A^\mu+IB^\mu)\Gamma_\mu + \frac{1}{2!} (A^{\mu\nu} + I
B^{\mu\nu}) \Gamma_\mu \Gamma_\nu + \cdots + p \Gamma_0 \Gamma_1 \Gamma_2
\Gamma_3 \nonumber \\ 
& & 
\end{eqnarray}
 $a,b,X^\mu, Y^\mu, \ldots, B^{\mu\nu}, \ldots \in I\!\!R$.

Moreover, the even sub-algebra of ${\cal C}\ell_{4,1}$ is isomorphic to
${\cal C}\ell_{1,3}$, {\em i.e.\/},
\[
{\cal C}\ell_{4,1}^+ \simeq {\cal C}\ell_{1,3}.
\]
${\cal C}\ell_{4,1}$, the complexified spacetime algebra is the well
known Dirac algebra studied in physics textbooks. Indeed
${\cal C}\ell_{4,1}$ is isomorphic to $M_4(\;\mbox{{\sf I}}\!\!\!C)$,
the algebra of $4\times 4$
matrices over the complex. One representation (the standard one) of
the $\Gamma_\mu$ defined by  eq.(\ref{be42}) is 
\begin{eqnarray} \label{be46} 
&& \Gamma_0 \ \leftrightarrow \ \left(\begin{array}{cccc}
1 & 0 &  0   & 0   \\
0 & 1 &  0   & 0   \\
0 & 0 & $-1$ & 0   \\
0 & 0 &  0   & $-1$
\end{array}\right);  \quad
\Gamma_1 \ \leftrightarrow \ \left(\begin{array}{cccc}
0 & 0 &  0   & $-1$   \\
0 & 0 & $-1$ & 0   \\
0 & 1 &  0   & 0   \\
1 & 0 &  0   & 0
\end{array}\right);  \nonumber \\ 
&& \\ 
&& \Gamma_2 \ \leftrightarrow \ \left(\begin{array}{cccc}
 0  &  0  &  0   & $i$   \\
 0  &  0  & $-{\rm i}$ & 0   \\
 0  &$-{\rm i}$ &  0   & 0   \\
$i$ & 0   &  0   & 0
\end{array}\right);  \quad
\Gamma_3 \ \leftrightarrow \ \left(\begin{array}{cccc}
0 &  0   & $-1$ & 0   \\
0 &  0   &  0   & 1   \\
1 &  0   &  0   & 0   \\
0 & $-1$ &  0   & 0
\end{array}\right). \nonumber 
\end{eqnarray}

Consider the idempotent $f = {\displaystyle\frac{1}2} (1+\Gamma_0)
{\displaystyle\frac{1}{2}} (1+I\Gamma_1\Gamma_2) = e_\Sigma
{\displaystyle\frac{1}2}
(1+I\Gamma_1\Gamma_2)$ where $e_\Sigma = {\displaystyle\frac{1}2} (1 +
\Gamma_0)$ is a
primitive idempotent of ${\cal C}\ell_{4,1}^+ \simeq {\cal
C}\ell_{1,3}$. It
generates the left minimal ideal $I^\Sigma_{4,1} = {\cal C}\ell_{4,1}
f$ and
we can easily verify by explicit computation that
\begin{equation} \label{be47} 
I^\Sigma_{4,1} = {\cal C}\ell_{4,1} f \simeq {\cal C}\ell_{4,1}^+ f.
\end{equation}
Consider now the usual Dirac spinor $|\Psi\rangle \in \;\mbox{{\sf
I}}\!\!\!C^4$. There is an
obvious isomorphism between $\;\mbox{{\sf I}}\!\!\!C^4$ and minimal
left ideals of
$M_4(\;\mbox{{\sf I}}\!\!\!C)$, given by
\begin{equation} \label{be48} 
\;\mbox{{\sf I}}\!\!\!C^4 \ni |\Psi\rangle = \left(\begin{array}{c}
\psi_1 \\ \psi_2 \\ \psi_3 \\ \psi_4 \end{array}\right)
\leftrightarrow  \left(\begin{array}{cccc}
\psi_1 & 0 & 0 & 0 \\
\psi_2 & 0 & 0 & 0 \\
\psi_3 & 0 & 0 & 0 \\
\psi_4 & 0 & 0 & 0
\end{array}\right)
= \Psi \in \ \mbox{minimal left ideal of $M_4(\;\mbox{{\sf
I}}\!\!\!C)$}
\end{equation} 

One can, of course, work with $\Psi$ instead of $|\Psi\rangle$ and
since
$M_4(\;\mbox{{\sf I}}\!\!\!C)$ is a representation of the Dirac algebra
${\cal C}\ell_{4,1}
\simeq \;\mbox{{\sf I}}\!\!\!C \otimes {\cal C}\ell_{1,3}$ we can work
with the corresponding ideal
$I^\Sigma_{4,1}$ of the Dirac algebra. The isomorphisms discussed above
tell us that
\begin{equation} \label{be49} 
I^\Sigma_{4,1} = {\cal C}\ell_{4,1} f = (\;\mbox{{\sf I}}\!\!\!C
\otimes {\cal C}\ell_{1,3}) f \simeq
{\cal C}\ell_{4,1}^+ f \simeq {\cal C}\ell_{1,3} f = ({\cal
C}\ell_{1,3} e_\Sigma)
\frac{1}2 (1+ I \Gamma_1 \Gamma_2).
\end{equation}
Note that in the last equality we have a minimal left ideal
${\cal C}\ell_{1,3} e_\Sigma$ of the spacetime algebra. Moreover we
have
\begin{equation} \label{be50} 
If = \Gamma_2 \Gamma_1 f.
\end{equation}
These results mean that we can work with the ideal ${\cal C}\ell_{1,3}
e_\Sigma$ once we identify $\Gamma_2 \Gamma_1$ as playing in ${\cal
C}\ell_{1,3}$
the role of the imaginary unit. We can verify by explicit
calculation that
\begin{equation}
{\cal C}\ell_{1,3} e_\Sigma = {\cal C}\ell_{1,3}^+ e_\Sigma .
\end{equation}
We see that what the idempotent makes is to ``kill" redundant degrees
of freedom.
Since $\dim {\cal C}\ell_{1,3}^+ = \dim {\cal C}\ell_{3,0} = 8$ we can
work
with ${\cal C}\ell_{1,3}^+$ instead of ${\cal C}\ell_{1,3} e_\Sigma$
(this is not
the case for ${\cal C}\ell_{1,3}$ or $\;\mbox{{\sf I}}\!\!\!C \otimes
{\cal C}\ell_{1,3}$ since $\dim
{\cal C}\ell_{1,3}=16$ and $\dim (\;\mbox{{\sf I}}\!\!\!C \otimes {\cal
C}\ell_{1,3})=32)$. We have thus
established the isomorphism
\begin{equation}
\;\mbox{{\sf I}}\!\!\!C^4 \simeq {\cal C}\ell_{1,3}^+ .
\end{equation}
We shall call $\Psi_\Sigma$ the representative of $|\Psi\rangle$ in
${\cal C}\ell_{4,1}$. It is related to $\psi_\Sigma \in ({\cal
C}\ell_{4,1}^+)^+
\simeq {\cal C}\ell_{1,3}^+$ by
\begin{equation}
\Psi_\Sigma = \psi_\Sigma \frac{1}2 (1+\Gamma_0) \frac{1}{2}
(1+I\Gamma_1
\Gamma_2).
\end{equation}

Such a $\psi_\Sigma$ will be called the ``representative" of a
Dirac-Hestenes spinor in the basis $(\Gamma_0, \Gamma_1, \Gamma_2,
\Gamma_3)$ of
${\cal C}\ell_{4,1}^+ \simeq {\cal C}\ell_{1,3}$. Its standard matrix
representation is
\begin{equation}
\psi_\Sigma \ \leftrightarrow \ \left(\begin{array}{cccc}
\psi_1 & - \psi^*_2 & \psi_3 & \psi^*_4 \\
\psi_2 &   \psi^*_1 & \psi_4 & \psi^*_3 \\
\psi_3 &   \psi^*_4 & \psi_1 & -\psi^*_2 \\
\psi_4 & - \psi^*_3 & \psi_2 & \psi^*_1
\end{array}\right) \; .
\end{equation}

A {\it Dirac-Hestenes} spinor is an element of the quotient set
${\cal C}\ell_{1,3}/{\cal R}$ such that given two orthonormal basis
$\Sigma =
\{\Gamma_\mu\}$, $\dot\Sigma = \{\Gamma_\mu'\}$ of $I\!\!R^{1,3}
\subset
{\cal C}\ell_{1,3}$,
 $\psi_\Sigma \in {\cal C}\ell_{1,3}^+$, then $\psi_\Sigma \sim
 \psi_{\dot\Sigma}
({\rm mod} {\cal R})$ if and only if $\psi_{\dot\Sigma} = \psi_\Sigma
U^{-1}$ with
$\Sigma = {\cal L}(\Sigma)=U \Sigma U^{-1}$, $U \in {\rm Spin}_+
(1,3)$, ${\cal L} \in
SO_+(1,3)$ and ${\cal H}(U)={\cal L}$ where ${\cal H} : {\rm Spin}_+
(1,3) \rightarrow
SO_+(1,3)$ is the universal double covering of $SO_+(1,3)$. We already
said that $\psi_\Sigma$ is the representative of the Dirac-Hestenes
spinors in the basis $\Sigma$. When no confusion arises we shall write
only $\psi$ instead of $\psi_\Sigma$.
$\psi_\Sigma {\displaystyle\frac{1}2} (1+\Gamma_0)$ is the Dirac
algebraic spinor
introduced in Section \ref{bs2}. From now on we work with ${\cal
C}\ell_{1,3}$.
Then
$\psi \in {\cal C}\ell_{1,3}^+$ can be written as
\begin{equation}
\psi = S + F + \Gamma_5 P, \ \Gamma_5=\Gamma_0\Gamma_1\Gamma_2\Gamma_3,
\end{equation}
$S,P \in I\!\!R$, and $F \in \mbox{$\bigwedge$}^2 (R^{1,3}) \subset
{\cal C}\ell_{1,3}$ is a
bivector.

Suppose now that $\psi$ is nonsingular, {\em i.e.\/}, $\psi \widetilde\psi \neq
0$.
Since $\psi \in {\cal C}\ell_{1,3}^+$ we have
\begin{equation}
\psi\widetilde\psi = \sigma + \Gamma_5 \omega, \ \sigma, \omega \in
I\!\!R.
\end{equation}
Define $\rho = \sqrt{\sigma^2+w^2}$, $\tan \beta = \omega/\sigma$. Then we
have
\begin{equation}
\psi = \sqrt{\rho} \e^{\Gamma_5 \beta/2} R,
\end{equation}
where $R \in {\rm Spin}_+(1,3)$, $\rho \in I\!\!R^+$ and $0 \leq \beta <
2\pi$ is the so called Ivon-Takabayasi angle. This is the canonical
decomposition of Dirac-Hestenes spinors and reveals the secret
geometrical meaning of spinors, for if $X \in I\!\!R^{1,3} \subset
{\cal C}\ell_{1,3}$
\begin{equation}
\psi X \widetilde \psi = \rho R X \widetilde R = \rho Y \ \ , \ \  Y =
R X
\widetilde R \in I\!\!R^{1,3} \subset {\cal C}\ell_{1,3} ,
\end{equation}
{\em i.e.\/}, a Dirac Hestenes  spinor acting on a vector produces a Lorentz
rotation plus a dilation of the vector.

A Weyl spinor $\psi \in {\cal C}\!\ell_{1,3} / {\cal R}$ is such that
its representative in a given frame $\Sigma$ satisfy the condition
\cite{10}
\begin{equation} \label{1en7}
\gamma_5 \psi = \pm \psi \gamma_{21} \, .
\end{equation}
Such spinors are called positive and negative eigenstates of $\gamma_5$
and are denoted by $\psi_+$ ($\psi_-$). For a general $\psi \in {\cal
C}\!\ell_{1,3}^{+}$ we can write
\begin{equation} \label{1en8}
\psi_\pm = \frac{1}{2} [ \psi \mp \gamma_5 \psi \gamma_{21} ] .
\end{equation}

\subsection{The Clifford bundle of differential forms and the
Spin-\-Clifford bundle} \label{bs4}

Let ${\cal M}=(M,g,D)$ be Minkowski spacetime, where $(M,g)$ is a four
dimensional time oriented and spacetime oriented Lorentzian manifold,
with $M \simeq I\!\!R^4$ and with $g \in \sec (T^* M \times T^* M)$ being a
Lorentzian metric of signature
$(1,3)$.\footnote{Here sec means Section of a given
bundle.} $T^*M[TM]$ is the cotangent [tangent] bundle.
$T^*M=\cup_{x\in M} T^*_x M \ [TM=\cup_{x\in M} T_x M]$ and $T_x M
\simeq T^*_x M \simeq I\!\!R^{1,3}$, the Minkowski vector space already
defined above. $D$ is the Levi-Civita connection of $g$, {\em i.e.\/}, $D(g) =
0$, $T(D)=0$. Also $I\!\!R(D)=0$, $T$ and $I\!\!R$ being respectively
the
torsion
and curvature tensors. Now, the {\it Clifford bundle} of differential
forms ${\cal C}\ell(M)$ is the vector bundle of algebras ${\cal
C}\ell(M) =
\cup_{x\in
M} {\cal C}\ell(T^*_x M)$ where $\forall x \in M$, ${\cal C}\ell (T^*_x
M) \simeq
{\cal C}\ell_{1,3}$, the spacetime algebra. As a linear space
${\cal C}\ell(T^*_x M)$ is isomorphic to the exterior algebra
$\mbox{$\bigwedge$} (T^*_x
M) \simeq \mbox{$\bigwedge$} (I\!\!R^{1,3*})$ of the space
$I\!\!R^{1,3*}$ dual of
$I\!\!R^{1,3}$. Then the so called Cartan bundle $\mbox{$\bigwedge$}(M)
= \cup_{x\in M}
\mbox{$\bigwedge$} (T^*_x M)$ can be thought as ``embedded" in ${\cal
C}\ell(M)$. In this
way sections of ${\cal C}\ell(M)$ can be represented as a sum of
inhomogeneous differential forms \cite{2,3}.

Let $\{e_\mu\}$, $\mu=0,1,2,3$, $e_\mu \in \sec TM$ be an orthonormal 
basis of $TM$, {\em i.e.\/}, $g(e_\mu, e_\nu) = \eta_{\mu\nu} = diag
(1,-1,-1,-1)$ and let its dual basis be $\{\gamma^\mu\}$, $\mu=0,1,2,3$,
$\gamma^\mu \in \sec \mbox{$\bigwedge$}^1(M) \subset \sec {\cal
C}\ell(M)$. Then if $g^{-1} \in \sec (TM \times TM)$ is the metric on
$T^* M$, we have $g^{-1} (\gamma^\mu, \gamma^\nu) = \eta^{\mu\nu} = diag
(1,-1,-1,-1)$. The fundamental Clifford product is generated by
\begin{equation}
\gamma^\mu \gamma^\nu + \gamma^\nu \gamma^\mu = 2\eta^{\mu\nu}.
\end{equation}
We introduce also the reciprocal basis $\{\gamma_\mu\}$, $\mu =
0,1,2,3$,
 $\gamma_\mu . \gamma^\nu = \delta^\nu_\mu$, $\gamma_\mu \in \sec
 \mbox{$\bigwedge$}^1(M) \subset
 \sec
{\cal C}\ell(M)$. Then ${\cal C} \in \sec {\cal C}\ell(M)$ can be
written as 
\begin{equation} \label{be63} 
{\cal C} = s + v^\mu \gamma_\mu + \frac{1}{2} b^{\mu\nu} \gamma_\mu
\gamma_\nu +
\frac{1}{3!} a^{\mu\nu \rho} \gamma_\mu \gamma_\nu \gamma_\rho + p
\gamma_5,
\end{equation}
where $\gamma_5 = \gamma_0 \gamma_1 \gamma_2 \gamma_3$ and $v$, $v^\mu$,
$b^{\mu\nu}$,
 $a^{\mu\nu\rho}$, $\rho \in \sec \mbox{$\bigwedge$}^0(M) \subset \sec
 {\cal C}\ell (M)$.

Besides ${\cal C}\ell(M)$ we need to introduce another vector bundle, 
${\cal C}\ell_{{\rm Spin}_+(1,3)} (M)$, called the Spin-Clifford
bundle \cite{2,3}, which is a quotient bundle,{\em i.e.\/}, ${\cal
C}\ell_{{\rm
Spin}_+(1,3)}
= {\cal C}\ell(M)/{\cal R}$. This means that the sections of ${\cal
C}\ell_{{\rm
Spin}_+(1,3)}
(M)$ are some special equivalence classes of sections of the Clifford
bundle, {\em i.e.\/}, they are equivalence sections of non-homogeneous
differential forms. A given Section $\psi \in \sec
{\cal C}\ell_{{\rm Spin}_+(1,3)}$ is then represented by $\psi_\Sigma$,
$\psi_{\dot\Sigma}, \ldots \in$ $\sec {\cal C}\ell(M)$ where $\Sigma,
\dot \Sigma,
\ldots$
 are  orthonormal bases of $\mbox{$\bigwedge$}^1(M) \subset {\cal
 C}\ell(M)$, with 
\begin{equation} \label{be64} 
\psi_{\dot\Sigma} = \psi_\Sigma R
\end{equation}
and $\forall x \in M$, $R(x) \in {\rm Spin}_+(1,3)$.

Dirac-Hestenes spinor fields (DHSF) are sections of
${\cal C}\ell_{{\rm Spin}_+(1,3)}^+ (M)$, the even sub-bundle of
${\cal C}\ell_{{\rm Spin}_+(1,3)} (M)$. The representative of a DHSF on
${\cal C}\ell(M)$ in the basis $\Sigma$ is then 
\begin{equation} \label{be65} 
\psi_\Sigma = s + \frac{1}2 b^{\mu\nu} \gamma_\mu \gamma_\nu + p
\gamma_5
\end{equation}
The Hodge star map $\star : \bigwedge^p(M) \rightarrow
\bigwedge^{4-p}(M)$ can be
represented in ${\cal C}\ell(M)$ by the following algebraic operation: 
\begin{equation} \label{be66} 
\star A = \widetilde A \gamma_5 ,
\end{equation}
for $A \in \sec \mbox{$\bigwedge$}^p(M) \subset {\cal C}\ell(M)$.

Let ${\rm d}$ and $\delta$ be respectively the {\it differential} and {\it
Hodge codifferential} operators acting on sections of
$\mbox{$\bigwedge$}(M) \subset
{\cal C}\ell(M)$. We have
\[
{\rm d} : \mbox{$\bigwedge$}^p(M) \rightarrow \mbox{$\bigwedge$}^{p+1} (M).
\]
If $\sec \mbox{$\bigwedge$}^p(M) \ni \omega_p =
{\displaystyle\frac{1}{p!}} \omega_{\alpha\beta} \ldots
\gamma^\alpha \wedge \gamma^\beta \wedge \ldots$ then, 
\begin{equation} \label{be67} 
{\rm d} \omega_p = \frac{1}{p!} e_\mu(\omega_{\alpha\beta} \ldots) \gamma^\mu
\wedge \gamma^\alpha \wedge
\gamma^\beta \wedge  \ldots
\end{equation}
and ${\rm d}^2=0$. Also,  
\begin{equation} \label{be68} 
\delta : \mbox{$\bigwedge$}^p(M) \rightarrow \mbox{$\bigwedge$}^{p-1}
(M), \quad
\delta \omega_p = (-)^p \star^{-1} {\rm d} \star \omega_p,
\end{equation}
where $\star\star^{-1} = \star^{-1}\star = \ \mbox{identity and
$\delta^2 =0$}$.

The Dirac operator acting on sections of ${\cal C}\ell(M)$ is the
invariant first order differential operator $\partial: {\cal C}\ell(M)
\rightarrow
{\cal C}\ell(M)$ 
\begin{equation} \label{be69} 
\partial = \gamma^\mu D_{e_\mu}
\end{equation}
and it holds the very important result (see {\em e.g.\/} \cite{1}) 
\begin{equation} \label{be70} 
\partial = \partial \wedge + \partial . = {\rm d}-\delta
\end{equation}

For $\omega_p \in \sec \mbox{$\bigwedge$}^p (M) \subset \sec {\cal
C}\ell(M)$ 
\begin{eqnarray} \label{be71} 
\partial \omega_p = \gamma^\mu D_{e_\mu} \omega_p & = & \gamma^\mu
\wedge (D_{e_\mu} \omega_p)
+
\gamma^\mu . (D_{e_\mu} \omega_p) \nonumber \\
& = & \partial \wedge \omega_p + \partial . \omega_p
\end{eqnarray}
The operator $\Box = \partial\partial = ({\rm d}-\delta)({\rm d} -\delta) =
-({\rm d}\delta+\delta {\rm d})$ is
called Hodge Laplacian.

\subsection{Maxwell theory in ${\cal C}\ell(M)$ and the Hertz
potential} \label{bs5}

We shall need the concepts of inertial reference frames ($I$), observers
and naturally adapted coordinate systems.

Let ${\cal M} = (M,g,D)$ be Minkowski spacetime. An {\it inertial
reference frame} (irf) $I$ is a timelike vector field $I \in \sec
TM$ pointing into the future such that $g(I,I)=1$ and $DI=0$. Each
integral line of $I$ is called an inertial {\it observer}.
The coordinate functions  $\langle x^\mu\rangle, \ \mu = 0,1,2,3$
of a chart of the maximal atlas of $M$ are said to be
a naturally adapted coordinate system to $I$ (nacs/$I$)  if $I =
\partial/\partial x^0$ \cite{11,12}. Putting $I = e_0$ we can find $e_i
= \partial/\partial
x^i, i=1,2,3$
such that $g(e_\mu, e_\nu)=\eta_{\mu\nu}$ and the coordinate functions
$x^{\mu}$ are the usual Einstein-Lorentz ones and have a precise
operational meaning: $x^0=ct$,\footnote{Here $c$ is the constant called
velocity of light in vacuum. In view of the superluminal and subluminal
solutions of Maxwell equations found in this paper we don't think the
terminology to be still satisfactory.} where $t$ is measured
by ``ideal clocks" at rest on $I$ and synchronized ``\`a la Einstein",
$x^i, i=1,2,3$ are determined with ideal rules \cite{13}. (We
use units where $c=1$.) 

Let $e_\mu \in \sec TM$ be an orthonormal basis $g(e_\mu,e_\nu) =
\eta_{\mu\nu}$ and $e_\mu=\partial/\partial x^\mu (\mu,\nu=0,1,2,3)$.
$e_0$
determines an IRF. Let $\gamma^\mu \in \sec \mbox{$\bigwedge$}^2(M)
\subset \sec
{\cal C}\ell(M)$ be the dual basis and let $\gamma_\mu = \eta_{\mu\nu}
\gamma^\nu$ be the reciprocal basis to $\gamma^\mu$, {\em i.e.\/}, $\gamma^\mu
.
\gamma_\nu = \delta^\mu_\nu$. We have $\gamma^\mu = {\rm d}x^\mu$.

As is well known the electromagnetic field is represented by a
two-form $F \in \sec \mbox{$\bigwedge$}^2(M) \subset \sec {\cal
C}\ell(M)$. We have
\begin{equation} \label{be72} 
F = \frac{1}2 F^{\mu\nu} \gamma_\mu \gamma_\nu, \ F^{\mu\nu} =
\left(\begin{array}{cccc}
 0  & -E^1 & -E^2 & -E^3 \\
E^1 &  0   & -B^3 & B^2 \\
E^2 & B^3  & 0    & -B^1 \\
E^3 & -B^2 & B^1  &  0
\end{array}\right) ,
\end{equation}
where $(E^1, E^2, E^3)$ and $(B^1, B^2, B^3)$ are respectively the
Cartesian components of the electric and magnetic fields.

Let $J \in \sec \mbox{$\bigwedge$}^1 (M) \subset \sec {\cal C}\ell(M)$
be such that
\begin{equation} \label{be73} 
J = J^\mu \gamma_\mu = \rho \gamma_0 + J^1 \gamma_1 + J^2 \gamma_2 +
J^3 \gamma_3 ,
\end{equation}
where $\rho$ and $(J^1, J^2, J^3)$ are the Cartesian components of the
charge and (3-dimensional) current densities. Recalling the
definition of the operators ${\rm d}$ (eq.(\ref{be67})) and $\delta$ (eq.(\ref{be68})) we
see
that we can write Maxwell equations as
\begin{equation} \label{be74} 
{\rm d}F = 0, \ \delta F = -J.
\end{equation}

Since ${\rm d}F$ and $\delta F$ are sections of ${\cal C}\ell(M)$ we can add
the
two equations in eq.(\ref{be74}) and get

\[
({\rm d}-\delta)F = J.
\]

But from eq.(\ref{be69}), ${\rm d}-\delta=\partial$, the Dirac operator acting on
sections
of ${\cal C}\ell(M)$, and we get
\begin{equation} \label{be75} 
\partial F = J
\end{equation}
which may now be called {\it Maxwell equation}, instead of Maxwell
equations.

We now write Maxwell equation in ${\cal C}\ell^+(M)$, the even
sub-algebra
of ${\cal C}\ell(M)$. The typical fiber of ${\cal C}\ell^+(M)$, which
is a
vector bundle, is isomorphic to
the Pauli algebra (see Section \ref{bs4}).

We put
\begin{equation} \label{be76} 
\vec\sigma_i = \gamma_i\gamma_0 , \ {\bf i} = \vec\sigma_1 \vec\sigma_2
\vec\sigma_3 =
\gamma_0 \gamma_1 \gamma_2 \gamma_3 = \gamma_5 .
\end{equation}
Recall that ${\bf i}$ commutes with bivectors and since ${\bf i}^2=-1$ it
acts like the imaginary unit ${\rm i}=\sqrt{-1}$ in ${\cal C}\ell^+(M)$. From
eq.(\ref{be69}), using eq.(\ref{be35}) we get
\begin{equation} \label{be77} 
F = \vec E + {\bf i} \vec B
\end{equation}
with $\vec E = E^i \vec\sigma_i$, $ \vec B = B^j \vec\sigma_j$,
$i,j=1,2,3$.

Now, since $\partial=\gamma_\mu\partial^\mu$ we get $\partial\gamma_0 =
\partial/\partial x^0 +
\vec\sigma_i \partial^i = \partial/\partial x^0 - \nabla$. Multiplying
eq.(\ref{be72}) on the
right by $\gamma_0$ we have 

\[
\partial\gamma_0 \gamma_0 F \gamma_0 = J \gamma_0 ,
\]
\begin{equation} \label{be78} 
(\partial/\partial x^0 - \nabla)(-\vec E+{\bf i}\vec B) = \rho + \vec J,
\end{equation}
where we used $\gamma^0 F \gamma_0 = -\vec E + {\bf i} \vec B$ and $\vec
J = J^i
\vec \sigma_i$.

From eq.(\ref{be78}) we have
\begin{eqnarray*}
&& - \partial_0 \vec E + {\bf i} \partial_0 \vec B + \nabla \vec E - {\bf
i} \nabla \vec B =
\rho
+ \vec J \\
&& - \partial_0 \vec E + {\bf i} \partial_0 \vec B + \nabla .\vec E +
\nabla \wedge \vec E -
{\bf i} \nabla . \vec B -
{\bf i}\nabla \wedge \vec B = \rho + \vec J
\end{eqnarray*}

Now we have
\begin{equation} \label{be79} 
-{\bf i} \nabla \wedge \vec A \equiv \nabla \times \vec A
\end{equation}
since the usual vector product between two vectors $\vec a = a^i
\vec\sigma_i$, $\vec b = b^i \vec\sigma_i$ can be identified  with the
dual
of the bivector $\vec a \wedge \vec b$ through the formula $\vec a
\times \vec b = -{\bf i} (\vec a \wedge \vec b)$. Observe that in this
formalism $\vec a \times \vec b$ is a true vector and not the
nonsense pseudo vector of the Gibbs vector calculus. Using eq.(\ref{be76})
and equating the terms with the same grade we have
\begin{equation} \label{be80} 
\begin{array}{c}
\nabla . \vec E = \rho \; ; \ \ \  \nabla \times  \vec B - \partial_0
\vec E =
\vec J \; ;  \\
\\
\nabla \times \vec E + \partial_0 \vec B = 0 \; ; \ \ \  \nabla . \vec
B = 0 \;
;
\end{array}
\end{equation}
which are Maxwell equations in the usual vector notation.

We now introduce the concept of Hertz potential \cite{14} which permits
us to find  nontrivial solutions of the free ``vacuum" Maxwell
equation
\begin{equation} \label{be81} 
\partial F = 0
\end{equation}
once we know nontrivial solutions of the scalar wave equation, 
\begin{equation} \label{be82} 
\Box \Phi = (\partial^2/\partial t^2 - \nabla^2) \Phi = 0 ; \ \Phi  \in
\sec
\mbox{$\mbox{$\bigwedge$}^0$} (M) \subset \sec {\cal C}\ell (M) \; .
\end{equation}

Let $A \in \sec \mbox{$\bigwedge$}^1(M) \subset \sec {\cal C}\ell(M)$ be
the vector
potential. We fix the Lorentz gauge, {\em i.e.\/}, $\partial. A = - \delta A
=0$
such that $F = \partial A = ({\rm d}-\delta)A = {\rm d}A$. We have the following
\\

{\sc Theorem.} {\sl Let $\pi \in \sec \mbox{$\bigwedge$}^2(M)
\subset \sec {\cal C}\ell(M)$ be
the so called Hertz potential. If $\pi$ satisfies the wave equation,
{\em i.e.\/}, $\Box\pi=\partial^2 \pi =
({\rm d}-\delta)({\rm d}-\delta)\pi=-({\rm d}\delta+\delta {\rm d})\pi=0$ and
if we take $A = -\delta\pi$, then $F = \partial A$ satisfies the
Maxwell
equation $\partial F = 0$.} \\

{\it Proof.\/} $A = -\delta\pi$ implies that $\delta A = -\delta^2
\pi=0$ and $F=\partial A = {\rm d}A$. Then $\partial F =({\rm
d}-\delta)({\rm d}-\delta)A = \delta {\rm d}(\delta\pi) = -\delta^2 {\rm d}\pi=0$,
since $\delta {\rm d}\pi=-{\rm d}\delta\pi$ from $\partial^2\pi=0$.

From the above we see that if $\Phi \in \sec \mbox{$\bigwedge$}^0 (M)
\subset \sec
{\cal C}\ell(M)$ satisfies $\partial^2 \Phi=0$, then we can find a non
trivial
solution of $\partial F = 0$, using a Hertz potential given, {\em e.g.\/}, by
\begin{equation} \label{be83} 
\pi = \Phi \gamma_1\gamma_2 \; .
\end{equation}
In Section \ref{3s3} this equation is used, {\em e.g.\/} to generate the
superluminal electromagnetic $X$-wave.

We now express the Hertz potential and its relation with the $\vec E$
and $\vec B$ fields, in order for our reader to see more familiar
formulas. We write $\pi$ as sum of electric and magnetic parts, {\em i.e.\/}, 
\begin{equation} \label{be84} 
\begin{array}{c}
\pi = \vec\pi_e + {\bf i} \vec\pi_m \\
\\
\vec\pi_e = - \pi^{0i} \vec\sigma_i , \ \vec\pi_m = -\pi^{23}
\vec\sigma_1 + \pi^{13} \vec\sigma_2 - \pi^{12} \vec\sigma_3
\end{array}
\end{equation}
Then, since $A = \partial\pi$ we have
\begin{eqnarray*}
A & = & \frac{1}2 (\partial\pi-\pi\stackrel{\leftarrow}{\partial}) \\
A\gamma_0 & = & -\partial_0 \vec\pi_e + \nabla . \vec \pi_e - (\nabla
\times
\vec\pi_m)
\end{eqnarray*}
and since $A = A^\mu \gamma_\mu$ we also have

\[
A^0 = \nabla . \vec\pi_e \; ; \qquad \vec A = A^i \vec\sigma_i =
-\frac{\partial}{\partial x^0} \vec\pi_e - \nabla \times \vec\pi_m \;
.
\]
Since $\vec E = -\nabla A^0 - {\displaystyle\frac{\partial}{\partial
x^0}} \vec A, \
\vec B = \nabla \times \vec A$, we obtain
\begin{eqnarray*}
&& \vec E = - \partial_0(\nabla \times \vec \pi_m) + \nabla \times
\nabla \times
\vec \pi_e \; ; \\
&& \vec B = \nabla \times (-\partial_0 \vec\pi_e - \nabla \times \vec
\pi_m) =
-\partial_0(\nabla \times \vec\pi_e) - \nabla \times \nabla \times
\vec\pi_m \; .
\end{eqnarray*}
We define $\vec E_e, \vec B_e, \vec E_m, \vec B_m$ by  
\begin{equation} \label{be85} 
\begin{array}{ll}
\vec E_e = \nabla \times \nabla \times \vec\pi_e \, ; & \vec B_e =
-\partial_0(\nabla \times \vec \pi_e) \, ; \\ 
\vec E_m = -\partial_0(\nabla \times \vec\pi_m) \, ; & \vec B_m =
-\nabla
\times
\nabla  \times \vec \pi_m \, .
\end{array}
\end{equation}

We now introduce the 1-forms of stress-energy. Since $\partial F=0$ we
have $\widetilde F \widetilde \partial = 0$. Multiplying the first
equation on the
left by $\widetilde F$ and the second on the right by $\widetilde F$
and summing
we have:
\begin{equation} \label{be86} 
 1/2(\widetilde F \partial F + \widetilde F \widetilde \partial F) =
 \partial_\mu ((1/2) \widetilde F \gamma^\mu
 F )=  \partial_\mu T^\mu =  0 ,
\end{equation}
where $\widetilde F \widetilde\partial \equiv - (\partial_\mu \frac{1}2
F_{\alpha\beta}\gamma^\alpha\gamma^\beta) \gamma^\mu$.

Now,
\begin{equation} \label{be87} 
-\frac{1}2 (F\gamma^\mu F)\gamma^\nu = -\frac{1}2 ( F\gamma^\mu
F\gamma^\nu\rangle
\end{equation}

Since $\gamma^\mu . F = {\displaystyle\frac{1}2} (\gamma^\mu F - F
\gamma^\mu) = F .
\gamma^\mu$, we have 
\begin{eqnarray} \label{be88} 
T^{\mu\nu} & = & -\langle(F.\gamma^\mu)F\gamma^\nu\rangle_0 -
\frac{1}{2} \langle
\gamma^\mu F^2 \gamma^\nu\rangle_0  \nonumber \\
& = & -(F.\gamma^\mu) . (F.\gamma^\nu) - \frac{1}2 (F.F)\gamma^\mu .
\gamma^\nu \\
& = & F^{\mu\alpha} F^{\lambda\nu} \eta_{\alpha\lambda} + \frac{1}4
\eta^{\mu\nu}
F_{\alpha\beta} F^{\alpha\beta}, \nonumber
\end{eqnarray}
which we recognize as the stress-energy momentum tensor of the
electromagnetic field, and $T^\mu = T^{\mu\nu}\gamma_\nu$.

By writing $F = \vec E + {\bf i}\vec B$ as before we can immediately
verify that 
\begin{eqnarray} \label{be89} 
T_0 & = & -\frac{1}2 F \gamma_0 F \nonumber \\
& = & \left[\frac{1}2 (\vec E^2 + \vec B^2) + (\vec E \times \vec
B)\right] \gamma_0.
\end{eqnarray}
We have already shown that $\partial_\mu T^\mu = 0$, and we can easily
show
that
\begin{equation} \label{be90} 
\partial . T^\mu = 0 \, . 
\end{equation}
We now define the density of {\it angular momentum}. Choose as before a
Lorentzian chart
$\langle x^\mu\rangle$ of the maximal atlas of $M$ and consider the
1-form $x
=
x^\mu \gamma_\mu = x_\mu \gamma^\mu$. Define

\[
M_\mu = x \wedge T_\mu = \frac{1}2 (x_\alpha T_{\mu\nu} - x_\nu
T_{\alpha\mu}) \gamma^\alpha \wedge \gamma^\nu \; .
\]
It is trivial to verify that as $T_{\mu\nu} = T_{\nu\mu}$ and
$\partial_\mu T^{\mu\nu} = 0$, it holds
\begin{equation} \label{be91} 
\partial^\mu M_\mu = 0.
\end{equation}

The {\it invariants} of the electromagnetic field $F$ are $F.F$, $F
\wedge
F$ and $F^2 = F.F + F\wedge F$ with 
\begin{equation} \label{be92} 
F.F = -\frac{1}2 F^{\mu\nu} F_{\mu\nu} \; ; \qquad F \wedge F =
-\gamma_5
F^{\mu\nu} F^{\alpha\beta} \varepsilon_{\mu\nu\alpha\beta} \; .
\end{equation}

Writing as before $F = \vec E + {\bf i} \vec B$ we have
\begin{equation} \label{be93} 
F^2 = (\vec E^2 - \vec B^2) + 2{\bf i} \vec E . \vec B = F.F + F \wedge
F.
\end{equation}

\subsection{Dirac theory in ${\cal C}\ell(M)$}
\label{bs6}

Let $\Sigma = \{\gamma^\mu\} \in \sec \mbox{$\mbox{$\bigwedge$}^1$} (M)
\subset \sec
{\cal C}\ell(M)$
be an orthonormal basis. Let $\psi_\Sigma \in \sec (\mbox{$\bigwedge$}^0
(M)+\mbox{$\bigwedge$}^2(M) +
\mbox{$\bigwedge$}^4(M)) \subset \sec {\cal C}\ell(M)$ be the
representative of a
Dirac-Hestenes Spinor field in the basis $\Sigma$. Then the
representative of Dirac equation in ${\cal C}\ell(M)$ is the following
equation ($\hbar=c=1$):
\begin{equation} \label{be94} 
\partial\psi_\Sigma \gamma_1\gamma_2 + m \psi_\Sigma \gamma_0 = 0 \; .
\end{equation}
To see that, consider the {\it complexification} ${\cal C}\ell_C(M)$ of ${\cal
C}\ell(M)$
called the {\it complex Clifford bundle}. Then
${\cal C}\ell_C(M) = \;\mbox{{\sf I}}\!\!\!C \otimes {\cal C}\ell(M)$
and by the results of Section \ref{bs4} 
we know that the typical fiber of ${\cal C}\ell_C(M)$ is ${\cal
C}\ell_{4,1}=\;\mbox{{\sf I}}\!\!\!C
\otimes {\cal C}\ell_{1,3}$, the Dirac
algebra.

Now let $\{\Gamma_0, \Gamma_1, \Gamma_2, \Gamma_3, \Gamma_4\} \subset
\sec \mbox{$\bigwedge$}^1
(M) \subset \sec {\cal C}\ell_C(M)$ be an orthonormal basis with 
\begin{eqnarray} \label{be95} 
&& \Gamma_a \Gamma_b + \Gamma_b \Gamma_a = 2 g_{ab} \; , \\
&& g_{ab} = diag(+1,+1,+1,+1,-1) \; . \nonumber
\end{eqnarray}
Let us identify $\gamma_\mu = \Gamma_\mu \Gamma_4$ and call $I =
\Gamma_0
\Gamma_1 \Gamma_2 \Gamma_3 \Gamma_4$. Since $I^2 = -1$ and $I$ commutes
with all
elements of ${\cal C}\ell_{4,1}$ we identify $I$ with ${\rm i}=\sqrt{-1}$ and
$\gamma_\mu$ with the fundamental set of ${\cal C}\ell(M)$. Then if
${\cal A} \in
\sec {\cal C}\ell_C(M)$ we have
\begin{equation} \label{be96} 
{\cal A} = \Phi_s + A^\mu_C \gamma_\mu + \frac{1}2 B^{\mu\nu}_C
\gamma_\mu
\gamma_\nu + \frac{1}{3!} \tau^{\mu\nu\rho}_C \gamma_\mu \gamma_\nu
\gamma_\nu+\Phi_p \gamma_5 ,
\end{equation}
where $\Phi_s, \Phi_p, A^\mu_C, B^{\mu\nu}_C, \tau^{\mu\nu\rho}_C \in
\sec \;\mbox{{\sf I}}\!\!\!C \otimes \mbox{$\bigwedge$}^0(M) \subset
\sec {\cal C}\ell_C(M)$, {\em i.e.\/}, $\forall x \in M$, $\Phi_s(x)$,
$\Phi_p(x)$, $A^\mu_C(x)$, $B^{\mu\nu}_C (x)$, $\tau^{\mu\nu\rho}_C(x)$
are complex numbers.

Now,
\[
f = \frac{1}2 (1+\gamma_0) \frac{1}2 (1+{\rm i} \gamma_1 \gamma_2)
\]
is a primitive idempotent field of ${\cal C}\ell_C(M)$. We recall that,
by
eq.(\ref{be50}), ${\rm i}f = \gamma_2 \gamma_1 f$. From (\ref{be91}) we can write the
following
equation in ${\cal C}\ell_C(M)$:  
\begin{eqnarray*}
&& \partial \psi_\Sigma \gamma_1 \gamma_2 f + m \psi_\Sigma \gamma_0 f
= 0 \\
&& \partial \psi_\Sigma {\rm i} f - m \psi_\Sigma f = 0
\end{eqnarray*}
and we have the following equation for $\Psi=\psi_{\Sigma} f$: 
\begin{equation} \label{be97} 
{\rm i} \partial \Psi - m \Psi = 0.
\end{equation}

By eq.(\ref{be48}) and using for $\gamma_\mu$ the matrix representation
eq.(\ref{be46})
(denoted here by $\underline\gamma_\mu$) we get that the matrix
representation
of eq.(\ref{be94}) is 
\begin{equation} \label{be98} 
{\rm i} \underline \gamma^\mu \partial_\mu |\Psi\rangle - m |\Psi\rangle = 0
\end{equation}
where now $|\Phi\rangle$ is a usual Dirac spinor field.

We now define a {\it potential} for the Dirac-Hestenes field
$\psi_\Sigma$. Since $\psi_\Sigma \in \sec {\cal C}\ell^+(M)$ it is
clear that
there exist $A,\, B \in \sec \bigwedge^1(M) \subset
\sec{\cal C}\ell(M)$ such
that 
\begin{equation} \label{be99} 
\psi_\Sigma = \partial(A + \gamma_5 B),
\end{equation}
since 
\begin{eqnarray*}
\partial(A+\gamma_5B) & = & \partial . A + \partial\wedge A - \gamma_5
\partial . B - \gamma_5 \partial
\wedge B \\
& = & S + B + \gamma_5 P \, ;
\end{eqnarray*}
\[
S = \partial.A; \quad B = \partial \wedge A - \gamma_5 \partial \wedge
B; \quad P = - \partial. B \;
.
\] 

We see that when $m=0$, $\psi_\Sigma$ satisfies the {\it Weyl
equation}\footnote{We recall again that a Weyl spinor must satisfy $\gamma_5
\psi = \pm \psi \gamma_{21}$ (see {\em e.g.\/} \cite{10}).}
\begin{equation} \label{be100}
\partial\psi_\Sigma = 0 \; .
\end{equation}
Using eq.(\ref{be100}) we see that
\begin{equation} \label{be101} 
\partial^2 A = \partial^2 B = 0.
\end{equation}
This last equation allows us to find UPWs solutions of arbitrary speeds
for the Weyl equation once we know UPWs solutions of the scalar wave
equation $\Box \Phi = 0$, $\Phi \in \sec \mbox{$\bigwedge$}^0(M) \subset
\sec {\cal C}\ell(M)$. Indeed it is enough to put ${\cal A}=(A+\gamma_5
B) = \Phi(1 + \gamma_5)v$, where $v$ is a constant 1-form field. This
result has been used in \cite{15} to present subluminal and superluminal
solutions of the Weyl equation. An example of a subluminal solution
(indeed a stationary one)  of the massless Dirac equation is obtained
with  the use of the ``superpotential'' ${\cal A}_0$:
\begin{equation} \label{ben102} 
{\cal A}_0 = \frac{C}{r} ( \sin \Omega r \cos \Omega t \gamma^0 - \sin \Omega r \sin \Omega t \gamma^1 \gamma^2 \gamma^3 ) . 
\end{equation} 
We have then 
\begin{eqnarray}
\psi_0 = \partial {\cal A}_0 &=&  \frac{C}{r^3} [
-\Omega r^2 \sin \Omega r \sin \Omega t  \nonumber \\
& + &\gamma^0\gamma^1 \lambda x \cos \Omega t
+ \gamma^0\gamma^2 \lambda y \cos \Omega t \nonumber \\
& +& \gamma^0\gamma^3 \lambda z \cos \Omega t
- \gamma^1\gamma^2 \lambda z \sin \Omega t \nonumber \\
& + &\gamma^1\gamma^3 \lambda y \sin \Omega t
- \gamma^2\gamma^3 \lambda x \sin \Omega t \nonumber \\
& +&\gamma^0\gamma^1\gamma^2\gamma^3 \Omega r^2 \sin \Omega r \cos
\Omega t ] ,
\end{eqnarray}
where $\lambda = \Omega r \cos \Omega r - \sin \Omega r $, $r =
\sqrt{x^2 + y^2 + z^2}$.

The above solution in the usual formalism reads
\begin{equation}
\Psi_0 = \left( \begin{array}{c}
\displaystyle{{\rm i}\sin{\Omega t}\left(\frac{\lambda z}{r^3} +
{\rm i}\frac{\Omega}{r}\sin{\Omega r}\right)} \\
\displaystyle{{\rm i}\sin{\Omega t} \left( \frac{x+{\rm i} y}{r^3}\right)\lambda}
\\
\displaystyle{-\cos{\Omega t}\left(\frac{\lambda z}{r^3} +
{\rm i}\frac{\Omega}{r}\sin{\Omega r}\right)} \\
\displaystyle{-\cos{\Omega t}\left( \frac{x+{\rm i} y}{r^3}\right)\lambda}
\end{array} \right) \, . 
\end{equation}
 
Another very interesting possibility for constructing solutions of Weyl equation is the following. Suppose that 
$F \in \sec \bigwedge^2(M) \subset \sec {\cal C}\ell (M)$, $F^2
\neq 0$, is a solution of $\partial F =0$. Then $\psi = \e^F$  is a massless Dirac spinor field satisfying 
$\partial \psi =0$.\footnote{A closed expression for $\e^F$ is given in \cite{16b}. In particular, writing $F = \vec{E} + {\bf i} \vec{B} = z \hat{F}$ (eq.(\ref{be77})), where $\hat{F} = z^* F /|z|^2$, $\hat{F}^2 = 1$, 
 $z \in \mbox{{\sf I}}\!\!\!C$, and $\hat{F}$ is a complex vector (in the Pauli algebra sense), then $\e^F = \e^{z\hat{F}} = \cosh z + \hat{F} \sinh z$.} Using this result and eq.(\ref{1en8}) we can
construct solutions of $\partial \psi_W =0$. 
To end this Section we show how to construct luminal or superluminal solutions of the Dirac equation. 

We know (see Section \ref{sec2}) that the Klein-Gordon equation
has besides the subluminal solutions also luminal and superluminal
solutions. Let $\Phi$ be a subluminal, luminal or superluminal solution of $\Box
\Phi + m \Phi = 0$. Suppose $\Phi$ is a section of
${\cal C}\ell_C(M)$. Then in ${\cal C}\ell_C(M)$ we have the following
factorization:
\begin{equation}
(\partial+{\rm i} m)(\partial-{\rm i} m) \Phi=0.
\end{equation}

Now
\begin{equation}
\Psi = (\partial-{\rm i} m) \Phi  f
\end{equation}
is a Dirac spinor field in ${\cal C}\ell_C(M)$, since
\begin{equation}
(\partial+{\rm i} m) \Psi = 0 \, . 
\end{equation}
$\Psi$ is then a subluminal, luminal or superluminal UPW solution of Dirac equation, depending on $\Phi$. 

\section{EXTRAORDINARY SOLUTIONS OF THE (SCALAR) HOMOGENEOUS
WAVE
EQUATION AND OF KLEIN-GORDON EQUATION}
\label{sec2}

\subsection{Subluminal and superluminal solutions of the HWE}
\label{as4}

Consider the HWE ($c=1$)
\begin{equation} \label{ae2p}
\frac{\partial^2}{\partial t^2} \Phi - \nabla^2 \Phi = 0
\;.
\end{equation}
We now present some subluminal and superluminal solutions of
this equation.

{\it Subluminal and Superluminal Spherical Bessel Beams.} To
introduce these beams we define the variables
\begin{equation} \label{3e2a} 
\xi_< = [x^2+y^2+\gamma^2_< (z-v_<t)^2]^{1/2} \; ;
\end{equation}  
\begin{equation} \label{3e2b}  
\gamma_< = {\displaystyle\frac{1}{\sqrt{1-v^2_<}}} \; ;
\ \ \omega^2_< - k^2_< =
\Omega^2_<  \; ; \ \ v_< = {\displaystyle\frac{{\rm d}\omega_<}{{\rm
d}k_<}} \; ; \end{equation}
\begin{equation} \label{3e2c} 
\xi_> = [-x^2-y^2+\gamma^2_> (z-v_> t)^2]^{1/2} \; ;
\end{equation} 
\begin{equation} \label{3e2d} 
\gamma_> = {\displaystyle\frac{1}{\sqrt{v^2_> - 1}}} \; ;
\ \ \omega^2_> - k^2_> =
-\Omega^2_> \; ; \ \ v_> = {\rm d}\omega_>/{\rm d}k_> \; . 
\end{equation}  

We can now easily verify that the functions $\Phi^{\ell_m}_<$ and
$\Phi^{\ell_m}_>$ below
are respectively subluminal and superluminal solutions of the HWE (see
example 3 below for how to obtain these solutions). We have
\begin{equation} \label{ae30}
\Phi^{\ell m}_p (t, \vec x) = C_\ell j_\ell (\Omega_p \xi_p) P^\ell_m
(\cos
\theta) \e^{{\rm i}m\theta} \e^{{\rm i} (\omega_pt - k_p z)}
\end{equation}
where the index $p = < , \; > $, $C_\ell$ are constants, $j_\ell$ are
the spherical
Bessel functions, $P^\ell_m$ are the Legendre functions and $(r,\theta,
\varphi)$ are the usual spherical coordinates. $\Phi^{\ell m}_<$ 
$[\Phi^{\ell m}_>]$ has phase velocity $(w_</k_<) < 1$ $ [(w_>/k_>) > 1]$
and the modulation function $j_\ell (\Omega_< \xi_<)$ $[j_\ell (\Omega_>
\xi_>)]$ moves with group velocity $v_<$ $[v_>]$, where $0 \leq v_< <
1$ $[1 <
v_> < \infty]$. Both $\Phi^{\ell m}_<$ and $\Phi^{\ell m}_>$ are {\it
undistorted progressive waves} (UPWs). This term has been introduced
by Courant and Hilbert \cite{4}. However, they didn't suspect of UPWs
moving with speeds greater than $c=1$. For use in the main text we
write
the explicit  form of $\Phi^{00}_<$ and $\Phi^{00}_>$, which we denote
simply by $\Phi_<$ and $\Phi_>$:
\begin{equation} \label{ae31}
\Phi_p(t,\vec x) = C \frac{\sin (\Omega_p \xi_p)}{\xi_p}
\e^{{\rm i}(\omega_pt-k_pz)}\ \ ; \ \ p=< \; \mbox{or} \; > \; .
\end{equation}
When $v_< = 0$, we have $\Phi_< \rightarrow \Phi_0$,
\begin{equation} \label{ae32}
\Phi_0(t,\vec x) = C \frac{\sin \Omega_< r}{r}
\e^{{\rm i}\Omega_< t}, \;  r =
(x^2+y^2+z^2)^{1/2} \; .
\end{equation}
This solution has been found by Bateman in 1915 \cite{16}. The
superluminal solution $\Phi_>$ was discovered by Barut and Chandola
 in 1993 \cite{17}. In what follows we show methods to obtain the
 solutions $\Phi_<$ and $\Phi_>$ for the HWE.
When $v_>=\infty$, $\omega_>=0$ and $\Phi^0_> \rightarrow \Phi_\infty$,
\begin{equation} \label{ae33}
\Phi_\infty(t,\vec x) = C_\infty {\displaystyle\frac{\sinh \rho}{\rho}}
\e^{i\Omega_> z}
\; , \ \  \rho = (x^2+y^2)^{1/2} \; .
\end{equation}  

We observe that if our interpretation of phase and group velocities is
correct, then there must be a Lorentz frame where $\Phi_p$ is at rest.
It is trivial to verify that in the coordinate chart $\langle x^{'\mu}
\rangle$ which is a (nacs/$I'$) (see Section \ref{sec5} for more details), where $I' =
(1-v_{<}^{2})^{-1/2} \partial/\partial t + (v_</\sqrt{1-v_{<}^{2}})
\partial/\partial z$ is a Lorentz frame moving with speed $v_<$ in the
$z$ direction relative to $I = \partial/\partial t$, $\Phi_p$ goes in
$\Phi_0 (t',\vec{x}')$ given by eq.(\ref{ae32}) with $t \mapsto t'$,
$\vec{x}
\mapsto \vec{x}'$. We can also verify that there is no Lorentz
frame with velocity parameter $0<v<1$ where $\Phi_>$ is at rest.

{\it Subluminal and Superluminal Bessel Beams.} The solutions of the
HWE in cylindrical coordinates are well known \cite{14}. Here we
recall how these solutions are obtained in order to present new
subluminal
and superluminal solutions of the HWE. In what follows the cylindrical
coordinate functions are denoted by $(\rho,\theta, z)$, $\rho =
(x^2+y^2)^{1/2}$, $x=\rho\cos\theta$, $y=\rho\sin\theta$. We write for
$\Phi$:
\begin{equation} \label{ae34}
\Phi(t,\rho,\theta,z)=f_1(\rho)f_2(\theta)f_3(t,z) \; .
\end{equation}
Inserting (\ref{ae34}) in (\ref{ae2p}) gives
\begin{equation} \label{ae35a} 
\rho^2 {\frac{{\rm d}^2}{{\rm d}\rho^2}} f_1 +
\rho{\frac{{\rm d}}{{\rm d}\rho}} f_1 +
(B\rho^2-\nu^2)f_1 = 0
\end{equation} 
\begin{equation} \label{ae35b} 
\left({\frac{{\rm d}^2}{{\rm d}\theta^2}} + \nu^2\right)f_2 =
0 ; 
\end{equation} 
\begin{equation} \label{ae35c} 
\left({\displaystyle\frac{{\rm d}^2}{{\rm d}t^2}} -
{\frac{\partial^2}{\partial z^2}}  +
B\right)f_3=0 . 
\end{equation} 

\noindent $B$ and $\nu$ are separation constants. Since we want $\Phi$
to be
periodic in $\theta$ we choose $\nu=n$ an integer. For $B$ we consider
two
cases: 
\smallskip 
(i) {\it Subluminal Bessel solution}, $B = \Omega^2_< > 0$

In this case (\ref{ae35a}) is a Bessel equation and we have
\begin{equation} \label{ae36}
\Phi^<_{J_n} (t,\rho,\theta,z) = C_n J_n (\rho\Omega_<) \e^{{\rm i}(k_< z -
w_< t +
n
\theta)}, \ \ n=0,1,2,\ldots ,
\end{equation}
where $C_n$ is a constant, $J_n$ is the $n$-th order Bessel function
and
\begin{equation} \label{ae37}
\omega^2_< - k^2_< = \Omega^2_< \; .
\end{equation}

In \cite{18} the $\Phi^<_{J_n}$ are called the
$n$-th order non-diffracting Bessel
beams.\footnote{The only difference is that $k_<$ is
denoted by $\beta = \sqrt{\omega^2_<-\Omega^{2}_<}$ and $\omega_<$ is
denoted by
$k'=\omega/c>0$. (We use units where $c=1$.)}
Bessel beams are examples of undistorted progressive waves
(UPWs). They are ``subluminal" waves. Indeed, the group velocity for
each wave is
\begin{equation} \label{ae38}
v_< = {\rm d}\omega_< / {\rm d}k_< , \ 0 < v_< < 1 \; ,
\end{equation}
but the phase velocity of the wave is $(\omega_</k_<) > 1$. That this
interpretation is correct follows from the results of the acoustic
experiment described in \cite{5,6}.

It is convenient for what follows to define the variable $\eta$,
called the axicon angle \cite{19},
\begin{equation} \label{ae39}
k_< = \overline k_< \cos \eta\;  , \ \ \Omega_< = \overline k_< \sin
\eta \; , \ \
0 < \eta < \pi/2 \; .
\end{equation}
Then
\begin{equation} \label{ae40}
\overline k_< = \omega_< > 0
\end{equation}
and eq.(\ref{ae36}) can be rewritten as $\Phi^<_{A_n} \equiv \Phi^<_{J_n}$,
with
\begin{equation} \label{ae41}
\Phi^<_{A_n} = C_n J_n (\overline k_< \rho \sin \eta) \e^{{\rm i} (\overline
k_< z
\cos\eta
- \omega_< t + n\theta)} .
\end{equation}
In this form the solution is called in \cite{18} the $n$-th order
non-diffracting portion of the  {\it Axicon Beam}.

Now, the phase velocity $v^{ph} = 1/\cos \eta$ is independent of
$\overline k_<$, but, of course, it is dependent on $k_<$.  We shall
show below that waves constructed from the $\Phi^<_{J_n}$ beams can be
{\it subluminal} or {\it superluminal} ! 

\smallskip 

(ii) {\it Superluminal (Modified) Bessel Solution}, $B = -\Omega^2_> <
0$

In this case (\ref{ae35a}) is the modified Bessel equation and
we denote the solutions by
\begin{equation} \label{ae42}
\Phi^>_{K_n} (t,\rho,\theta,z) = C_n K_n (\Omega_> \rho) \e^{{\rm i}(k_> z -
\omega_> t
+
n\theta)}, \ n=0,1, \ldots ,
\end{equation}
where $K_n$ are the modified Bessel functions, $C_n$ are constants
and
\begin{equation} \label{ae43}
\omega^2_> - k^2_> = -\Omega^2_> \;.
\end{equation}
We see that $\Phi^>_{K_n}$ are also examples of UPWs, each of which has
group velocity $v_> = {\rm d}\omega_> /{\rm d}k_>$ such that $1 < v_> <
\infty$ and phase velocity  $(\omega_>/k_>) < 1$. As in the case of the
spherical Bessel beam (eq.(\ref{ae31})) we see again that our
interpretation of phase and group velocities is correct. Indeed, for the
superluminal (modified) Bessel beam there is no Lorentz frame where the
wave is stationary.

The $\Phi^>_{K_0}$ beam was discussed by Band \cite{20,21} in 1988 as
an example of superluminal motion. Band proposed to launch the
$\Phi^>_{K_0}$ beam in the exterior of a cylinder of radius
$r_1$
on which there is an appropriate superficial charge density.
Since $K_0 (\Omega_> r_1)$ is non singular, his solution works. In
Section \ref{sec3} we discuss some of Band's statements.

We are now prepared to present some other very interesting solutions
of the HWE, in particular the so called $X$-waves, which are
superluminal, as proved by the acoustical experiments described in
\cite{5,6}.  \\

{\sc Theorem} (Lu and Greenleaf) {\sl The three
functions below
are
families of exact solutions of the HWE [eq.(\ref{ae2p})] in cylindrical
coordinates:
\begin{eqnarray} \label{ae44} 
&& \Phi_\eta(s) = \int^\infty_0 T(\overline k_<) \left[ \frac{1}{2\pi}
\int^\pi_{-\pi} A(\phi)f(s){\rm d}\phi \right] {\rm d}\overline k_< \; ; \\ 
\label{ae45} 
&& \Phi_K(s) = \int^{\pi}_{-\pi} D(\eta) \left[ \frac{1}{2\pi}
\int^{\pi}_{-\pi} A(\phi)f(s){\rm d}\phi \right] {\rm d}\eta \; ; \\ 
\label{ae46} 
&& \Phi_L (\rho,\theta,z-t) = \Phi_1(\rho,\theta) \Phi_2(z-t) \; ;
\end{eqnarray}
where
\begin{equation} \label{ae47}
s = \alpha_0 (\overline k_<, \eta)\rho \cos (\theta-\phi) + b(\overline
k_<, \eta) [z\pm
c_1(\overline k_<,\eta)t]
\end{equation}
and
\begin{equation} \label{ae48}
c_1(\overline k_<, \eta) = \sqrt{1+[\alpha_0 (\overline k_<, \eta)/b
(\overline
k_<,\eta)]^2} \; .
\end{equation}
} 

In these formulas $T(\overline k_<)$ is any complex function (well
behaved) of $\overline k_<$ and could include the {\it temporal
frequency transfer function} of a radiator system, $A(\phi)$ is any
complex function (well behaved) of $\phi$ and represents a {\it
weighting function} of the integration with respect to $\phi, f(s)$ is
any complex function (well behaved) of $s$ (solution of
eq.(\ref{ae2p})), $D(\eta)$ is any complex function (well behaved) of
$\eta$ and represents a weighting function of the integration with
respect to $\eta$, called the axicon angle, $\alpha_0(\overline k_<,
\eta)$ is any complex function of $\overline k_<$ and $\eta$,
$b(\overline k_<, \eta)$ is any complex function of $\overline k_<$ and
$\eta$.

As in the previous solutions, we take $c = 1$. Note that $\overline
k_<$, $\eta$ and the wave vector $k_<$ of the $f(s)$ solution of
eq.(\ref{ae2p}) are related by eq.(\ref{ae39}). Also $\Phi_2(z-t)$ is
any complex function of $(z-t)$ and $\Phi_1(\rho,\theta)$ is any
solution of the transverse Laplace equation, {\em i.e.\/},
\begin{equation} \label{ae49}
\left[\frac{1}\rho \frac{\partial}{\partial \rho} \left(\rho
\frac{\partial}{\partial
\rho}\right)
+
\frac{1}{\rho^2} \frac{\partial^2}{\partial\theta^2} \right] \Phi_1
(\rho,\theta)=0.
\end{equation}
The proof is obtained by direct substitution of $\Phi_\eta, \Phi_K$ and
$\Phi_L$ in the HWE. Obviously,  the exact solution $\Phi_L$ is an
example of a luminal UPW, because if one ``travels"  with the speed $c
= 1$,
{\em i.e.\/}, with $z-t = \mbox{constant}$, both the lateral and axial
components, $\Phi_1 (\rho, \theta)$ and $\Phi_2(z-t)$ will be the same
for
all time $t$ and distance $z$.
When $c_1(\overline k, \eta)$ in eq.(\ref{ae47}) is real, ($\pm$) represent
respectively backward and forward propagating waves.

We recall that $\Phi_\eta (s)$ and $\Phi_K(s)$ represent families of
UPWs if $c_1(\overline k_<, \eta)$ is independent of $\overline k_<$ and
$\eta$ respectively. These waves travel to infinity at speed $c_1$.
$\Phi_\eta(s)$ is a generalized function that contains some of the UPWs
solutions of the HWE derived previously. In particular, if $T(\overline
k_<) = \delta(\overline k_< - \overline k_<')$,  $\overline k_<' =
\omega > 0$ is a constant and if $f(s)=\e^s$, $\alpha_0(\overline k_<,
\eta) = -{\rm i}\Omega_<$, $b(\overline k_<, \eta) ={\rm i}\beta = {\rm
i}\omega/c_1$, one obtains  Durnin's UPW beam \cite{20}
\begin{equation} \label{ae50}
\Phi_{Durnin} (s) = \left[ \frac{1}{2\pi} \int^\pi_{-\pi} A(\phi)
\e^{-{\rm i}\Omega_< \rho \cos (\theta-\phi)} {\rm d}\phi \right]
\e^{{\rm i}(\beta z-\omega t)}.  
\end{equation}

If $A(\phi) = {\rm i}^n \e^{{\rm i}n\phi}$, we obtain the $n$-th order UPW Bessel
beam  $\Phi^<_{J_n}$ given by eq.(\ref{ae36}). $\Phi^<_{A_n}(s)$ is obtained
in the same way with the transformation $k_< = \overline k_< \cos
\eta$;
$\Omega_< = \overline k_< \sin \eta$. 

\smallskip 

\noindent {\bf The {\em X}-waves}. We now present a superluminal UPW
discovered in 1992 by Lu and Greenleaf \cite{18}
which,
as discussed in \cite{5,19}, is physically realizable in an approximate
way  in the
acoustic case and  can be used to generate Hertz potentials for the
electromagnetic field (see Section \ref{sec3}). We take in
eq.(\ref{ae44}):
\begin{eqnarray} \label{ae51}
&& T(\overline k_<) = B(\overline k_<) \e^{-a_0 \overline k_<};
\ \ A(\phi) = {\rm i}^n
\e^{{\rm i}n\phi};
\ \  f(s)=\e^s \nonumber \\
&&\alpha_0 (\overline k_<, \eta) = -{\rm i}\overline k_< \sin \eta \ \
 b(\overline k_<, \eta) = {\rm i}\overline{k}_< \cos\eta; \ \ .
\end{eqnarray}
Then we get
\begin{equation} \label{ae52}
\Phi^>_{X_n} = \e^{{\rm i}n\theta} \int^\infty_0 B(\overline k_<) J_n
(\overline k_< \rho \sin\eta) \e^{-\overline k_< [a_0 - {\rm
i}(z\cos\eta-t)]} {\rm d}\overline k_< .
\end{equation}
In eq.(\ref{ae52}) $B(\overline k_<)$ is any well behaved complex
function of $\overline k_<$ and represents a {\it transfer function of a
practical radiator}, $\overline k_< = \omega$ and $a_0$ is a constant,
and $\eta$ is again called the axicon angle. Equation (\ref{ae52}) shows
that $\Phi^>_{X_n}$ is represented by a Laplace transform of the
function $B(\overline k_<) J_n (\overline k_< \rho \sin \eta)$, and an
azimuthal phase term $\e^{{\rm i}n \theta}$.  The name {\it X-waves} for
the $\Phi^>_{X_n}$ comes from the fact that these waves have an $X$-like
shape in a plane containing the axis of symmetry of the waves
\cite{18,5,6}. 

\smallskip 

\noindent{\bf The $\Phi^>_{XBB_n}$ waves}. This wave is obtained from
eq.(\ref{ae52}) putting $B(\overline k_<) = a_0$. It is called the
$X$-wave produced by an {\it infinite} aperture and {\it broad
bandwidth}. We use in this case the notation $\Phi^>_{XBB_n}$. Under
these conditions we get
\begin{equation} \label{ae53}
\Phi^>_{XBB_n} = \frac{a_0(\rho\sin \eta )^n
\e^{{\rm i}n\theta}}{\sqrt{M}(\tau+\sqrt{M})^n} \ \ , \ \ (n=0,1,2,\ldots)
\end{equation}
where the subscript denotes ``broadband". Also
\begin{equation} \label{ae54}
M = (\rho\sin \eta)^2+\tau^2 \; ;
\end{equation}
\begin{equation} \label{ae55}
\tau = [a_0 - {\rm i}(z\cos\eta-t)]
\end{equation}

For $n=0$ we get $\Phi^>_{XBB_0}$ and
\begin{equation} \label{ae56}
\Phi^>_{XBB_0} = \frac{a_0}{\sqrt{(\rho\sin\eta)^2 +
[a_0-{\rm i}(z\cos\eta-t)]^2}} \; .
\end{equation}
It is clear that all $\Phi^>_{XBB_n}$ are UPWs which propagate with
speed $c_1 = 1/\cos\eta > 1$ in the $z$-direction. Our statement is
justified for as can be easily seen (as in the modified superluminal
Bessel beam) there is no Lorentz frame where $\Phi_{XBB_n}^{>}$ is at
rest.  Observe that this is the real speed of the wave; phase and group
velocity concepts are not applicable here. Equation (\ref{ae56}) does
not give any dispersion relation.  The $\Phi^>_{XBB_n}$ waves cannot be
produced in practice as they have infinite energy (see Section
\ref{as7}), but   a good approximation for them has been realized with
{\it finite aperture} radiators \cite{22,5}. 

\smallskip 

Recall that if in eq.(\ref{ae52}) we put $B(\overline{k}_<) \e^{-a_0
\overline{k}_<} = \overline{A}(\overline{k}_<)$ and if we take into
account that for each component Bessel beam in the packet the following
dispersion relation holds:
\begin{equation} \label{aen57}
\omega = k_</\cos \eta \equiv k/\cos\eta \; (k_< =k)\, ,
\end{equation}
then for both the broad band $X$-waves (as {\em e.g.\/} eq.(\ref{ae56})) and
the limited band $X$-waves where $A(k) \equiv \overline{A}
(\overline{k}_<)$ is centered in $k_0$ we can write
\begin{equation} \label{aen58}
\omega (k) = \omega (k_0) + \left. \frac{{\rm d}\omega}{{\rm d}k} \right|_{k_0}
(k-k_0) =
\omega_0 + \frac{{\rm d}\omega}{{\rm d}k_0} (k-k_0) \, ,
\end{equation}
where $\omega_0 = 1/\cos\eta$, ${\rm d}\omega/{\rm d}k_0 = 1/\cos\eta$. It follows
that eq.(\ref{ae52}) can be written
\begin{eqnarray} \label{aen59}
\Phi_{X_n}(t,x,y,z) &=&\frac{\e^{{\rm i}n\theta}}{\cos\eta} \int {\rm d}k \, A(k)
\, J_n (k \rho \tan \eta) \e^{{\rm i} (z- \frac{1}{\cos \eta} t)k} \nonumber
\\
&=& \Phi \left( 0, x,y,z-\frac{1}{\cos \eta}t \right)
\end{eqnarray}
showing that the $X$-waves propagate without distortion with speed
${\rm d}\omega/{\rm d}k_0 = 1/\cos \eta$.

We end this section with the commentary that in \cite{6} we develop 
methods for projecting ``finite aperture approximations'' to the exact 
acoustic and electromagnetic solutions discussed in this paper. 


\subsection{Donnelly-Ziolkowski method  for designing
subluminal, luminal
and superluminal UPW solutions of the HWE and the Klein-Gordon
equation (KGE) \protect\cite{23,24}}

Consider first the HWE for $\Phi$ (eq.(\ref{ae2p})) in a homogeneous
medium.
Let $\widetilde \Phi(\omega,\vec k)$ be the Fourier transform of
$\Phi(t,\vec
x)$, {\em i.e.\/},
\begin{equation} \label{ae69a}
\widetilde \Phi(\omega,\vec k) =
{\int_{R^3} {\rm d}^3 x
\int^{+\infty}_{-\infty}} {\rm d}t \, \Phi(t,\vec x) \e^{-{\rm i}(\vec k \vec x -
\omega
t)} ,
\end{equation}
\begin{equation} \label{ae69b}
\Phi(t,\vec x) = {\frac{1}{(2\pi)^4} \int_{R^3} {\rm d}^3 \vec k
\int^{+\infty}_{-\infty}} {\rm d}\omega \, \widetilde \Phi (\omega,\vec k)
\e^{{\rm i}(\vec k \vec x - \omega t)}.  
\end{equation}

Inserting (\ref{ae69a}) in the HWE we get
\begin{equation} \label{ae70}
(\omega^2-\vec k^2) \widetilde \Phi (\omega, \vec k) = 0
\end{equation}
and we are going to look for solutions of the HWE and eq.(\ref{ae70})
in the
sense of distributions. We rewrite eq.(\ref{ae70}) as
\begin{equation} \label{ae71}
(\omega^2 - k^2_z - \Omega^2) \widetilde \Phi (\omega, \vec k) = 0 .
\end{equation}
It is then obvious that any $\Phi(\omega, \vec k)$ of the form
\begin{equation} \label{ae72}
\widetilde \Phi(\omega,\vec k) = \Xi (\Omega,\beta) \, \delta [\omega -
(\beta+\Omega^2/4\beta)] \, 
\delta[k_z - (\beta-\Omega^2/4\beta)] \; ,
\end{equation}
where $\Xi(\Omega, \beta)$ is an arbitrary weighting function,  is a
solution of eq.(\ref{ae71}) since the $\delta$-functions imply that
\begin{equation} \label{ae73}
\omega^2-k^2_z = \Omega^2 \; .
\end{equation}

In 1985 Ziolkowski \cite{25} found a {\it luminal} solution of
the HWE called the Focus Wave Mode. To obtain this solution we
choose, {\em e.g.\/},
\begin{equation} \label{ae74}
\Xi_{FWM} (\Omega, \beta) = \frac{\pi^2}{{\rm i}\beta} \exp (-\Omega^2
z_0/4\beta) ,
\end{equation}
whence we get, assuming $\beta > 0$ and $z_0 > 0$,
\begin{equation} \label{ae75}
\Phi_{FWM} (t,\vec x) = \e^{{\rm i}\beta(z+t)} \frac{\exp\{-\rho^2\beta/[z_0 +
{\rm i}(z-t)]\}}{4\pi {\rm i} [z_0+{\rm i}(z-t)]} .
\end{equation}
Despite the velocities $v_1 = + 1$ and $v_2 = -1$ appearing in the
phase, the modulation function of $\Phi_{FWM}$ has very
interesting properties, as discussed in details in \cite{25}. It
remains
to observe that eq.(\ref{ae75}) is a special case of Brittingham's
formula \cite{26}.

Returning to eq.(\ref{ae72}) we see that the $\delta$-functions make
any
function of the Fourier transform variables $\omega, k_z$ and $\Omega$
to
lie in a line on the surface $\omega^2 - k^2_z - \Omega^2 = 0$
(eq.(\ref{ae71})).
Then,
the support of the $\delta$-functions is the line
\begin{equation}
\omega= \beta + \Omega^2/4\beta ; \ k_z = \beta - \Omega^2/4\beta\; .
\end{equation}
The projection of this line in the $(\omega, k_z)$ plane is a straight
line of slope $-1$ ending at the point $(\beta, \beta)$. When $\beta =
0$
we must have $\Omega=0$, and in this case the line is $\omega = k_z$
and
$\Phi(t,\vec x)$ is simply a superposition of plane waves, each one
having frequency $\omega$ and traveling with speed $c=1$ in the
positive
$z$ direction.

Luminal UPWs solutions can be easily constructed by the ZM,
but will not be discussed here. Instead, we now show how to use
ZM to construct subluminal and superluminal solutions of the HWE. 

\smallskip

\noindent {\bf First Example:} Reconstruction of the subluminal Bessel
Beams
$\Phi^<_{J_0}$ and the superluminal $\Phi^>_{XBB_0}$ ($X$-wave). 
Starting from the ``dispersion relation" $\omega^2-k^2_z - \Omega^2 =
0$, we
define
\begin{equation} \label{ae77}
\widetilde \Phi (\omega, \overline k) = \Xi (\overline k, \eta) \delta
(k_z - \overline k \cos\eta)
\delta (\omega-\overline k).
\end{equation}
This implies that
\begin{equation} \label{ae78}
k_z = \overline k \cos\eta; \quad \cos \eta = k_z / \omega , \ \ \omega
> 0, \ \ -1
< \cos\eta < 1\, .
\end{equation}
We take moreover
\begin{equation}
\Omega = \overline k \sin \eta ; \ \ \ \overline k > 0 \; .
\end{equation}

We recall that $\vec\Omega = (k_x, k_y)$, $\vec\rho = (x,y)$ and we
choose
$\vec
\Omega . \vec \rho = \Omega \rho \cos \theta$. Now, putting
eq.(\ref{ae77}) in
eq.(\ref{ae69b})
we get
{\small
\begin{equation} \label{ae80}
\Phi(t, \vec x) = \frac{1}{(2\pi)^4} \int^\infty_0 {\rm d}\overline k \;
\overline k
\sin^2
\eta \left[\int^{2\pi}_0 {\rm d}\theta \, \Xi (\overline k, \eta)
\e^{{\rm i}\overline k \rho
\sin \eta \cos \theta} \right] \e^{{\rm i}(\overline k \cos\eta z - \overline
k t)} .
\end{equation}
}

Choosing
\begin{equation} \label{ae81}
\Xi (\overline k, \eta) = (2\pi)^3 \frac{z_0 \e^{-\overline k z_0
\sin\eta}}{\overline k
\sin \eta },
\end{equation}
where $z_0 > 0$ is a constant, we obtain
\begin{equation} \label{aen81}
\Phi(t, \vec x) = z_0 \sin \eta \int^\infty_0 {\rm d}\overline k \,
\e^{-\overline k z_0 \sin \eta} \left[\frac{1}{2\pi} \int^{2\pi}_0 {\rm
d}\theta \, \e^{{\rm i}\overline k \rho \sin\eta\cos\theta} \right]
\e^{{\rm i} \overline k (\cos\eta \, z - t)} \; .
\end{equation}
Calling $z_0 \sin\eta = a_0 > 0$, the last equation becomes
\begin{equation} \label{ae82}
\Phi^>_{X_0}(t,\vec x) = a_0 \int^\infty_0 {\rm d}\overline k \,
\e^{-\overline k a_0} J_0 (\overline k \rho \sin \eta) \e^{{\rm
i}\overline k(\cos\eta \, z-t)} .
\end{equation}
Writing $\overline k = \overline k_<$ and taking into account
eq.(\ref{ae41}) we see that
\begin{equation} \label{ae83}
J_0 (\overline k_< \rho\sin\eta) \e^{{\rm i}\overline k_< (z\cos\eta-t)}
\end{equation} is a subluminal Bessel beam, a solution of the HWE moving
in the positive $z$ direction. Moreover, a comparison of eq.(\ref{ae82})
with eq.(\ref{ae52}) shows that (\ref{ae82}) is a particular
superluminal $X$-wave, with $B(\overline k_<)=\e^{-a_0\overline k_<}$.
In fact it is the $\Phi^>_{XBB_0}$ UPW given by eq.(\ref{ae56}). 

\smallskip 

\noindent {\bf Second Example:} Choosing in (\ref{ae80})
\begin{equation} \label{ae84}
\Xi (\overline k, \eta) = (2\pi)^3 \e^{-z_0|\cos\eta|\overline k} \cot
\eta
\end{equation}
gives
\begin{eqnarray} \label{ae85a}
\Phi^> (t,\vec x) &=& \cos^2 \eta {\displaystyle\int^\infty_0}
{\rm d}\overline k \;\overline k
\e^{-z_0|\cos\eta|\overline k} J_0(\overline k \rho\sin\eta)
\e^{-{\rm i}\overline k(\cos\eta
z-t)} \\
& & \nonumber \\
&=&
\frac{[z_0-{\rm i}\,\mbox{sgn}(\cos\eta)(z-t/\cos\eta)]}{[\rho^2
\tan^2\eta + [z_0 + {\rm i} \,
\mbox{sgn}(\cos\eta)(z-t/\cos\eta)]^2]^{3/2}} . \label{ae85b}
\end{eqnarray}

Comparing eq.(\ref{ae85a}) with eq.(\ref{ae52}) we discover that ZM
produced in this example a more general $\Phi^>_{X_0}$ wave where
$B(\overline k_<)=\e^{-z_0|\cos\eta|\overline k_<}$. Obviously $\Phi^>
(t,\vec x)$ given by eq.(\ref{ae85b}) moves with superluminal speed
$(1/\!\cos\eta)$ in the positive or negative $z$-direction depending on
the sign of $\cos \eta$, denoted $\mbox{sgn}(\cos \eta)$.

In both examples studied above we see that the projection of the
supporting line of eq.(\ref{ae77}) in the $(\omega,k_z)$ plane is the
straight
line
$k_z/\omega = \cos\eta$, and $\cos\eta$ is its reciprocal slope. This
line
is inside the ``light cone" in the $(\omega,k_z)$ plane. 

\smallskip 

\noindent {\bf Third Example:} Consider two arbitrary lines with the
same
reciprocal slope that we denote by $v>1$, both running between the
lines $\omega=\pm k_z$ in the upper half plane $\omega > 0$ and each
cutting
the
$\omega$-axis at  different  values $\beta_1$ and $\beta_2$.

The two lines  are projections of members of a family of HWE solution
lines and each one can be represented as a portion of the straight
lines (between the lines $\omega=\pm k_z)$
\begin{equation} \label{ae86}
k_z = v (\omega-\beta_1) , \ k_z=v(\omega-\beta_2).
\end{equation}

It is clear that on the solution line of the HWE, $\Omega$ takes values
from zero up to a maximum value that depends on $v$ and $\beta$ and
then back to zero.

We see also that the  maximum value of $\Omega$, given by $\beta
v/\sqrt{v^2-1}$, on any HWE solution line occurs for those values of
$\omega$ and $k_z$ where the corresponding projection lines cut the
line
$\omega=v k_z$. It is clear that there are two points on any HWE
solution
line with the same value of $\Omega$ in the interval
\begin{equation} \label{ae87}
0 < \Omega < v \beta / \sqrt{v^2-1} = \Omega_0.
\end{equation}
It follows that in this case the HWE solution line breaks into two
segments, as is the case of the projection lines. We can then
associate two different weighting functions, one for each segment. We
write
\begin{eqnarray} \label{ae88}
\widetilde \Phi (\Omega, \omega, k_z) & = & \Xi_1
(\Omega,v,\beta)\delta \left[k_z -
\frac{v[\beta+\sqrt{\beta^2 v^2 - \Omega^2(v^2-1)]}}{(v^2-1)} \right]
\times
\nonumber \\
&& \times  \delta \left[ \omega - \frac{[\beta v^2 + \sqrt{v^2
\beta^2-\Omega^2(v^2-1)]}}{(v^2-1)} \right] + \nonumber \\
&& + \  \Xi_2 (\Omega,v,\beta) \delta \left\{ k_z - \frac{v[\beta -
\sqrt{\beta^2
v^2 - \Omega^2 (v^2-1)]}}{(v^2-1)} \right\} \times \nonumber \\
&& \times  \delta \left\{ \omega - \frac{[\beta v^2 - \sqrt{v^2 \beta^2
- \Omega^2 (v^2-1)]}}{(v^2-1)} \right\} .
\end{eqnarray}

Now, choosing
\begin{equation} \label{aen89}
\Xi_1 (\Omega, v, \beta) = \Xi_2 (\Omega,v,\beta) =
(2\pi)^3/2 \sqrt{\Omega^2_0-\Omega^2}
\end{equation}
we get
\begin{equation} \label{aen90}
\Phi_{v,\beta} (t,\rho,z) = \Omega_0 \exp \left(\frac{{\rm i}\beta
v(z-vt)}{\sqrt{v^2-1}} \right) \int^\infty_0 {\rm d}\raisebox{2.2pt}{$\chi$}
\; \raisebox{2.2pt}{$\chi$} J_0 (\Omega_0
\rho
\raisebox{2.2pt}{$\chi$}) \cos \left\{ \frac{\Omega_0 v}{\sqrt{v^2-1}}
\frac{(z-t/v)}{\sqrt{1-\raisebox{2.2pt}{$\chi$}^2}}\right\}.
\end{equation}
Then 
\begin{equation} \label{ae89}
\Phi_{v,\beta} (t,\rho,z) = \exp \left[ {\rm i} \beta
\frac{v(z-vt)}{\sqrt{v^2-1}} \right] \frac{\sin \left\{
\Omega_0\sqrt{\frac{v^2}{(v^2-1)}(z-t/v)^2+\rho^2}\right\} }{ \left\{
\Omega_0\sqrt{\frac{v^2}{(v^2-1)}(z-t/v)^2+\rho^2}\right\} } \, . 
\end{equation}
If we call $v_< = {\displaystyle\frac{1}v} < 1$ and take into account
the value of $\Omega_0$ given by eq.(\ref{ae87}), we can  write
eq.(\ref{ae89}) as
\begin{eqnarray} \label{aen91}
&& \Phi_{v_<} (t,\rho,z) = \frac{\sin(\Omega_0 \xi_<)}{\xi_<} \e^{{\rm
i} \Omega_0(z-vt)} \; ;  \\ \label{ae90} 
&&  \xi_< = \left[x^2+y^2 + \frac{1}{1-v^2_<} (z-v_< t)^2\right]^{1/2}
\; ;
\end{eqnarray} 
which we recognize as the subluminal spherical Bessel beam of Section
\ref{as4} (eq.(\ref{ae31})).

\subsection{Klein-Gordon equation (KGE)}

We show here the existence of subluminal, luminal and superluminal 
UPW solutions of the KGE. We want to solve
\begin{equation}  \label{ae91}
\left(\frac{\partial^2}{\partial t^2} - \nabla^2 + m^2 \right)
\Phi^{KG} (t,\vec x)
= 0, \qquad m > 0,
\end{equation}
with the Fourier transform method. We obtain for $\widetilde
\Phi^{KG}(\omega,\vec
k)$  (a generalized function) the equation
\begin{equation} \label{ae92}
\{\omega^2 - k^2_z - (\Omega^2+m^2)\} \widetilde\Phi^{KG} (\omega,\vec
k)=0 .
\end{equation}

As in the case of the HWE, any solution of the KGE will have a transform
$\widetilde \Phi (\omega,\vec k)$ such that its support line lies on the
surface

\begin{equation} \label{ae93}
\omega^2 - k^2_z - (\Omega^2 + m^2) = 0 \; .
\end{equation}
From eq.(\ref{ae93}), calling $\Omega^2 + m^2 = K^2$, we see that we
are in a
situation identical to the HWE for which we showed the existence of
subluminal, superluminal and luminal solutions. We write down as
examples one solution of each kind. 

\smallskip 

{\it Subluminal UPW solution of the KGE\/}. To obtain this solution it
is
enough to change in eq.(\ref{ae89}) $\Omega_0 = v\beta/\sqrt{v^2-1}
\rightarrow
\Omega^{KG}_0 =
\left[\left({\displaystyle\frac{v\beta}{\sqrt{v^2-1}}}\right)^2-m^2
\right]^{1/2}$. We have,
\begin{equation} \label{ae94a}
\Phi^{KG}_< (t,\rho,z)  =  \exp \left\{\frac{{\rm i}\beta
v(z-vt)}{\sqrt{v^2-1}}\right\} \frac{\sin (\Omega^{KG}_0
\xi_<)}{\xi_<}
\; ;
\end{equation} 
\begin{equation} \label{ae94b}
\xi_<  =  \left[x^2+y^2 + \frac{1}{1-v^2_<} (z-v_< t)^2 
\right]^{1/2} , \ v_< = 1/v .
\end{equation} 

\smallskip 

{\it Luminal UPW solution of the KGE\/}. To obtain a solution of this
type it is enough, as in eq.(\ref{ae72}), to write
\begin{equation} \label{ae95}
\widetilde\Phi^{KG} = \Xi (\Omega, \beta) \delta [k_z - (\Omega^2 +
(m^2-\beta^2)/2\beta)]
\delta [\omega-(\Omega^2 + (m^2+\beta^2)/2\beta)] \, .
\end{equation}
Choosing
\begin{equation} \label{ae96}
\Xi (\Omega, \beta) = \frac{(2\pi)^2}{\beta} \exp (-z_0
\Omega^2/2\beta),  \; z_0 >
0,
\end{equation}
gives
%
\begin{equation} 
\Phi^{KG}_\beta
(t,\vec x) = 
\e^{{\rm i}z(m^2-\beta^2)/2
\beta} \e^{-{\rm i}t(m^2+\beta)/2\beta} \frac{\exp\{-\rho^2
\beta/2[z_0-{\rm i}(z-t)]\}}{[z_0
- {\rm i}(z-t)]} .
\end{equation} 

{\it Superluminal UPW solution of the KGE}. To obtain a solution of this
kind we introduce a parameter $v$ such that $0 < v < 1$ and write for
$\widetilde \Phi^{KG}$ in (\ref{ae92})
\begin{eqnarray} \label{ae98}
\widetilde\Phi^{KG}_{v,\beta} (\omega,\Omega,k_z) &=& \Xi(\Omega,v,
\beta)
\delta \left[\omega-\frac{\left(-\beta v^2 + \sqrt{(\Omega^2 +
m^2)(1-v^2)+v^2\beta^2}\,\right)}{1-v^2} \right] \times   \nonumber \\
& \times &  \delta \left[k_z-\frac{v\left(-\beta + \sqrt{(\Omega^2 +
m^2)(1-v^2)+v^2\beta^2}\,\right)}{1-v^2} \right] \; . 
\end{eqnarray}
Next we choose
\begin{equation} \label{ae99}
\Xi (\Omega, v, \beta) = \frac{(2\pi)^3
\exp(-z_0\sqrt{\Omega^2_0+\Omega^2})}{\sqrt{\Omega^2_0 + \Omega^2}},
\end{equation}
where $z_0 > 0$ is an arbitrary parameter, and where
\begin{equation} \label{ae100}
\Omega^2_0 = \frac{\beta^2 v^2}{1-v^2} + m^2 \; .
\end{equation}
Then introducing $v_> = 1/v > 1$ and $\gamma_> =
{\displaystyle\frac{1}{\sqrt{v^2_>-1}}}$, we get
\begin{equation} \label{ae101} 
\Phi^{KG_>}_{v,\beta} (t,\vec x) = \e^{\frac{{\rm i}(\Omega^2_0 -
m^2)(z-v t)}{\beta v}} \frac{\exp\left\{-\Omega_0 \sqrt{[z_0-{\rm i} 
\gamma_> (z-v_> t)]^2 + x^2 + y^2}\right\}}{\sqrt{[z_0-{\rm i} 
\gamma_> (z-v_>t)]^2+x^2+y^2}} \; ,
\end{equation}
which is a superluminal UPW solution of the KGE moving with speed $v_>$
in the $z$ direction.  From eq.(\ref{ae101}) it is an easy task to
reproduce the superluminal spherical Bessel beam which is solution of
the HWE. 

\subsection{On the energy of the UPWs and the velocity of transport of
energy} \label{as7} 

Let $\Phi_r (t,\vec x)$ be a real solution of the HWE. Then, as it is
well known \cite{27a}, the energy of the solution is given by
\begin{equation} \label{ae102}
\varepsilon = \int\int\int_{I\!\!R^3} {\rm d}{\bf v}
\left[\left(\frac{\partial\Phi_r}{\partial t}
\right)^2 - \Phi_r \nabla^2 \Phi_r\right] + \lim_{R\rightarrow\infty}
\int\int_{S(R)} {\rm d} S \Phi_r \vec n . \nabla \Phi_r \; ,
\end{equation}
where $S(R)$ is the 2-sphere of radius $R$ .

We can easily verify that the real or imaginary parts of all UPWs
solutions of the HWE presented above have infinite energy. The
question arises of how to project superluminal waves, solutions of
the HWE, with finite energy. This can
be done if we recall that all UPWs discussed above can be indexed by
at least one parameter that here we call $\alpha$. Then,
calling $\Phi_\alpha (t,\vec
x)$ the real or imaginary parts of a given UPW solution we may form
``packets" of these solutions as
\begin{equation} \label{ae103}
\Phi(t,\vec x) = \int {\rm d}\alpha \, F(\alpha) \Phi_\alpha (t,\vec x)
\end{equation}

We now may test for a given solution $\Phi_\alpha$ and for weighting
function $F(\alpha)$ if the integral in eq.(\ref{ae102}) is convergent.
We can
explicitly show that for some (but not all) of the solutions showed
above
(subluminal, luminal and superluminal) that for weighting functions
satisfying certain integrability conditions the energy $\varepsilon$
results
finite.

It is particularly important in this context to quote that the finite
aperture approximations for all UPWs discussed in this paper have, of
course, finite energy.
For the case in which $\Phi$ given by eq.(\ref{ae103}) is used to
generate
solutions for, {\em e.g.\/}, Maxwell of Dirac fields,
the conditions for the energy
of these fields to be finite will in general be different from the
condition that gives for $\Phi$ a finite energy. This problem will be
discussed with more details in another paper.

To finish we remark that  for a scalar field satisfying
\begin{equation} \label{aen104}
\left(\frac{1}{c_*^2} \frac{\partial^2}{\partial t^2} -
\nabla^2 \right)\Phi = 0
\end{equation}
we have as is well known that the flux of momentum is given by
\begin{equation} \label{aen105}
\vec{S} = \nabla \Phi \frac{\partial \Phi}{\partial t}
\end{equation}
and
\begin{equation} \label{aen106}
u = \frac{1}{2} \left[ ( \nabla\Phi)^2 + \frac{1}{c_*^2} \left(
\frac{\partial \Phi}{\partial t} \right)^2 \right] \, .
\end{equation}

We can immediately verify that if the speed of transport of energy is
defined as $v_{\varepsilon} = |\vec{S}|/u$ then $v_{\varepsilon} \leq
c_*$. The acoustic experiments reported in \cite{5,6} show nevertheless
that for the $X$-wave the energy travels with speed $c_s/\cos\eta$ ($c_*
= c_s$). We thus see that the usual definitions of  magnitudes such as
density of energy  and momentum and the velocity of transport of energy
demand a careful revision. (See in this context also the discussion of
Section \ref{3s4}.)

\section{SUBLUMINAL AND SUPERLUMINAL UPW SOLUTIONS OF
MAXWELL
EQUATIONS (ME)} \label{sec3}

In this Section we make full use of the Clifford bundle formalism (CBF)
summarized in Section \ref{sec1}, but we translate  all the main results
into the standard vector formalism used by physicists. We start by
reanalyzing in Section \ref{3s1} the plane wave solutions (PWS) of ME
with the CBF. We clarify some misconceptions and explain the fundamental
role of the duality operator $\gamma_5$ and the meaning of ${\rm
i}=\sqrt{-1}$ in standard formulations of electromagnetic theory.  Next,
in Section \ref{3s2} we discuss subluminal UPWs solutions of ME and an
unexpected relation between these solutions and the possible 
existence of purely electromagnetic particles (PEPs) envisaged by
Einstein \cite{27b}, Poincar\'e \cite{28}, Ehrenfest \cite{29} and
recently discussed by Waite, Barut and Zeni \cite{30,31}.  In Section
\ref{3s3} we discuss the theory of superluminal electromagnetic
$X$-waves (SEXWs). In \cite{5,6} we present simulations of the motions
of the SEXWs and of their finite aperture approximations, which can
eventually be launched by appropriate physical devices.

\subsection{Plane wave solutions of Maxwell equations} \label{3s1} 

We recall that ME in vacuum can be written as [eq.(\ref{be75})]
\begin{equation} \label{3e1}
\partial F = 0,
\end{equation}
where $F \sec \mbox{$\bigwedge$}^2(M) \subset \sec {\cal C}\ell(M)$. The
well known PWS of
eq.(\ref{3e1}) are obtained as follows. We write in a given Lorentzian
chart
$\langle x^\mu\rangle$ of the maximal atlas of $M$ a PWS
moving in the $z$-direction  
\begin{equation} \label{3e2}
F = f \e^{\gamma_5 kx}
\end{equation}  
\begin{equation}\label{3e3}
k = k^\mu \gamma_\mu,\;  k^1=k^2=0, \; x = x^\mu\gamma_\mu ,
\end{equation}
where $k, x \in \sec \mbox{$\bigwedge$}^1(M) \subset \sec {\cal
C}\ell(M)$ and where $f$
is a constant 2-form. From eqs.(\ref{3e1}) and (\ref{3e2}) we obtain 
\begin{equation} \label{3e4}
kF=0
\end{equation}
Multiplying eq.(\ref{3e4}) by $k$ we get  
\begin{equation} \label{3e5}
k^2F=0
\end{equation}
and since $k \in \sec \mbox{$\bigwedge$}^1(M) \subset \sec {\cal
C}\ell(M)$ then
\begin{equation} \label{3e6}
k^2 = 0 \ \leftrightarrow \ k_0 = \pm |\vec k|=k^3 ,
\end{equation}
{\em i.e.\/}, the propagation vector is light-like. Also
\begin{equation} \label{3e7}
F^2 = F.F + F \wedge F = 0
\end{equation}
as can be easily seen by multiplying both members of eq.(\ref{3e4}) by $F$
and
taking into account that $k \neq 0$. Eq(\ref{3e7}) says that the field
invariants are null.

It is interesting to understand the fundamental role of the volume
element $\gamma_5$ (duality operator) in electromagnetic theory. In
particular since $\e^{\gamma_5 kx} = \cos kx + \gamma_5 \sin kx$, we see
that
\begin{equation} \label{3e8}
F = f \cos k x + \gamma_5 f \sin kx.
\end{equation}
Writing $F = \vec E + {\bf i} \vec B$, (see eq.(\ref{be72})) with ${\bf
i} \equiv \gamma_5$ and choosing $f = \vec{e}_1 + {\bf i} \vec{e}_2$,
$\vec{e}_1 . \vec{e}_2 =0$, $\vec{e}_1$, $\vec{e}_2$ constant vectors
in the Pauli subalgebra sense, eq.(\ref{3e8}) becomes 
\begin{equation} \label{3e9} 
\vec E + {\bf i}\vec B = \vec{e}_1 \cos kx - \vec{e}_2 \sin kx +  
{\bf i} ( \vec{e}_1 \sin k x + \vec{e}_2 \cos kx) .  
\end{equation} 
This equation is important because it shows that we must take care with
the ${\rm i}=\sqrt{-1}$ that appears in usual formulations of Maxwell
theory using complex electric and magnetic fields. The ${\rm i} =
\sqrt{-1}$ in many cases unfolds a secret that can only be known through
eq.(\ref{3e9}).  From eq.(\ref{3e4}) we can also easily show that $\vec
k . \vec E = \vec k .  \vec B = 0$, {\em i.e.\/}, PWS of ME are {\it
transverse} waves.

We can rewrite eq.(\ref{3e4}) as
\begin{equation} \label{3e10}
k\gamma_0 \gamma_0 F \gamma_0 = 0
\end{equation}
and since $k \gamma_0 = k_0 + \vec k, \ \gamma_0 F \gamma_0 = - \vec E
+ {\bf i}
\vec B$ we have
\begin{equation} \label{3e11}
\vec k f = k_0 f .
\end{equation}
Now, we recall that in ${\cal C}\ell^+(M)$
(where, as we said in Section \ref{sec1}, the typical fiber is
isomorphic to
the
Pauli algebra ${\cal C}\ell_{3,0}$) we can introduce \cite{32} the
operator of
space
conjugation denoted by $*$ such that writing $f = \vec e + {\bf i} \vec
b$ we have
\begin{equation}  \label{3e12}
f^* = - \vec{e} + {\bf i} \vec{b} \ \ ; \ \ k^*_0 = k_0 \ \ ; \ \ \vec k^* = - \vec k .
\end{equation}
We can now interpret the two solutions of $k^2 = 0$, {\em i.e.\/}, $k_0=| 
\vec k |$ and $k_0 = -|\vec k|$ as corresponding to the solutions $k_0 f =
\vec kf$ and $k_0 f^* = - \vec kf^*$; $f$ and $f^*$ correspond in
quantum theory to ``photons" which are of positive or negative
helicities. We can interpret $k_0 = |\vec k|$ as a particle and $k_0 =
-|\vec k|$ as an antiparticle.

Summarizing we have the following  important facts concerning PWS of
ME: (i) the propagation vector is light-like, $k^2=0$; (ii) the
field invariants are null, $F^2=0$; (iii) the PWS are transverse
waves, {\em i.e.\/}, $\vec k . \vec E = \vec k . \vec B = 0$. 

\subsection{Subluminal solutions of Maxwell equations and purely
electromagnetic particles} \label{3s2} 

We take $\Phi \in \sec (\mbox{$\bigwedge$}^0(M) \oplus
\mbox{$\bigwedge$}^4(M)) \subset \sec {\cal C}\ell(M)$ and consider the
following Hertz potential $\pi \in \sec \mbox{$\bigwedge$}^2(M) \subset
\sec {\cal C}\ell(M)$ [eq.(\ref{be83})]
\begin{equation} \label{3e13}
\pi = \Phi \gamma^1 \gamma^2 .
\end{equation}
We now write
\begin{equation} \label{3e14}
\Phi(t, \vec x) = \phi(\vec x) \e^{\gamma_5 \Omega t} .
\end{equation}
Since $\pi$ satisfies the wave equation, we have
\begin{equation} \label{3e15}
\nabla^2 \phi(\vec x) + \Omega^2 \phi(\vec x) = 0
\end{equation}
Solutions of eq.(3.15) (the Helmholtz equation) are well known. Here 
we consider the simplest solution in spherical coordinates,
\begin{equation} \label{3e16}
\phi(\vec x) = C \frac{\sin\Omega r}{r} \ \ , \ \ r =
\sqrt{x^2+y^2+z^2},
\end{equation}
where $C$ is an arbitrary real constant. From the results of Section
\ref{sec1} we obtain the  following stationary electromagnetic field,
which
is at rest in the reference frame $Z$ where $\langle x^\mu\rangle$ are
naturally adapted coordinates to $Z$ (see Section \ref{sec4} for the definition of these concepts). 

\begin{eqnarray} \label{3e17} 
&& F_0 = \frac{C}{r^3} [\sin \Omega t(\alpha \Omega r \sin \theta \sin
\varphi - \beta
\cos \theta\sin \theta \cos \varphi) \gamma_0 \gamma_1 \nonumber \\
&& - \sin \Omega t (\alpha\Omega r\sin\theta \cos \varphi + \beta
\sin\theta \cos\theta
\sin \varphi)\gamma_0 \gamma_2 \nonumber \\
&& + \sin \Omega t (\beta \sin^2 \theta - 2\alpha)\gamma_0 \gamma_3
+ \cos \Omega t
(\beta \sin^2 \theta - 2 \alpha ) \gamma_1 \gamma_2 \nonumber \\
&& + \cos\Omega t (\beta\sin\theta \cos\theta \sin\varphi +
\alpha\Omega r \sin \theta \cos
\varphi) \gamma_1 \gamma_3 \nonumber \\
&& + \cos \Omega t (-\beta \sin\theta \cos\theta \cos \varphi +
\alpha \Omega r \sin \theta
\sin\varphi)  \gamma_2 \gamma_3]
\end{eqnarray}
with $\alpha=\Omega r \cos\Omega r - \sin\Omega r$ and $\beta = 3\alpha
+ \Omega^2 r^2
\sin \Omega r$. Observe that $F_0$ is regular at the origin and
vanishes
at infinity. Let us rewrite the solution using the Pauli-algebra in
${\cal C}\ell^+(M)$. Writing $({\bf i} \equiv \gamma_5)$
\begin{equation} \label{3e18}
F_0 = \vec E_0 + {\bf i} \vec B_0
\end{equation}
we get
\begin{equation} \label{3e19}
\vec E_0 = \vec W \sin \Omega t \ \ , \ \ \vec B_0=\vec W \cos\Omega
t
\end{equation}
with
\begin{equation} \label{3e20}
\vec W = - C \left( \frac{\alpha\Omega y}{r^3} - \frac{\beta x z}{r^5}
,
-
\frac{\alpha\Omega x}{r^3} - \frac{\beta y z}{r^5},
\frac{\beta(x^2+y^2)}{r^5}
- \frac{2\alpha}{r^3}\right) .
\end{equation}
We verify that ${\rm div} \vec W = 0$, ${\rm div} \vec E_0 = {\rm div}
\vec B_0 =
0$, ${\rm rot} \vec E_0 + \partial \vec B_0/\partial t = 0$, ${\rm rot}
\vec B_0 - \partial \vec
E_0/\partial t = 0$, and
\begin{equation} \label{3e21} 
{\rm rot} \vec W =  \Omega \vec W.
\end{equation}
Now, from eq.(\ref{be89}) we know that $T_0 = {\displaystyle -
\frac{1}2} F \gamma_0 F$ is the 1-form representing the energy density
and the Poynting vector. It follows that $\vec E_0 \times \vec B_0 = 0$,
{\em i.e.\/}, the solution has zero angular momentum. The energy density $u =
S^{00}$ is given by
\begin{equation} \label{3e22} 
u = \frac{1}{r^6} [\sin^2 \theta (\Omega^2 r^2 \alpha^2 + \cos^2 \theta
\beta^2) +
(\beta\sin^2 \theta - 2\alpha)^2] . 
\end{equation}
Then $\int\!\int\!\int_{I\!\!R^3} u\,{\rm d}{\bf v} = \infty$.  
As for the case of the scalar field (see Section \ref{3s4}) 
a finite
energy solution can be constructed by considering ``wave packets"
with a distribution of intrinsic frequencies $F(\Omega)$ satisfying
appropriate conditions. Many possibilities exist, but they will not be
discussed here. Instead, we prefer to direct our attention to
eq.(\ref{3e21}). As it is well known, this is a very important equation
(called the force free equation \cite{30}) that appears {\em e.g.\/} in
hydrodynamics and in several different situations in plasma
physics \cite{33a}. The following considerations are more important.

Einstein \cite{27b} among others (see \cite{30} for a review) studied
the possibility of constructing PEPs. He started from Maxwell
equations for a PEP configuration described by an electromagnetic
field $F_p$ and a current density $J_p$, where
\begin{equation} \label{3e23}
\partial F_p = J_p
\end{equation}
and rightly concluded that the condition for existence of PEPs is
\begin{equation} \label{3e24}
J_p . F_p = 0.
\end{equation}
This condition implies in vector notation
\begin{equation} \label{3e25}
\rho_p \vec E_p = 0 \ , \ \ \vec\jmath_p . \vec E_p = 0 \ , \ \ 
\vec\jmath_p \times \vec B_p = 0
\end{equation}
From eq.(\ref{3e25}) Einstein concluded that the only possible solution
of eq.(\ref{3e23}) with the subsidiary condition given by
eq.(\ref{3e24}) is $J_p = 0$. However, this conclusion is correct, as
pointed in \cite{30,31}, only if $J^2_p > 0$, {\em i.e.\/}, if $J_p$ is a
time-like current density. If we suppose that $J_p$ can be
spacelike, {\em i.e.\/}, $J^2_p < 0$, there exists a reference frame where
$\rho_p = 0$ and a possible solution of eq.(3.24) is
\begin{equation}
\rho_p = 0 \ \ , \ \ \vec E_p . \vec B_p = 0 \ \ , \ \ \vec\jmath_p =
KC\vec B_p ,
\end{equation}
where $K = \pm 1$ is called the chirality of the solution and $C$ is
a real constant. In \cite{30,31} static solutions of eq.(\ref{3e23})
and (\ref{3e24})
are
exhibited where $\vec E_p = 0$. In this case we can verify that $\vec
B_p$ satisfies
\begin{equation}
\nabla \times \vec B_p = KC\vec B_p .
\end{equation}
Now, if we choose $F \in \sec \mbox{$\bigwedge$}^2 (M) \subset \sec
{\cal C}\ell(M)$ such
that
\begin{equation} \label{3e28}
\begin{array}{c}
F_0 = \vec E_0 + {\bf i} \vec B_0 \, , \\
\vec E_0 = \vec B_p \cos \Omega t \, , \quad  \vec B_0 = \vec B_p \sin
\Omega t
\end{array}
\end{equation}
and $\Omega = KC > 0$, we immediately realize that
\begin{equation} \label{3e29}
\partial F_0 = 0 \, . 
\end{equation}
This is an amazing result, since it means that the free Maxwell
equations may have stationary solutions that model PEPs. In such
solutions the structure of the field $F_0$ is such that we can write
\begin{equation} \label{3e30}
\begin{array}{c}
F_0 = F_{p}^{'} + \overline F = {\bf i} \vec{W} \cos \Omega t - \vec{W}
\sin
\Omega t  \, , \\
\partial F_{p}^{'} = -\partial \overline F = J_{p}^{'} \, , 
\end{array}
\end{equation}
{\em i.e.\/}, $\partial F_0=0$ is equivalent to a field plus a current. This
opens several interesting possibilities for modeling PEPs (see also
\cite{33b}) and we discuss more this issue in another publication.

We observe that moving subluminal solutions of ME can be easily
obtained
choosing as Hertz potential, {\em e.g.\/},
\begin{eqnarray} \label{3e31}
&& \pi^<(t,\vec x) = C \frac{\sin\Omega\xi_<}{\xi_<} \exp
[\gamma_5(\omega_<
t-k_< z)]\gamma_1 \gamma_2 \, , \\
&& \hspace*{7mm} \omega^2_< - k^2_< = \Omega^{2}_{<} \, , 
\nonumber \\
&& \xi_< = [x^2+y^2+ \gamma^2_< (z-v_< t)^2] \, , \\
&&  \gamma_< = \frac{1}{\sqrt{1-v^2_<}} \ , \ \ v_< =
{\rm d}\omega_</ {\rm d}k_< \, . \nonumber
\end{eqnarray}
We are not going to write explicitly the expression for $F^<$
corresponding to $\pi^<$ because it is very long and will not be used
in what follows.

We end this Section with the following observations: (i) In general
for subluminal solutions of ME (SSME) the propagation vector
satisfies an equation like eq.(\ref{3e30}).  (ii) As can be easily
verified,
for a SSME the field invariants are non-null. (iii) A SSME is not a
transverse wave. This can be seen explicitly from eq.(\ref{3e20}).

Conditions (i), (ii), (iii) are in contrast with the case of the PWS of
ME. In \cite{34,35} Rodrigues and Vaz showed that for free
electromagnetic
fields $(\partial F=0)$ such that $F^2 \neq 0$, there exists a
Dirac-Hestenes
equation
for $\psi \in \sec (\mbox{$\bigwedge$}^0(M) +
\mbox{$\bigwedge$}^2(M) +
\mbox{$\bigwedge$}^4(M))
\subset \sec {\cal C}\ell(M)$ where $F = \psi \gamma_1 \gamma_2
\widetilde \psi$. This was
the reason why Rodrigues and Vaz discovered subluminal and
superluminal solutions of Maxwell equations (and also of Weyl
equation \cite{15}) which solve the Dirac-Hestenes equation
[eq.(\ref{be94})]. An explicit superluminal solution of Maxwell equations
is given in \cite{15} using as Hertz potential $\Pi_> = \Phi_> \gamma^{12}$ 
where $\Phi_>$ is given by eq.(\ref{ae30}).

\subsection{The superluminal electromagnetic $X$-wave  
(SEXW)} 
\label{3s3} 

In this Section we present a family of solutions of Maxwell equations called 
the superluminal electromagnetic  $X$-waves $F_{XBB_n}$. A solution 
dual to $F_{XBB_n}$, called $\star F_{XBB_n}$ has been first presented 
by Lu and Greenleaf in an unpublished paper \cite{35b}. Later the solutions 
$F_{XBB_n}$ and others associated with it have been studied in detail  
\cite{37b,6}. 

To simplify the matter in what follows we now suppose that the functions
$\Phi_{X_n}$ [eq.(\ref{ae52})] and $\Phi_{XBB_n}$ [eq.(\ref{ae53})]
which are superluminal solutions of the scalar wave equation are 0-forms
sections of the complexified Clifford bundle ${\cal C}\ell_C(M) =
\;\mbox{{\sf I}}\!\!\!C \otimes {\cal C}\ell(M)$ (see Section
\ref{bs6}). We rewrite eqs.(\ref{ae52}) and
(\ref{ae56}) as ($n=0,1,2,\ldots$)
\begin{equation} \label{3e33}
\Phi_{X_n} (t, \vec x) = \e^{{\rm i}n\theta} \int^\infty_0 B(\overline
k) J_n (\overline k \rho\sin\eta) \e^{-\overline k[a_0 - {\rm i}(z
\cos\eta-t)]} {\rm d}\overline k
\end{equation}
and choosing $B(\overline k) = a_0$ we have
\begin{eqnarray} \label{3e34}
&& \hspace*{7mm} \Phi_{XBB_n} (t,\vec x) = \frac{a_0(\rho\sin \eta)^n
\e^{{\rm i}n\theta}}{\sqrt{M}(\tau+\sqrt{M})^n} \\  \label{3e35}
&& M = (\rho\sin\eta)^2 + \tau^2 \ \ \ ; \ \ \  \tau = [a_0 -
{\rm i}(z\cos\eta-t)].
\end{eqnarray}   
 
Further, we suppose now that the Hertz potential
$\pi$, the vector potential A and the corresponding electromagnetic
field $F$ are appropriate sections of ${\cal C}\ell_C(M)$. We
take
\begin{equation} \label{3e36}
\pi=\Phi \gamma_1 \gamma_2 \in \sec \;\mbox{{\sf I}}\!\!\!C \otimes
\mbox{$\bigwedge$}^2(M) \subset \sec
{\cal C}\ell_C(M),
\end{equation}
where $\Phi$ can be $\Phi_{X_n}, \Phi_{XBB_n}, \Phi_{XBL_n}$. 
Let us start by giving the
explicit form of the $F_{XBB_n}$ {\em i.e.\/}, the SEXWs. In this case
eq.(\ref{be84})
gives $\pi= {\bf i}\,\vec \pi_m$ and
\begin{equation} \label{3e37}
\vec \pi_m = \Phi_{XBB_n} \mbox{\boldmath $z$}
\end{equation}
where \mbox{\boldmath $z$} is the versor of the $z$-axis. Also, let
$\mbox{\boldmath $\rho$},
\mbox{\boldmath $\theta$}$  be respectively the versors of the $\rho$
and $\theta$ directions
where $(\rho, \theta, z)$ are the usual cylindrical coordinates.
Writing
\begin{equation} \label{3e38} 
F_{XBB_n} = \vec E_{XBB_n} + \gamma_5 \vec B_{XBB_n}
\end{equation}
we obtain from equations (\ref{ae53}) and (\ref{be85}):
\begin{equation} \label{3e39} 
\vec E_{XBB_n} = -\frac{\mbox{\boldmath $\rho$}}{\rho}
\frac{\partial^2}{\partial t\partial\theta}
\Phi_{XBB_n} + \mbox{\boldmath $\theta$} \frac{\partial^2}{\partial
t\partial \rho} \Phi_{XBB_n} 
\end{equation}  
%
\begin{equation} \label{3e40} 
\vec B_{XBB_n} = \mbox{\boldmath $\rho$} \frac{\partial^2}{\partial
\rho\partial z}
\Phi_{XBB_n} + \mbox{\boldmath $\theta$} \frac{1}{\rho}
\frac{\partial^2}{\partial \theta \partial z}
\Phi_{XBB_n} +
\mbox{\boldmath $z$} \left(\frac{\partial^2}{\partial z^2} 
- \frac{\partial^2}{\partial
t^2}  \right) \Phi_{XBB_n}
\end{equation} 

Explicitly we get for the components in cylindrical coordinates,

\hspace*{7mm} $(\vec E_{XBB_n})_\rho = -{\displaystyle\frac{1}{\rho}
n\frac{M_3}{\sqrt{M}}}
\Phi_{XBB_n}$ \hfill (4.41a)

\hspace*{7mm} $(\vec E_{XBB_n})_\theta = {\displaystyle\frac{1}{\rho}
{\rm i}\frac{M_6}{\sqrt{M}
M_2}
\Phi_{XBB_n}}$ \hfill (4.41b)

\hspace*{7mm} $(\vec B_{XBB_n})_\rho = \cos\eta (\vec
E_{XBB_n})_\theta$ \hfill
(4.41c)

\hspace*{7mm} $(\vec B_{XBB_n})_\theta = - \cos\eta (\vec
E_{XBB_n})_\rho$ \hfill
(4.41d)

\hspace*{7mm} $(\vec B_{XBB_n})_z = - \sin^2\eta {\displaystyle
\frac{M_7}{\sqrt{M}}}
\Phi_{XBB_n}$. \hfill
(4.41e)

The functions $M_i (i=2,\ldots,7)$ in (4.41) are

\hspace*{7mm} $M_2 = \tau + \sqrt{M}$ \hfill (4.42a)

\hspace*{7mm} $M_3 = n + {\displaystyle\frac{1}{\sqrt{M}}} \tau$ \hfill
(4.42b)

\hspace*{7mm} $M_4 = 2n + {\displaystyle\frac{3}{\sqrt{M}}}\tau$ \hfill
(4.42c)

\hspace*{7mm} $M_5 = \tau + n\sqrt{M}$  \hfill (4.42d)

\hspace*{7mm} $M_6 = (\rho^2 \sin^2\eta {\displaystyle\frac{M_4}{M}} -
n M_3)M_2 + n \rho^2
 {\displaystyle\frac{M_5}{M}} \sin^2\eta$  \hfill (4.42e)

\hspace*{7mm} $M_7 = (n^2-1) {\displaystyle\frac{1}{\sqrt{M}} + 3 n
\frac{1}{M}} \tau +
3  {\displaystyle\frac{1}{\sqrt{M^3}}}\tau^2$  \hfill (4.42f)

We immediately see from eqs.(4.41) that the $F_{XBB_n}$ are indeed
superluminal UPWs solutions of ME, propagating with speed $1/\cos\eta$
in the $z$-direction. That $F_{XBB_n}$ are UPWs is trivial and that
they propagate with speed $c_1=1/\cos\eta$ follows because
$F_{XBB_n}$ depends only on the combination of variables $(z-c_1t)$
and any  derivatives of $\Phi_{XBB_n}$ will keep the
$(z-c_1t)$ dependence structure.

Now, the Poynting vector $\vec P_{XBB_n}$ and the energy density
$u_{XBB_n}$ for $F_{XBB_n}$ are obtained by considering the real
parts of $\vec E_{XBB_n}$ and $\vec B_{XBB_n}$. We have, \\

\hspace*{7mm} $(\vec P_{XBB_n})_\rho = - Re\{(\vec E_{XBB_n})_\theta\}
Re\{(\vec
B_{XBB_n})_z\}$ \hfill (4.43a)

\hspace*{7mm} $(\vec P_{XBB_n})_\theta =  Re\{(\vec E_{XBB_n})_\rho\}
Re\{(\vec
B_{XBB_n})_z\}$ \hfill (4.43b)

\hspace*{7mm} $(\vec P_{XBB_n})_z = \cos\eta \left[|Re\{(\vec
E_{XBB_n})_\rho\}|^2 + |
Re\{(\vec
E_{XBB_n})_\theta\}|^2 \right]$ \hfill (4.43c)

\setcounter{equation}{43}
\begin{eqnarray} \label{4e44} 
u_{XBB_n} &=& (1+\cos^2 \eta) \left[|Re\{(\vec E_{XBB_n})_\rho\}|^2 + |Re
\{(\vec E_{XBB_n})_\theta\}|^2\right] \nonumber \\ 
&+& |Re\{(\vec B_{XBB_n})_z\}|^2 \, . 
\end{eqnarray}

The total energy of $F_{XBB_n}$ is then
\begin{equation}
\varepsilon_{XBB_n} = \int^\pi_{-\pi} {\rm d}\theta \int^{+\infty}_{-\infty}
{\rm d}z
\int^\infty_0 \rho \, {\rm d}\rho \, u_{XBB_n}
\end{equation}

Since as $z \rightarrow \infty \ \vec E_{XBB_n}$ decrease as
$1/|z-t\cos\eta|^{1/2}$ which occurs for the $X$-branches of
$F_{XBB_n}$,  $\varepsilon_{XBB_n}$ may not be finite.  Nevertheless, as
in the case of the acoustic $X$-waves, which experiments have shown to
travel with $v>c_s$ \cite{6},  we are quite sure that a finite aperture
approximation to $F_{XBL_n}$ (FAA$F_{XBL_n}$) can be launched over a
large distance. Indeed in \cite{6} computer simulations for the motion
of FAA$F_{XBB_n}$ are exhibited showing that with an antenna of 20~m of
diameter a FAA$F_{XBB_n}$ centered at a frequency of 700~GHz propagates
with superluminal speed without appreciable distortion up to 100~Km. See
also \cite{37b}. Obviously in this case the total energy of the
FAA$F_{XBL_n}$ is finite.

We conclude this Section observing that in general both subluminal and
superluminal UPW solutions of ME have non-null field invariants and are
not transverse waves. In particular our solutions have a longitudinal
component along the $z$-axis. This result is important because it shows
that, contrary to the speculations of Evans \cite{36}, we do not need an
electromagnetic theory with a non zero  photon-mass, {\em i.e.\/}, with
$F$ satisfying Proca's  equation (as proposed also by de Broglie
\cite{37}) in order to have an electromagnetic wave with a longitudinal
component.  Since Evans presents evidence \cite{36} of the existence of
longitudinal magnetic fields in many different physical situations, we
conclude that the theoretical and experimental study of subluminal and
superluminal UPW solutions of ME must be continued.

\subsection{The velocity of transport of energy of the UPW
solutions of Maxwell equations} \label{3s4} 

Since we found in this paper UPWs solutions of Maxwell equations with
speeds $0 \leq v < \infty$, the following question arises naturally:
Which is the velocity of transport of the energy of a superluminal UPW
(or quasi UPW) solution of ME?

We can find in many physics textbooks ({\em e.g.\/} \cite{38}) and in
scientific papers \cite{20,21} the following argument. Consider  an
arbitrary solution of ME in vacuum $\partial F = 0$. Then if $F = \vec E
+ {\bf i} \vec B$ (see eq.(\ref{be77})) it follows that the Poynting
vector and the energy density of the field are
\begin{equation} \label{3e46}
\vec P = \vec E \times \vec B \, , \quad  u=\frac{1}2 (\vec E^2 + \vec
B^2) \, . 
\end{equation}
It is obvious that the following inequality always holds:
\begin{equation} \label{3e47}
v_\varepsilon = \frac{|\vec P|}{u} \leq 1.
\end{equation}

Now,  the conservation of energy-momentum reads in integral form over
a finite volume $V$ with boundary $S = \partial V$: 
\begin{equation} \label{3e48}
\frac{\partial}{\partial t} \left\{ \int\!\!\int\!\!\int_V {\rm d}{\mbox{\bf
v}}
\frac{1}2 (\vec E^2 + \vec
B^2)\right\} = \oint_S d\vec S . \vec P
\end{equation}
Eq.(\ref{3e48}) is interpreted saying that $\oint_S d\vec S . \vec P$ is
the field energy flux across the surface $S = \partial V$, so that $\vec
P$ is the flux density --- the amount of field energy passing through a
unit area of the surface in unit time.
Now, for  plane wave solutions of Maxwell equations,
\begin{equation} \label{3e49}
v_\varepsilon = 1
\end{equation}
and this result gives origin to the ``dogma'' that free electromagnetic
fields
transport energy at speed $v_\varepsilon=c=1$.

However $v_\varepsilon \leq 1$ is true even for subluminal and
superluminal solutions of ME, as the ones discussed in Sections
\ref{3s2} and \ref{3s3}.  The same is true for the superluminal modified
Bessel beam found by Band \cite{20} in 1987. There he claims that since
$v_\varepsilon \leq 1$ there is no conflict between superluminal
solutions of ME and Relativity Theory since what Relativity forbids is
the propagation of energy with speed greater than $c$.

Here we challenge this conclusion. The fact is that as well known
$\vec P$ is not uniquely defined. Eq.(\ref{3e48}) continues
to hold true if we substitute $\vec P \mapsto \vec P
+ \vec P'$ with $\nabla . \vec P' = 0$. But of course we can easily
find for subluminal, luminal or superluminal solutions of Maxwell
equations a $\vec P'$ such that
\begin{equation} \label{3e50}
\frac{|\vec P + \vec P'|}{u} \geq 1 .
\end{equation}
We arrive at the conclusion that the question of the transport of
energy in superluminal UPWs solutions of ME is an experimental
question. For the acoustic superluminal $X$-wave solution of the HWE (see
\cite{5,6}) the energy around the peak area flows together with the
wave, {\em i.e.\/}, with speed $c_1 = c_s/\cos \eta$ while (as we said in
Section \ref{as7}) the usual theory predicts for the speed of
propagation of sound waves that $|\vec{S}|/u < c_s$, where $\vec{S}$ is
the flux of momentum and $u$ is the energy density (eqs. (\ref{aen105})
and (\ref{aen106})). Since we can see no possibility of the field
energy of the superluminal electromagnetic wave to travel outside the
wave we are confident to state that the velocity of energy transport of
superluminal electromagnetic waves is superluminal.

Before ending we give another example to illustrate that eq.(\ref{3e47}) is
devoid of physical meaning. Consider a spherical conductor in
electrostatic equilibrium with uniform superficial charge density
(total charge $Q$) and with a dipole magnetic moment. Then we have
\begin{equation} \label{3e51}
\vec E = Q\frac{\mbox{\bf r}}{r^2} \ \ ; \ \ \vec B = \frac{C}{r^3}
(2\cos\theta \, {\bf r} + \sin \theta \, \mbox{\boldmath $\theta$})
\end{equation}
and
\begin{equation} \label{3e52}
\vec P = \vec E \times \vec B = \frac{CQ}{r^5} \sin\theta \,
\mbox{\boldmath $\varphi$} \ \ , \ \ u = \frac{1}{2}
\left[\frac{Q^2}{r^4} + \frac{C^2}{r^6} (3\cos^2 \theta + 1) \right].
\end{equation} 
Thus
\begin{equation} \label{3e53}
\frac{|\vec P|}{u} = \frac{2rCQ \sin\theta}{r^2 Q^2+C^2(3\cos^2
\theta+1)} \neq 0, \quad \mbox{for $r\neq 0$.} 
\end{equation} 
Since the fields are static the conservation law eq.(4.3) continues to
hold true, as there is no motion of charges and for any closed surface
containing the spherical conductor we have
\begin{equation} \label{3e54}
\oint_S d\vec S . \vec P = 0.
\end{equation}
But {\it nothing} is in motion! In view of these results we must
investigate whether the existence of superluminal UPWs solutions of ME
is compatible or not with the
Principle of Relativity. We analyze this question in  detail in the
next Section.

To end this Section we recall that in Section~2.19 of his book \cite{14} 
Stratton presents a discussion of the Poynting vector and energy
transfer which essentially agrees with the view presented above. Indeed
he finished that Section with the words: ``By this standard there is
every reason to retain the Poynting-Heaviside viewpoint until a clash
with new experimental evidence shall call for its
revision.''  

\section{SUPERLUMINAL SOLUTIONS OF MAXWELL EQUATIONS AND
THE
PRINCIPLE OF RELATIVITY} \label{sec4}

In \cite{6} it was shown that it seems possible with present technology
to launch in free space finite aperture approximations to the
superluminal electromagnetic waves (SEXWs).  We show in the following
that the physical existence of SEXWs implies a breakdown of the
Principle of Relativity (PR). Since this is a fundamental issue, with
implications for all branches of theoretical physics, we will examine
the problem with great care.  In Section \ref{4s1} we give a rigorous
mathematical definition of the PR and in Section \ref{4s2} we present
the proof of the above statement.

\subsection{Mathematical formulation of the Principle of
Relativity
and its physical meaning} \label{4s1} 

In Section \ref{sec1} we defined Minkowski spacetime as the triple
$\langle M,g,D\rangle$, where $M \simeq I\!\!R^4$, $g$ is a Lorentzian
metric and $D$ is the Levi-Civita connection of $g$.
Consider now $G_M$, the group of all diffeomorphisms of $M$, called
the manifold mapping group. Let ${\bf T}$ be a geometrical
object defined in $A\subseteq M$. The diffeomorphism
$h\in G_M$ induces a deforming mapping
$
h_*: {\bf T} \rightarrow h_* {\bf T} = {\bf T}
$
such that: 

\begin{description}

\item[(i)] If $f: M\supseteq A \rightarrow I\!\!R$, then $h_*f = f\circ
h^{-1}: h(A)
\rightarrow
I\!\!R$. 

\item[(ii)] If ${\bf T} \in \sec T^{(r,s)}(A) \subset \sec T(M)$, where
$T^{(r,s)}(A)$ is the sub-bundle of tensors of type $(r,s)$ of the
tensor bundle $T(M)$, then
\begin{equation} \label{4e1}
(h_*{\bf T})_{h_e} (h_* \omega_1,\ldots, h_* \omega_r, h_*
X_1,\ldots,h_*X_s)  
 = {\bf T}_e(\omega_1,\ldots,\omega_r, X_1,\ldots,X_s)
\end{equation} 
$\forall X_i \in T_e A$, $i=1,\ldots,s$, $\forall \omega_j \in T^*_e A$,
$j=1,\ldots,r$, 
$\forall e \in A $. 

\item[(iii)] If $D$ is the Levi-Civita connection and $X,Y \in \sec
TM$, then
\begin{equation} \label{4e2}
(h_*D_{h*x} h_* Y)_{he} h_{*f} = (D_xY)_e f \ \ \ \forall e \in M \, . 
\end{equation}

\end{description}

If $\{f_\mu = \partial/\partial x^\mu\}$ is a coordinate basis for $TA$
and $\{\theta^\mu ={\rm d}x^\mu\}$ is the corresponding dual basis for
$T^*A$ and if
\begin{equation} \label{4e3}
{\bf T} = T^{\mu_1\ldots \mu_r}_{\nu_1\ldots\nu_s} \theta^{\nu_1}
\otimes\ldots
\otimes \theta^{\nu_s} \otimes f_{\mu_1}\otimes\ldots\otimes f_{\mu_r}
,
\end{equation}
then
\begin{equation} \label{4e4}
h_*{\bf T} = [T^{\mu_1\ldots \mu_r}_{\nu_1\ldots\nu_s}\circ h^{-1}] h_*
\theta^{\nu_1} \otimes\ldots \otimes h_* \theta^{\nu_r} \otimes h_*
f_{\mu_1} \otimes\ldots\otimes
h_*f_{\mu_s}
\end{equation}
Suppose now that $A$ and $h(A)$ can be covered by the local chart
$(U,\eta)$ of the maximal atlas of $M$, and $A \subseteq U, h(A)
\subseteq
U$. Let $\langle x^\mu\rangle$ be the coordinate functions associated
with
$(U,\eta)$. The mapping
\begin{equation} \label{4e5}
x^{'\mu} = x^{\mu} \circ h^{-1} \ : \ h(U) \rightarrow I\!\!R
\end{equation}
defines a coordinate transformation $\langle x^\mu\rangle \mapsto
\langle x^{'\mu}\rangle$ if $h(U) \supseteq A \cup h(A)$. Indeed
$\langle x^{'\mu}\rangle$ are the coordinate functions associated with
the local chart $(V,\varphi)$ where $h(U)\subseteq V$ and $U\cap V \neq
\phi$. Now, since it is well known that under the above conditions $h_*
\partial/\partial x^\mu \equiv \partial/\partial x^{'\mu}$ and $h_*{\rm
d}x^\mu \equiv {\rm d}x^{'\mu}$, eqs.(\ref{4e1}), (\ref{4e3}) and
(\ref{4e4}) imply that
\begin{equation} \label{4e6}
(h_*{\bf T})_{\langle x^{'\mu} \rangle}(he) = {\bf T}_{\langle x^\mu
\rangle}(e) \, , 
\end{equation}
where ${\bf T}_{\langle x^\mu\rangle}(e)$ means the components of ${\bf
T}$ in the chart $\langle x^\mu\rangle$ at the event $e \in M$, {\em i.e.\/} 
${\bf T}_{\langle x^\mu \rangle}(e) = T^{\mu_1\ldots
\mu_r}_{\nu_1\ldots\nu_s} (x^\mu(e))$ and where
$\overline{T}^{'\mu_1\ldots \mu_r}_{\nu_1\ldots\nu_s} (x^{'\mu}(he))$
are the components of $\overline{{\bf T}}= h_* {\bf T}$ in the basis
$\{h_*\partial/\partial x^\mu = \partial/\partial x^{'\mu}\}$, $\{h_*
{\rm d}x^\mu={\rm d}x^{'\mu}\}$, at the point $h(e)$.
Then eq.(\ref{4e6}) reads
\begin{equation} \label{4e7}
{\overline{T}}^{'\mu_1\ldots \mu_r}_{\nu_1\ldots\nu_s} (x^{'\mu} (he))
=
T^{\mu_1\ldots
\mu_r}_{\nu_1\ldots\nu_s}
(x^\mu(e))
\end{equation}
or using eq.(\ref{4e5})
\begin{equation} \label{4e8}
{\overline{T}}^{'\mu_1\ldots \mu_r}_{\nu_1\ldots\nu_s}(x^{'\mu}(e)) =
(\Lambda^{-1})^{\mu_1}_{\alpha_1} \ldots
\Lambda^{\beta_s}_{\nu_s}
T^{'\alpha_1\ldots
\alpha_r}_{\beta_1\ldots\beta_s} (x^{'\mu}(h^{-1}e))
\end{equation}
where $\Lambda^{\mu}_{\alpha}= \partial x^{'\mu}/\partial
 x^{\alpha}$, etc.

In Section \ref{sec1} we already introduced the concept of inertial reference
frames $I \in \sec TU$, $U\subseteq M$ by
\begin{equation} \label{4e9}
g(I,I) =1 \ \ {\rm and} \ \ DI =0
\end{equation}
A general frame $Z$ satisfies $g(Z,Z)=1$, with $DZ \neq 0$. If $\alpha
=
g(Z,\ ) \in \sec T^*U$, it holds
\begin{equation} \label{4e10}
(D\alpha)_e = a_e \otimes \alpha_e + \sigma_e + \omega_e + \frac{1}{3}
\theta_e h_e, e
\in U
\subseteq M ,
\end{equation}
where $a = g(\mbox{A}, \ )$, $\mbox{A}=D_Z Z$ is the acceleration and
where $\omega_e$ is the
rotation tensor, $\sigma_e$ is the shear tensor, $\theta_e$ is the
expansion and $h_e = g_{|H_e}$ where
\begin{equation} \label{4e11}
T_eM = [Z_e]\oplus [H_e] \,  .
\end{equation}
$H_e$ is the rest space of an {\it instantaneous  observer} at $e$, {\em
i.e.\/} the pair $(e, Z_e)$. Also $h_e(X,Y) = g_e(p(p X, pY), \forall
X,Y \in T_e M$ and $p: T_eM \rightarrow H_e$. (For the explicit form of
$\omega, \sigma, \theta$, see \cite{11,39}).  From eqs.(\ref{4e9}) and
(\ref{4e10}) we see that an inertial reference frame has no
acceleration, has no rotation, no shear and no expansion.

We introduced also in Section \ref{sec1} the concept of a (nacs/$I$). A
(nacs/$I$)  $\langle x^\mu\rangle$ is said to be in the Lorentz gauge if
$x^\mu, \mu=0,1,2,3$ are the usual Lorentz coordinates and $I
=\partial/\partial x^0 \in \sec TM$.
We recall that it is a theorem that putting $I=e_0 =\partial/\partial
x^0$, there exist three other fields $e_i \in \sec TM$, such that
$g(e_i, e_i) =-1, \ i=1,2,3$, and $e_i =\partial/\partial x^i$. 

Now, let $\langle x^\mu \rangle$ be Lorentz coordinate functions as
above. We say that $\ell \in G_M$ is a {\it Lorentz mapping} if and only
if
\begin{equation} \label{4e12}
x^{'\mu}(e) = \Lambda^\mu_\nu x^\mu(e),
\end{equation}
where $\Lambda^\mu_\nu \in {\cal L}^\uparrow_+$ is a Lorentz
transformation. For
abuse of notation we denote the subset $\{\ell\}$ of $G_M$ such that
eq.(\ref{4e12}) holds true also by ${\cal L}^\uparrow_+ \subset G_M$.

When $\langle x^{\mu} \rangle$ are Lorentz coordinate functions,
$\langle x^{'\mu} \rangle$ are also Lorentz coordinate functions. In
this case we denote
\begin{equation} \label{4e13}
e_\mu = \partial/\partial x^\mu, \ e'_\mu = \partial/\partial x^{'\mu}
, \
\gamma_\mu = {\rm d}x^\mu , \ \gamma'_\mu = {\rm d}x^{'\mu} \; ;
\end{equation}
when $\ell \in {\cal L}^\uparrow_+ \subset G_M$ we say that $\ell_*{\bf
T}$ is the {\it Lorentz deformed version} of ${\bf T}$.

Let $h \in G_M$. If for a geometrical object ${\bf T}$ we have
\begin{equation} \label{4e14}
h_* {\bf T} ={\bf T},
\end{equation}
then $h$ is said to be a symmetry of ${\bf T}$ and the set of all $h
\in G_M$ such that eq.(\ref{4e13}) holds is said to be the symmetry
group of ${\bf T}$.
We can immediately verify that for $\ell \in {\cal L}^\uparrow_+
\subset G_M$
\begin{equation} \label{4e15}
\ell_* g = g , \ \ell_*D = D,
\end{equation}
{\em i.e.\/}, the special restricted orthochronous Lorentz group ${\cal
L}^\uparrow_+ $ is a symmetry group of $g$ and $D$.

In \cite{12} we maintain that a physical theory $\tau$ is
characterized by: 

\begin{description}
\item{(i)} the theory of a certain ``species of structure" in the
sense of Boubarki \cite{40};
\item{(ii)} its physical interpretation;
\item{(iii)} its present meaning and present applications.
\end{description}

We recall that in the mathematical exposition of a given physical
theory
$\tau$, the postulates or basic axioms are presented as definitions.
Such
definitions mean that the physical phenomena described by $\tau$ behave
in a certain way. Then, the definitions require more motivation
than the pure mathematical definitions. We call coordinative
definitions the physical definitions, a term introduced by
Reichenbach \cite{41}. It is necessary also to make clear that
completely convincing and genuine motivations for the coordinative
definitions cannot be given, since they refer to nature as a whole
and to the physical theory as a whole.

The theoretical approach to physics behind (i), (ii) and (iii) above is
then to admit the mathematical concepts of the ``species of structure"
defining $\tau$ as primitives, and define coordinatively the observation
entities from them.  Reichenbach assumes that ``{\it physical knowledge}
is characterized by the fact that concepts are not only defined by other
concepts, but are also coordinated to real objects". However, in our
approach, each physical theory, when characterized as a species of
structure, contains some implicit geometric objects, like some of the
reference frame fields defined above, that cannot in general  be
coordinated to real objects. Indeed it would be an absurd to suppose
that all the infinity of IRF that exist in $M$ must have a material
support.

We define a {\it spacetime} theory as a theory of a species of
structure such that, if Mod $\tau$ is the class of models of $\tau$,
then
each $\Upsilon \in$ Mod $\tau$ contains a substructure called spacetime
(ST). 
More precisely, we have

\begin{equation} \label{4e16}
\Upsilon = (\mbox{ST}, {\bf T}_1, \ldots, {\bf T}_m\}\; ,
\end{equation}
where ST can be a very general structure \cite{12}.  For what follows we
suppose that $\mbox{ST} = {\cal M} = (M, g, D)$ {\em i.e.\/}, that ST is
Minkowski spacetime. The ${\bf T}_i, i=1,\ldots,m$ are (explicit)
geometrical objects defined in $U \subseteq M$ characterizing the
physical fields and particle trajectories that cannot be geometrized in
$\Upsilon$. Here, to be geometrizable means to be a metric field or a
connection on $M$ or objects derived from these concepts, as {\em
e.g.\/}, the Riemann tensor or the torsion tensor in more general
theories.  The reference frame fields will be called the {\it implicit}
geometrical objects of $\tau$, since they are mathematical objects that
do not necessarily correspond to properties of a physical system
described by $\tau$.

Now, with the Clifford bundle formalism we can formulate in
${\cal C}\ell(M)$
all modern physical theories (see Section \ref{sec1}) including 
Einstein's gravitational theory \cite{1}. We introduce now the
Lorentz-Maxwell electrodynamics (LME) in ${\cal C}\ell(M)$ as a theory
of a
species
of structure. We say that LME has as model
\begin{equation} \label{4e17}
\Upsilon_{LME} = \langle
M,g,D,F,J,\{\varphi_i,m_i,e_i\}\rangle\end{equation}
where $(M,g,D)$ is Minkowski spacetime, $\{\varphi_i,m_i,e_i\}$,
$i=1,2,\ldots,N$ is the set of all charged particles, $m_i$ and $e_i$
being
the masses and charges of the  particles and $\varphi_i: I\!\!R \supset
I
\rightarrow M$ being the world lines of the particles characterized by
the
fact that if $\varphi_{i\ast} \in \sec TM$ is the velocity vector,
then $\check\varphi_i =
g(\varphi_{i*},\ ) \in \sec \Lambda^1(M) \subset \sec {\cal C}\ell(M)$
and
$\check\varphi_i.\check\varphi_i =1$. $F \in
\sec \Lambda^2(M) \subset\sec {\cal C}\ell(M)$ is the electromagnetic
field and $J
\in \sec \Lambda^1(M)\subset \sec {\cal C}\ell(M)$ is the current
density. The
proper
axioms of the theory are
\begin{equation} \label{4e18}
\begin{array}{c}
\partial F = J \, , \\
m_i D_{\varphi_{i*}} \check\varphi_i = e_i \check\varphi_i \cdot F \, . 
\end{array} 
\end{equation}
From a mathematical point of view it is a trivial result that
$\tau_{LME}$ has the following property: If $h \in G_M$ and if the set
of eqs.(5.16) has a solution $\langle F,J, (\varphi_i,m_i,e_i)\rangle$
in $U \subseteq M$ then $\langle h_*F, h_*J, (h_* \varphi_i, m_i, e_i)
\rangle $ is also a solution of eqs.(5.16) in $h(U)$. Since the result
is true for any $h \in G_M$ it is true for $\ell \in {\cal L}^\uparrow_+
\subset G_M$, {\em i.e.\/} for any Lorentz mapping.

We must now make it clear that $\langle F,J,
\{\varphi_i,m_i,e_i\}\rangle$ which
is
a solution of eq.(5.16) in $U$ can be obtained only by imposing
{\it mathematical boundary conditions} which we denote by $BU$. The
solution will be realizable in nature if and only if the mathematical
boundary conditions can be physically realizable. This is indeed a
nontrivial point \cite{12} for in particular it says to us that
even if $\langle h_*F, h_*J, \{h_*\varphi_i,m_i,e_i\}\rangle$ can be a
solution
of eqs.(5.16) with mathematical boundary conditions $Bh(U)$, it may
happen that $Bh(U)$ cannot be physically realizable in nature. The
following statement, denoted $PR_1$,  is usually presented \cite{12}
as the Principle of (Special) Relativity in active form:

\smallskip 

$PR_1$: Let $\ell \in {\cal L}^\uparrow_+ \subset G_M$. If for a
physical theory $\tau$ we have $\Upsilon \in {\rm Mod}\, \tau $, where
$\Upsilon = \langle M,g,D,{\bf T}_1,\ldots,{\bf T}_m\rangle$ is a
possible physical phenomenon, then $\ell_* \Upsilon = \langle M,g,D,
l_*{\bf T}_1$,\linebreak $\ldots,l_*{\bf T}_m\rangle$ is  also a
possible physical phenomenon.

\smallskip 

It is clear that {\em hidden\/} in $PR_1$ is the assumption that the
boundary conditions that determine $\ell_*\Upsilon$ are physically
realizable.  Before we continue we introduce the statement denoted
$PR_2$ known as the Principle of (Special) Relativity in passive form
\cite{12}.

\smallskip

$PR_2$: All inertial reference frames are physically equivalent
or indistinguishable.

\smallskip 

We now give a precise mathematical meaning to the above statement.

Let $\tau$ be a spacetime and let $\mbox{ST} = \langle M,g,D\rangle$ be
a substructure of Mod $\tau$ representing spacetime. Let $I \in \sec TU$
and $I' \in \sec TV, U, V \subseteq M$ be two inertial reference frames.
Let $(U,\eta)$ and $(V,\varphi)$ be two Lorentz charts of the maximal
atlas of $M$ that are naturally adapted respectively to $I$ and $I'$.
For $\langle x^\mu\rangle$ and $\langle x^{'\mu}\rangle$ the coordinate
functions associated with $(U,\eta)$ and $(V,\varphi)$ we have $I
=\partial/\partial x^0, I' = \partial/\partial x^{'0}$. 

\medskip 

{\em Definition.} Two inertial reference frames $I$ and $I'$ as
above are said to be physically equivalent according to $\tau$ if and
only if the following conditions are satisfied:

(i) $G_M \supset {\cal L}^\uparrow_+ \ni \ell: U \rightarrow \ell(U)
\subseteq V\, ,  \
x^{'\mu} = x^{\mu} \circ \ell^{-1} \Rightarrow I' = \ell_* I$

When $\Upsilon \in$ Mod $\tau$, $\Upsilon = \langle M,g,D, {\bf
T}_1,\ldots {\bf T}_m\rangle$, 
is such that $g$ and $D$ are defined over all $M$ and ${\bf T}_i 
\in \sec {\cal C}\ell (U) \subset \sec {\cal C}\ell (M)$, 
$i=1,\ldots,m$,  calling $o= \langle
g,D,{\bf T}_1,\ldots {\bf T}_m\rangle$, $o$ solves a set of differential
equations in $\eta(U) \subset I\!\!R^4$ with a given set of boundary
conditions denoted $b^{o\langle x^{\mu}\rangle}$, which we write as
\begin{equation} \label{4e19}
D^\alpha_{\langle x^\mu \rangle} (o_{\langle x^\mu \rangle})_e =0 \ ; \
b^{o\langle x^{\mu}\rangle} \ ; \ e \in U
\end{equation}
and we must have:

(ii) If $\Upsilon \in$ Mod $\tau \Leftrightarrow \ell_* \Upsilon \in $
Mod $\tau$,
then
necessarily
\begin{equation} \label{4e20}
\ell_*\Upsilon = \langle M,g,D, \ell_*{\bf T}_1,\ldots \ell_* {\bf
T}_m\rangle
\end{equation}
is defined in $\ell(U) \subseteq V$ and calling $\ell_* o\equiv
\{g,D,\ell_* {\bf T}_1,\ldots, \ell_* {\bf T}_m\}$ we must have
\begin{equation} \label{4e21}
D^\alpha_{\langle x^{'\mu} \rangle} (\ell_* o_{\langle x^{'\mu}
\rangle})_{|\ell e} =0 \ ; \  b^{\ell_* o\langle x^{'\mu}\rangle}
\ \ \ell e \in
\ell (U) \subseteq V .
\end{equation}
The system of differential equations (5.19) must have the same
functional form as  the system of differential equations (5.17) and
$b^{\ell_* o\langle x^{'\mu}\rangle}$ must be relative to $\langle
x^{'\mu}\rangle$  the same  as $b^{o\langle x^{\mu}\rangle}$ is relative
to $\langle x^{\mu}\rangle$ and if $b^{o\langle x^{\mu}\rangle}$ is
physically realizable then $b^{\ell_* o\langle x^{'\mu}\rangle}$ must
also be physically realizable.  We say under these conditions that $I
\sim I'$ and that $\ell_* o$ is the Lorentz version of  the phenomenon
described by $o$.

\medskip 

Since in the above definition $\ell_* \Upsilon= \langle M,g,D, \ell_*
T_1,\ldots,\ell_*T_m\rangle$, it follows that when $I \sim I'$, then
$\ell_* g=g, \ell_* D = D$ (as we already know) and this means that the
spacetime structure does not give a preferred status to $I$ or $I'$
according to $\tau$. 

\subsection{Proof that the existence of SEXWs implies a
breakdown of PR$_1$ and PR$_2$} \label{4s2}

We are now able to prove the statement presented in the beginning of
this Section, that the existence of SEXWs implies a breakdown of the
Principle of Relativity in both  its active ($PR_1$) and passive
($PR_2$) versions.

Let  $\ell \in {\cal L}^\uparrow_+ \subset G_M$ and let $F$,
$\overline{F} \in \sec \Lambda^2(M) \subset \sec {\cal C}\ell(M)$,
$\overline{F} = \ell_* F$.  Let $\overline F = \ell_*
F=R\check{F}R^{-1}$, where $\check{F}_e = (1/2) F_{\mu\nu}
(x^\mu(\ell^{-1}e)) \gamma^\mu \gamma^\nu$ and where $R \in \sec $
Spin$_+(1,3) \subset \sec {\cal C}\ell(M)$ is a Lorentz mapping, such
that $\gamma^{'\mu} =R\gamma^\mu R^{-1} = \Lambda^\mu_\alpha
\gamma^\alpha$, $\Lambda^\mu_\alpha \in {\cal L}^\uparrow_+$ and let
$\langle x^\mu\rangle$ and $\langle x^{'\mu}\rangle$ be Lorentz
coordinate functions as before such that $\gamma^\mu = {\rm d}x^\mu$,
$\gamma^{'\mu} = {\rm d}x^{'\mu}$ and $x^{'\mu} = x^\mu \circ
\ell^{-1}$.  We write
$$F_e = {\displaystyle\frac{1}{2}} F_{\mu\nu}(x^\mu(e))
\gamma^\mu\gamma^\nu\eqno{(5.22\mbox{a})}$$ \label{4e22a}
$$F_e = {\displaystyle\frac{1}{2}} F'_{\mu\nu}(x^{'\mu}(e))
\gamma^{'\mu}\gamma^{'\nu}\eqno{(5.22\mbox{b})}$$ \label{4e22b}
$$\overline F_e = {\displaystyle\frac{1}{2}} \overline
F_{\mu\nu}(x^\mu(e))
\gamma^\mu\gamma^\nu\eqno{(5.23\mbox{a})}$$ \label{4e23a}
$$\overline F_e = {\displaystyle\frac{1}{2}} \overline
F'_{\mu\nu}(x^{'\mu}(e))
\gamma^{'\mu}\gamma^{'\nu}. \eqno{(5.23\mbox{b})}$$ \label{4e23b}
From (5.22a) and (5.22b) we get that
\setcounter{equation}{23}
\begin{equation} \label{4e24}
F'_{\alpha\beta} (x^{'\mu}(e)) =
(\Lambda^{-1})^\mu_\alpha(\Lambda^{-1})^\nu_\beta
F_{\mu\nu} (x^\mu(e)) .
\end{equation}
From (5.23a) and (5.23b) we also get
\begin{equation} \label{4e25}
\overline F_{\alpha\beta} (x^{\mu}(e)) = \Lambda^\mu_\alpha
\Lambda^\nu_{\beta} F_{\mu\nu}
(x^\mu(\ell^{-1}e)) \, .
\end{equation}

Now, suppose that $F$ is a superluminal solution of Maxwell equation, in
particular a SEXW as discussed in Section 3.  Suppose that $F$ has been
produced in the inertial frame $I$ with $\langle x^\mu\rangle$ as
(nacs/$I$), with the physical device described in Section 3. $F$ is then
traveling with speed $c_1=1/\cos \eta$ in the negative $z$-direction and
being generated  in the plane $z =0$, will travel to the future in
spacetime, according to the observers in $I$.  Now, there exists $\ell
\in {\cal L}^\uparrow_+$ such that $\ell_* F = \overline F = RFR^{-1}$
will be a solution of Maxwell equations and such that if the velocity
1-form of $F$ is  $v_F=(c_1^2 -1)^{-1/2}(1,0,0,-c_1)$, then the velocity
 1-form of $\overline F$ is $v_{\overline{F}} = (c^{'2}_{1}
-1)^{-1/2}(-1,0,0,-c'_1)$, with $c'_1 >1$, i.e, $v_{\overline{F}}$ is
pointing to the past. As its is well known $\overline{F}$ carries
negative energy density according to the observers in the $I$ frame.

We then arrive at the conclusion that to assume the  validity of $PR_1$
is to assume the physical possibility of sending to the past waves
carrying negative energy. This seems to the authors an impossible task,
and the reason is that there do not exist physically realizable boundary
conditions that permit the observers in $I$ to launch $\overline F$ in
spacetime and such that it travels to its own past.

We now show that there is also a breakdown of $PR_2$, {\em i.e.}, that
it is not true that all inertial frames are physically equivalent.
Suppose we have two inertial frames $I$ and $I'$ as above i.e, $I
=\partial/\partial x^0$, $I' = \partial/\partial x^{'0}$.  Suppose that
$F$ is a SEXW which can be launched in $I$ with velocity 1-form as above
and suppose $\overline F$ is a SEXW built in $I'$ at the plane $z'=0$
and with velocity 1-form relative to $\langle x^{'\mu} \rangle$ given by
$v_{\overline F} = v^{'\mu} \gamma'_{\mu}$ and
\begin{equation} \label{4e26}
v_{\overline F} = \biggl(\frac{1}{\sqrt{c^{2}_1-1}}, 0,0, -
\frac{c_1}{\sqrt{c^{2}_1-1}}\biggr)
\end{equation}
If $F$ and $\overline F$ are related as above we see  that $\overline
F$, which has positive energy and is traveling to the future according
to $I'$, {\it can be sent} to the past of the observers at rest in the
$I$ frame. Obviously this is impossible and we conclude that $\overline
F$ is not a physically realizable phenomenon in nature.  It cannot be
realized in $I'$ but $F$ can be realized in $I$. It follows that $PR_2$
does not hold.

If the set of inertial reference frames are not equivalent then there
must exist a fundamental reference frame. Let
$I \in \sec TM$ be the fundamental frame. If
$I'$ is moving with speed $V$ relative to $I$, {\em i.e.\/} 
\begin{equation} \label{4e27}
I' = \frac{1}{\sqrt{1-V^2}} \frac{\partial}{\partial t} -
\frac{V}{\sqrt{1-V^2}}\frac{\partial}{\partial z} \; ,
\end{equation}
then, if observers in $I'$ are equipped with a generator of SEXWs and if
they prepare their apparatus in order to send SEXWs with different
velocity 1-forms in all possible directions in spacetime, they will find
a particular velocity 1-form in a given spacetime direction in which the
device stops working. A simple calculation yields then, for the
observers in $I'$, the value of $V$! 

In \cite{42} Recami argued that the Principle of Relativity continues to
hold true even if superluminal phenomena exist in nature. In this theory
of tachyons there exists, of course, a situation completely analogous to
the one described  above (called the Tolman-Regge paradox), and
according to Recami's view $PR_2$ is valid because $I'$ must interpret
$\overline F$ a being an anti-SEXW carrying positive energy and going
into the future according to him.  In his theory of tachyons Recami was
able to show that  the dynamics of tachyons implies that no detector
{\it at rest} in  $I$ can detect a tachyon (the same would be valid for
a SEXW like $\overline{F}$) sent by $I'$ with velocity 1-form given by
eq.(\ref{4e26}). Thus he claimed that $PR_2$ is true. At first sight the
argument seems good, but it is at least incomplete. Indeed, a detector
in $I$ does not need to be at rest in $I$.  We can imagine a detector in
periodic motion in $I$ which can absorb the $\overline F$ wave generated
by $I'$ if this was indeed possible.  It is enough for the detector to
have relative to $I$ the speed $V$ of the $I'$ frame in the appropriate
direction at the moment of absorption.  This simple argument shows that
there is no salvation for $PR_2$ (and for $PR_1$) if superluminal
phenomena exist in nature. Our argumentation is endorsed by Barashenkov
and Yur'iev \cite{44}.

The attentive reader at this point probably has the following question
in his/her mind: How could the authors start with Minkowski spacetime,
with equations carrying the Lorentz symmetry and yet arrive at the
conclusion that $PR_1$ and $PR_2$ do not hold?  The reason is that the
Lorentzian structure of $\langle M,g,D\rangle$ can be seen to exist
directly from the Newtonian spacetime structure as proved in \cite{45}.
In this paper, Rodrigues and  collaborators show that even if ${\cal
L}^\uparrow_+$ is not a symmetry group of Newtonian dynamics it is a
symmetry group of the only possible coherent formulation of
Lorentz-Maxwell electrodynamic theory compatible with experimental
results that is possible to formulate in the Newtonian
spacetime.\footnote{We recall that Maxwell equations have, as  is well
known, many symmetry groups besides ${\cal L}^\uparrow_+$.}

We finish calling to the reader's attention that there are some
experiments reported in the literature which suggest also a breakdown of
$PR_2$ for the roto-translational motion of solid bodies. A discussion
and references can be found in \cite{13}. A coherent spacetime model
which can accommodate superluminal phenomena has been recently proposed
by Matolcsi and Rodrigues \cite{46}.

\section{CONCLUSIONS} \label{sec5} 

In this paper we presented a unified theory showing that the homogeneous
wave equation, the Klein-Gordon equation, Maxwell equations and the
Dirac and Weyl equations have solutions with the form of  undistorted
progressive waves (UPWs) of arbitrary speeds \protect{$0 \leq v
<\infty$}.  We exhibit also some subluminal and superluminal solutions
of Maxwell equations. We showed that subluminal solution can in
principle be used to model purely electromagnetic particles.

The possible existence of superluminal electromagnetic waves implies in
a breakdown of the Principle of Relativity. It is important to recall
here that exact Lorentz symmetry can be preserved in an abstract
mathematical level through the ingenious construction of Santilli's
isominkowskian spaces (see [51-55]). Santilli's theory is important,
{\em e.g.\/} for situations involving the hadronic medium, where
superluminal velocities can occur. We observe that besides the
fundamental theoretical implications, the practical implications of the
existence of UPWs solutions of the main field equations of theoretical
physics (and their finite aperture realizations) are very important.
This practical importance ranges from applications in ultrasound medical
imaging to the project of electromagnetic bullets and new communication
devices \cite{54b}. Also, we would like to conjecture that the existence
of subluminal and superluminal solutions of the Weyl equation may be
important to solve some of the mysteries associated with neutrinos.
Indeed, if neutrinos can be produced in subluminal or superluminal modes
(see \cite{53,54}  for some experimental evidence concerning
superluminal neutrinos)  they can eventually escape detection on earth
after leaving the sun. Moreover, for neutrinos in a subluminal or
superluminal mode it would be possible to define a kind of ``effective
mass". Recently some cosmological evidences that neutrinos have a
non-vanishing mass have been discussed, {\em e.g.\/}, by Primack et al
\cite{55}. One such ``effective mass" could be responsible for those
cosmological evidences, and in such a way we can still have a
left-handed neutrino since it would satisfy the Weyl equation. We shall
discuss more this issue in another publication.

\bigskip 

\noindent {\bf Acknowledgments}
The authors are grateful  for CNPq, FAPESP and FINEP for partial
financial support. We would like also to thank  Professors  U.
Barttocci, V. Barashenkov,  G. Nimtz, E. Recami and R.M. Santilli and
 Drs. M. C. Duffy, E.C. de Oliveira,  Q.A.G. de Souza,  J. Vaz Jr.  and
W. Vieira for many valuable discussions. A special acknowledgment is due
to Dr. J.-Y. Lu who did the measurements of the speeds of the acoustic
Bessel pulses and the $X$-waves discussed in \cite{5,6}.


\begin{thebibliography}{99} 
 
\bibitem{1} W.A. Rodrigues Jr., Q.A.G. de Souza. The Clifford bundle
and the nature of the gravitational field, {\em Found. of Phys.\/} {\bf
23}, 1465--1490 (1993).

\bibitem{2} W.A. Rodrigues Jr., Q.A.G. de Souza, J. Vaz Jr..  
Spinor fields and superfields as equivalence classes of exterior 
algebra fields, {\em in\/} R. Ablamowicz and P. Lounesto (eds.), {\em 
Clifford algebras and spinor structures\/}, 177--198, Kluwer Acad. Pub.,
Dordrecht (1995).

\bibitem{3} W.A. Rodrigues Jr., Q.A.G. de Souza, J. Vaz Jr., P.
Lounesto. Dirac Hestenes spinor fields in Riemann-Cartan spacetime,
{\em Int. J. of Theor. Phys.\/} {\bf 35}, 1849--1900 (1996).

\bibitem{4} R. Courant, D. Hilbert. {\it Methods of mathematical
physics\/}. Vol. II. John Wiley and Sons, New York (1966).

\bibitem{5} J.-Y. Lu, W. A. Rodrigues Jr.. What is the speed of a sound
wave in a homogeneous medium, RP~25/96, IMECC-UNICAMP, subm. for
publication.

\bibitem{6} W. A. Rodrigues Jr., J.-Y. Lu. On the existence of
undistorted progressive waves (UPWs) of arbitrary speeds $0 \leq v <
\infty$ in nature, RP~12/96, IMECC-UNICAMP, in publication in {\em
Found. of Phys.\/} {\bf 27}, no. 3 (1997). 

\bibitem{7} J.-Y. Lu, J.F. Greenleaf. Ultrasonic nondiffracting
transducer for medical imaging, {\em IEEE Trans. Ultrason. Ferroelec.
Freq. Contr.\/} {\bf 37}, 438--477 (1990).

\bibitem{8} J.-Y. Lu, J.F. Greenleaf. Pulse-echo imaging using a
nondiffracting beam transducer, {\em Ultrasound Med. Biol.\/} {\bf 17},
265--281 (1991).

\bibitem{9} I. Porteous. {\em Topological geometry}. Van Nostrand,
London (1969).

\bibitem{10} P. Lounesto. Clifford algebras and Hestenes spinors,
{\em Found. Phys.\/} {\bf 23}, 1203--1237 (1993).

\bibitem{11} W.A. Rodrigues Jr., M.A.F. Rosa. The meaning of time in
relativity and Einstein's later view of the twin paradox, {\em Found.
Phys.\/} {\bf 19}, 705--724 (1989).

\bibitem{12} W.A. Rodrigues Jr., M.E.F. Scanavini, L.P. de Alc\^antara.
Formal structures, the concepts of covariance, invariance, equivalent
reference frames and the principle of relativity, {\em Found. Phys.
Lett.\/} {\bf 3}, 59--79 (1990).

\bibitem{13} W.A. Rodrigues Jr., J. Tiomno. On experiments to detect
possible failures of relativity theory, {\em Found. Phys.\/} {\bf 15},
995--961 (1985).

\bibitem{14} J.A. Stratton. {\it Electromagnetic theory\/}. McGraw-Hill,
New York (1941).

\bibitem{15} W.A. Rodrigues Jr., J. Vaz Jr.. Subluminal and superluminal
solutions in vacuum of the maxwell equations and the massless Dirac
equation. RP 44/95 IMECC-UNICAMP, in publication in {\em Advances in
Appl. Clifford Algebras}.

\bibitem{16b} J.R.R. Zeni, W.A. Rodrigues Jr.. A thoughtful study of
Lorentz transformations by Clifford algebras, {\em Int. J. Mod. Phys.\/}
{\bf A7}, 1793--1817 (1992).

\bibitem{16} H. Bateman. {\it Electrical and optical motion}. Cambridge
Univ. Press, Cambridge (1915).

\bibitem{17} A.O. Barut, H.C. Chandola. Localized tachyonic wavelet
solutions of the wave equation, {\em Phys. Lett.\/} {\bf A180}, 5--8
(1993).

\bibitem{18} J.-Y. Lu, J.F. Greenleaf. Nondiffracting $X$-waves -
exact solutions to free-space scalar wave equation  and their finite
aperture realizations, {\em IEEE Transact. Ultrason. Ferroelec. Freq.
Contr.\/} {\bf 39}, 19--31 (1992).

\bibitem{19} J.-Y. Lu, Z. Hehong, J.F. Greenleaf. Biomedical ultrasound
beam forming, {\em Ultrasound in Med. \& Biol.\/} {\bf 20}, 403--428
(1994).

\bibitem{20} W. Band. Can information be transfered faster than light~?
I. A gedanken device for generating electromagnetic wave packets with
superoptic group velocity, {\em Found Phys.\/} {\bf 18}, 549--562
(1988).

\bibitem{21} W. Band. Can information be transfered faster than light~?
II. The relativistic Doppler effect on electromagnetic wave packets with
suboptic and superoptic group velocities, {\em Found Phys.\/} {\bf 18},
625--638 (1988).

\bibitem{22} J.-Y. Lu, J.F. Greenleaf. Experimental verification of
nondiffracting $X$-wave, {\em IEEE Trans. Ultrason. Ferroelec. Freq.
Contr.\/} {\bf 39}, 441--446, (1992).

\bibitem{23} R. Donnelly, R. Ziolkowski. A method for constructing
solutions of homogeneous partial differential equations: localized
waves, {\em Proc. R. Soc. London\/} {\bf A437}, 673--692 (1992).

\bibitem{24} R. Donnelly, R. Ziolkowski. Designing localized waves, {\em
Proc. R. Soc. London\/} {\bf A460}, 541--565 (1993).

\bibitem{25} R.W. Ziolkowski. Exact solutions of the wave equation with
complex source locations, {\em J. Math. Phys.\/} {\bf 26}, 861--863
(1985).

\bibitem{26} J.N. Brittingham. Focus waves modes in homogeneous
Maxwell's equations: transverse electric mode, {\em J. Appl. Phys.\/}
{\bf 54}, 1179 (1983).

\bibitem{27a} A.O. Barut. {\em Electrodynamics and classical theory of
fields and particles\/}, Macmillan, NY (1964), Dover, NY (1980).

\bibitem{27b} A. Einstein. {\em Sitzungsberichte der Preussischen Akad.
D.  Wissenschaften\/} (1919), translated in H.A. Lorentz, A. Einstein,
H. Minkowski and H. Weyl, {\it The principle of relativity}, Dover, N.Y.
(1952).

\bibitem{28} H. Poincar\'e. Sur la dynamique de l'electron, 
{\em R.C. Circ. Mat. Palermo\/} {\bf 21}, 129--175 (1906).

\bibitem{29} P. Ehrenfest. Die translation deformierbarer elektron und
der fl\"achensatz, {\em Ann. Phys. (Leipzig)\/} {\bf 23}, 204--205
(1907).

\bibitem{30} T. Waite. The relativistic Helmholtz theorem and solitons,
{\em Phys. Essays\/} {\bf 8}, 60--70 (1995).

\bibitem{31} T.Waite, A.O.Barut, J.R. Zeni. The purely electromagnetic
electron re-visited, in publ. in J. Dowling (ed.) {\it Electron theory
and quantum electrodynamics}. Nato Asi Series Volume, Plenum Press,
London (1995).

\bibitem{32} D. Hestenes. {\em Spacetime algebra\/}, Gordon and 
Breach, NY (1984). 

\bibitem{33a} A.M. Shaarawi. An electromagnetic charge-current basis for
the de Broglie double solution, preprint Dep. Eng. Phys. and Math.,
Cairo Univ., Egypt (1995).

\bibitem{33b} D. Reed. Archetypal vortex topology in nature, {\em Spec.
Sci. and Tech.\/} {\bf 17}, 205--224 (1994).

\bibitem{34} J. Vaz Jr., W.A. Rodrigues Jr.. On the equivalence of
Maxwell and Dirac equations and quantum mechanics, {\em Int. J. Theor.
Phys.\/} {\bf 32}, 945--958 (1993).

\bibitem{35} J. Vaz Jr., W.A. Rodrigues Jr.. Maxwell and Dirac theories
as an  already unified theory, RP 45/95 IMECC-UNICAMP, in publ. in {\em
Advances in Appl. Clifford Algebras\/}.

\bibitem{35b} J.Y. Lu, J.F. Greenleaf. Non diffracting electromagnetic
$X$-waves, preprint Biodynamics Research Unit, Mayo Clinic and
Foundation, Rochester (1991).  

\bibitem{37b} J.Y. Lu, J.F. Greenleaf. Limited diffraction solutions to
Maxwell and Schr\"odinger equations, preprint Biodynamics Research Unit,
Mayo Clinic and Foundation, Rochester (1995).

\bibitem{36} M.W. Evans. Classical relativistic theory of the
longitudinal ghost fields in electromagnetism, {\em Found. Phys.\/} {\bf
24}, 1671--1688 (1994).

\bibitem{37} L. de Broglie. {\em Ondes electromagn\'etiques et
photons\/}, Gauthier-Villars, Paris (1968).

\bibitem{38} B. Jancewicz. {\it Multivectors and Clifford algebras in
electrodynamics}, World Sci., Singapore (1988).

\bibitem{39} R.K. Sachs, H. Wu. {\it General relativity for
mathematicians}, Springer, New York (1977).

\bibitem{40} N. Bourbaki. {\it Th\'eorie des ensembles}, 
Hermann, Paris (1957).

\bibitem{41} H. Reichenbach. {\it The philosophy of space and time},
Dover, New York, 1958.

\bibitem{42} E. Recami. Classical tachyons and applications, {\em
Riv. N. Cimento\/} {\bf 9}, 1--178 (1986).

\bibitem{44} V.S. Barashenkov, M.Z. Yur'iev. {\em Hadronic J.\/} {\bf
18}, 433--450 (1996).

\bibitem{45} W.A. Rodrigues Jr., Q.A.G. de Souza, Y. Bozhkov. The
mathematical structure of newtonian spacetime: classical dynamics and
gravitation, {\em Found. Phys.\/} {\bf 25}, 871--924 (1995).

\bibitem{46} T. Matolcsi, W.A. Rodrigues Jr.. Spacetime model 
with superluminal phenomena, RP 27/96, IMECC-UNICAMP, subm. for 
publication. 

\bibitem{47} R.M. Santilli. Lie isotopic lifting of special relativity
for extended particles, {\em Lett. N. Cimento\/} {\bf 37}, 545--555
(1983).

\bibitem{48} R.M. Santilli. Nonlinear, nonlocal and noncanonical
isotopies of the Poincar\'e symmetry, {\em J. Moscow Phys. Soc.\/} {\bf
3}, 255--280 (1993).

\bibitem{49} R.M. Santilli. Limitations of the special and general
relativities and their isotopic generalizations, {\em Chinese J. of
Syst. Eng. \& Electr.\/} {\bf 6}, 157--176 (1995).


\bibitem{51} R.M. Santilli. {\em Elements of hadronic mechanics\/},
Vols. I and II (second ed.), Naukora Dumka Publ., Ukraine Acad. Sci.,
Kiev (1995).

\bibitem{52} R.M. Santilli. {\em Isospecial relativity with applications
to quantum gravity, antigravity and cosmology\/}, Balkan Geom. Press,
Budapest (in press).

\bibitem{54b} R.W. Ziolkowski. Localized transmission of electromagnetic
energy, {\em Phys. Rev.\/} {\bf A39}, 2005--2033 (1989).

\bibitem{53} E.W. Otten. Squeezing the neutrino mass with new
instruments, {\em Nucl. Phys. News\/} {\bf 5}, 11--16 (1995).

\bibitem{54} E. Gianetto {\em et al\/}, Are neutrinos faster than light
particles~?, {\em Phys. Lett.\/} {\bf B178}, 115--118 (1986).

\bibitem{55} J.R. Primack, J. Holtzman, A. Klypin, D.O.  Caldwell,
Cold+hot dark matter cosmology with $m(\nu_\mu) \simeq m(\nu_\tau)
\simeq 2.4 \,\mbox{eV}$, {\em Phys. Rev. Lett.\/} {\bf 74}, 2160 (1995).

\end{thebibliography}
\end{document}